\documentclass[reqno]{amsart}
\usepackage[lite]{amsrefs}
\usepackage{amssymb}
\usepackage{bm}
\usepackage[all,cmtip]{xy}

\newcommand{\T}{\mathbb{T}}

\newcommand{\R}{\mathbb{R}}
\newcommand{\C}{\mathbb{C}}
\newcommand{\Z}{\mathbb{Z}}
\newcommand{\N}{\mathbb{N}}

\newcommand{\E}{\mathbb{E}}

\newcommand{\Om}{\Omega}
\newcommand{\eps}{\varepsilon}

\newcommand{\mcN}{\mathcal{N}}
\newcommand{\mcNt}{\mcN_\tau}
\newcommand{\mcA}{\mathcal{A}}
\newcommand{\mcF}{\mathcal{F}}
\newcommand{\mcW}{\mathcal{W}}
\newcommand{\mcM}{\mathcal{M}} 
\newcommand{\mcWt}{\mathcal{W}_\tau}
\newcommand{\mfh}{\mathfrak{h}}
\newcommand{\mcDe}{\mathcal{D}_{\varepsilon}}
\newcommand{\mcTe}{\mathcal{T}_{\varepsilon}}
\newcommand{\mcD}{\mathcal{D}}

\newcommand{\mcS}{\mathcal{S}}
\newcommand{\mbt}{\mathbf{t}}
\newcommand{\mbx}{\mathbf{x}}
\newcommand{\mby}{\mathbf{y}}
\newcommand{\mcVd}{\mathcal{V}_\delta}
\newcommand{\mcV}{\mathcal{V}}

\numberwithin{equation}{section}

\theoremstyle{plain}

\theoremstyle{definition}

\theoremstyle{remark}

\usepackage{graphicx, color}

\definecolor{darkred}{rgb}{0.9,0,0.3}
\definecolor{darkblue}{rgb}{0,0.3,0.9}

\newcommand{\vertiii}[1]{{\left\vert\kern-0.25ex\left\vert\kern-0.25ex\left\vert #1 
		\right\vert\kern-0.25ex\right\vert\kern-0.25ex\right\vert}}

\theoremstyle{plain} 
\newtheorem{theorem}{Theorem}[section]
\newtheorem*{theorem*}{Theorem}
\newtheorem{lemma}[theorem]{Lemma}
\newtheorem*{lemma*}{Lemma}
\newtheorem{corollary}[theorem]{Corollary}
\newtheorem*{corollary*}{Corollary}
\newtheorem{proposition}[theorem]{Proposition}
\newtheorem*{proposition*}{Proposition}

\newtheorem*{conjecture*}{Conjecture}

\theoremstyle{definition} 
\newtheorem{definition}[theorem]{Definition}
\newtheorem*{definition*}{Definition}

\newtheorem*{example*}{Example}
\newtheorem{remark}[theorem]{Remark}
\newtheorem*{remark*}{Remark}
\newtheorem{assumption}[theorem]{Assumption}
\newtheorem*{assumption*}{Assumption}

\begin{document}
\title[Derivation of Gibbs measures for the 1D focusing cubic NLS]{A microscopic derivation of Gibbs measures for the 1D focusing cubic nonlinear Schr\"odinger equation}
\author{Andrew Rout}
\address{University of Warwick, Mathematics Institute, Zeeman Building, Coventry CV4 7AL, United Kingdom.}
\email{A.Rout@warwick.ac.uk.}
\author{Vedran Sohinger}
\address{University of Warwick, Mathematics Institute, Zeeman Building, Coventry CV4 7AL, United Kingdom.}
\email{V.Sohinger@warwick.ac.uk.}

\maketitle

\begin{abstract}
In this paper, we give a microscopic derivation of Gibbs measures for the focusing cubic nonlinear Schr\"{o}dinger equation on the one-dimensional torus from many-body quantum Gibbs states. Since we are not making any positivity assumptions on the interaction, it is necessary to introduce a truncation of the mass in the classical setting and of the rescaled particle number in the quantum setting. Our methods are based on a perturbative expansion of the interaction, similarly as in \cite{FKSS17}. Due to the presence of the truncation, the obtained series have infinite radius of convergence. We treat the case of bounded, $L^1$ and delta function interaction potentials, without any sign assumptions. Within this framework, we also study time-dependent correlation functions. This is the first such known result in the focusing regime.
\end{abstract}


\section{Introduction}
\subsection{Setup}
Let $\mathfrak{h}$ be a Hilbert space with a given Hamiltonian function $H \in C^{\infty}(\mathfrak{h})$ and Poisson bracket $
\{\cdot,\cdot\}:C^{\infty}(\mathfrak{h}) \times C^{\infty}(\mathfrak{h}) \rightarrow C^{\infty}(\mathfrak{h})$. The \emph{Gibbs measure} associated with the Hamiltonian $H$ and Poisson bracket $\{\cdot,\cdot\}$ is formally-defined as the probability measure on $\mathfrak{h}$ given by
\begin{equation}
\label{Gibbs_formal}
d \mathbb{P}_{\mathrm{Gibbs}}(\varphi):=\frac{1}{z_{\mathrm{Gibbs}}}\,e^{-H(\varphi)}\,d\varphi\,.
\end{equation}
Here, $d \varphi$ denotes Lebesgue measure on $\mathfrak{h}$ (which is ill-defined when $\mathfrak{h}$ is infinite-dimensional) and $z_{\mathrm{Gibbs}}$ is the \emph{partition function}, i.e.\ the normalisation constant which makes $d \mathbb{P}_{\mathrm{Gibbs}}$ into a probability measure on $\mathfrak{h}$. The problem of the rigorous construction of measures of type \eqref{Gibbs_formal} was first considered in the constructive quantum field theory literature. For an overview, see the classical works \cite{Glimm_Jaffe,Nelson,Nelson2,Simon74} and the later works \cite{Barashkov_Gubinelli,Brydges_Frohlich_Sokal,Brydges_Slade,Carlen_Frohlich_Lebowitz,Gubinelli_Hofmanova,LRS,McKean_Vaninsky1,McKean_Vaninsky2,Oh_Sosoe_Tolomeo}, as well as the references therein.

When $H$ is not positive-definite, it is sometimes not possible to define \eqref{Gibbs_formal} with finite normalisation constant $z_{\mathrm{Gibbs}}$, as formally one can have $\int e^{-H(\varphi)}\,d\varphi=\infty$. Instead, one considers a modification of \eqref{Gibbs_formal} given by
\begin{equation}
\label{Gibbs_formal_truncated}
d \mathbb{P}_{\mathrm{Gibbs}}^{f}(\varphi):=\frac{1}{z_{\mathrm{Gibbs}}^{f}}\,e^{-H(\varphi)}\,f(\|\varphi\|_{\mathfrak{h}}^2)\,d\varphi\,,
\end{equation}
where $f$ is a suitable cut-off function in $C_c^{\infty}(\R)$, and $z_{\mathrm{Gibbs}}^{f}$ is a normalisation constant that makes \eqref{Gibbs_formal_truncated} a probability measure on $\mathfrak{h}$. In general, when $H$ is not positive-definite, we say that we are in the \emph{focusing} (or \emph{thermodynamically unstable}) regime.

In this paper, we fix the spatial domain $\T \equiv \T^1 \equiv \R/\Z \equiv [-\frac{1}{2},\frac{1}{2})$ to be the one-dimensional torus\footnote{Some of our results generalise to other domains; see Remark \ref{Remark_AppendixB} (3) below. For simplicity, we work on $\T^1$.}. We henceforth consider the Hilbert space $\mathfrak{h}:=L^2(\T; \C) \equiv L^2(\T)$. Let us now define the precise Hamiltonian. We make the following assumption.
\begin{assumption}[The interaction potential]
\label{w_definition}
We consider an {\it interaction potential} which is of one of the following types. 
\begin{itemize}
\item[(i)] $w: \T \rightarrow \R$ is even and belongs to $L^1(\T)$.
\item[(ii)] $w=-\delta$, where $\delta$ is the Dirac delta function.
\end{itemize}
Let us note that, in Assumption \ref{w_definition}, we do not assume any conditions on the sign of $w$ or the sign of $\hat{w}$ (pointwise almost everywhere). 
\end{assumption}

With $w$ as in Assumption \ref{w_definition}, the Hamiltonian that we consider is given by
\begin{equation}
\label{integral_Hamiltonian_definition}
H(\varphi) := \int dx \, \left(|\nabla \varphi(x)|^2 + \kappa |\varphi(x)|^2\right) + \frac{1}{2}\int dx\,dy\, |\varphi(x)|^2\,w(x-y)|\varphi(y)|^2\,.
\end{equation}
In \eqref{integral_Hamiltonian_definition}, and throughout the sequel, we fix $\kappa>0$ to be the \emph{(negative) chemical potential} and we write $\int dx := \int_{\T} dx$. 
On the space of fields $\varphi: \T \rightarrow \C$, we consider a Poisson bracket defined by
\begin{equation}
\label{Poisson_bracket}
\{\varphi(x),\overline{\varphi}(y)\} = i\delta(x-y), \quad \{\varphi(x),\varphi(y)\} = \{\overline{\varphi}(x),\overline{\varphi}(y)\} = 0\,.
\end{equation}

We note that, by Assumption \ref{w_definition}, the Hamiltonian \eqref{integral_Hamiltonian_definition} is not necessarily positive-definite. Hence, when studying the associated Gibbs measure, one has to use the modification given by \eqref{Gibbs_formal_truncated}, instead of \eqref{Gibbs_formal}. This setup was previously used in \cite{Bou94,Carlen_Frohlich_Lebowitz,LRS}.

The Hamiltonian equation of motion associated with Hamiltonian \eqref{integral_Hamiltonian_definition} and Poisson bracket 
\eqref{Poisson_bracket} is the time-dependent nonlocal {\it nonlinear Schr\"odinger equation} (NLS)
\begin{equation}
\label{NLS1}
i \partial_t \varphi(x) = (-\Delta +\kappa) \varphi(x) + \int dy \, |\varphi(y)|^2 \, w(x-y) \varphi(x)\,.
\end{equation}
Here, we abbreviate the notation $\varphi(x) \equiv \varphi(x,t)$ with $\varphi: \T \times \R \rightarrow \C$.
For $w \in L^1$, as in Assumption \ref{w_definition} (i), one usually refers to \eqref{NLS1} as the {\it Hartree equation}. We will also consider the {\it focusing local cubic NLS}
\begin{equation}
\label{NLS_local_focus}
i \partial_t \varphi(x) = (-\Delta+\kappa) \varphi(x) - |\varphi(x)|^2\varphi(x)\,,
\end{equation}
which corresponds to \eqref{NLS1} with $w = -\delta$, as in Assumption \ref{w_definition} (ii). 
We refer to\footnote{When one has suitable positivity (in other words \emph{defocusing}) assumptions on $w$, the analysis of the problem we are considering for \eqref{NLS1} has already been done in \cite{FKSS17}; see Section \ref{Previously known results} below for an overview. Our main interest lies in the case when these assumptions are relaxed, which we refer to as the \emph{focusing} regime.} \eqref{NLS1} and \eqref{NLS_local_focus} as the \emph{focusing cubic nonlinear Schr\"{o}dinger equation (NLS)}.

The arguments in \cite{Bou93} show that the focusing cubic NLS \eqref{NLS1}--\eqref{NLS_local_focus} is globally well-posed for initial data in $\mathfrak{h} \equiv L^2(\T)$. 
In particular, there exists a well-defined solution map $S_t$ that maps any initial data $\varphi_0 \in \mathfrak{h}$ to the solution at time $t$ given by
\begin{equation}
\label{flow_map}
\varphi (\cdot) \equiv \varphi(\cdot, t) := S_t \varphi_0 (\cdot)  \in \mathfrak{h}\,.
\end{equation}
Moreover, $\|S_t \varphi_0\|_{\mathfrak{h}}=\|\varphi_0\|_{\mathfrak{h}}$.

Given the measure $d \mathbb{P}_{\mathrm{Gibbs}}^{f}$ as in \eqref{Gibbs_formal_truncated} and the time evolution as in \eqref{flow_map}, one can consider the corresponding \emph{time-dependent correlation functions}. Namely, for $m \in \N^*$, times $t_1,\ldots,t_m \in \R$, and  functions $X^1, \ldots, X^m \in C^{\infty}(\mathfrak{h})$, we define
\begin{equation}
\label{time_dependent_correlations}
\mathcal{Q}_{\mathrm{Gibbs}}^{f}(X^1,\ldots,X^m;t_1,\ldots,t_m):=\int d \mathbb{P}_{\mathrm{Gibbs}}^{f}(\varphi)\,X^1(S_{t_1}\varphi)\,\cdots \, X^m(S_{t_m}\varphi)\,,
\end{equation}
which we call the \emph{$m$-particle time-dependent correlation associated with $H$ and $X^j,t_j$, $j=1,\ldots,m$}.

The main goal of this paper is to show that one can obtain \eqref{time_dependent_correlations} as a \emph{mean-field} limit of corresponding many-body quantum objects, which we henceforth refer to as a \emph{microscopic derivation}. We do this in two steps.
\begin{itemize}
\item[(i)] \textbf{Step 1:} Analysis of the \emph{time-independent problem}, i.e.\ when 
\\ $t_1=\cdots=t_m=0$.
\item[(ii)] \textbf{Step 2:} Analysis of the \emph{time-dependent problem}. This is the general case.
\end{itemize}

The precise results are stated in Section \ref{Statement of the results}. In Section \ref{The Classical Problem}, we define the objects with which we work in the classical setting. In Section \ref{The Quantum Problem}, we define the objects with which we work in the quantum setting.

\subsection{The Classical Problem}
\label{The Classical Problem}

The {\it one-particle space} on which we work is $\mathfrak{h} = L^2(\T)$. We use the following convention for the scalar product.
\begin{equation*}
\langle g_1,g_2 \rangle_{\mathfrak{h}} := \int dx \, \overline{g_1}(x)\,g_2(x)\,.
\end{equation*} 
We consider the {\it one-body Hamiltonian} given by
\begin{equation}
\label{one_body_Hamiltonian}
h:= -\Delta + \kappa,
\end{equation} 
where $\kappa > 0$ is as in \eqref{integral_Hamiltonian_definition}. This is a positive self-adjoint densely defined operator on $\mathfrak{h}$. 
We can write $h$ spectrally as
\begin{equation}
\label{spectral_hamiltonian}
h:= \sum_{k \in \N} \lambda_k u_ku^*_k,
\end{equation}
where 
\begin{equation}
\label{hamiltonian_eigenvalue_definition}
\lambda_k := 4\pi^2|k|^2 +\kappa
\end{equation}
are the eigenvalues of $h$ and
\begin{equation}
\label{hamiltonian_eigenfunction_definition}
u_k:=e^{2\pi ikx}
\end{equation}
are the normalised eigenvalues of $h$ on $\mathfrak{h}$. 
Since we are working on $\T$, we have
\begin{equation}
\label{trace_hamiltonian}
\mathrm{Tr}(h^{-1})=\sum_{k \in \N} \frac{1}{4\pi^2|k|^2 +\kappa} < \infty,
\end{equation}
where the trace is taken over $\mathfrak{h}$.

For each $k \in \N$, we define $\mu_k$ to be a standard complex Gaussian measure. In other words, $\mu_k := \frac{1}{\pi}e^{-|z|^2}dz$, where $dz$ is the Lebesgue measure on $\C$. Let $(\C^{\N},\mathcal{G},\mu)$ be the product probability space with
\begin{equation}
\label{Wiener_measure_defn}
\mu := \bigotimes_{k \in \N} \mu_k.
\end{equation}
We denote elements of the probability space $\C^\N$ by $\omega = (\omega_k)_{k \in \N}$. Let the {\it classical free field} $\varphi \equiv \varphi^\omega$ be defined by
\begin{equation}
\label{random_classical_initial}
\varphi := \sum_{k \in \N} \frac{\omega_k}{\sqrt{\lambda_k}}u_k.
\end{equation}
Note that \eqref{trace_hamiltonian} implies \eqref{random_classical_initial} converges almost surely in $H^{\frac{1}{2}-\eps}(\T)$ for $\eps>0$ arbitrarily small.
Here $H^s(\T)$ denotes the $L^2$-based Sobolev space of order $s$ on $\T$ with norm given by
\begin{equation*}
\|g\|_{H^s(\T)}:=\Biggl(\sum_{k \in \Z} (1+|k|)^{2s} |\widehat{g}(k)|^2\Biggr)^{1/2}\,.
\end{equation*}
We take the following convention for the Fourier transform on $L^1(\T)$.
\begin{equation}
\label{Fourier_transform}
\hat{g}(k) := \int dx \, g(x) e^{-2\pi i kx}\,,\quad k \in \Z\,.
\end{equation}

The measure $\mu$ satisfies the following Wick theorem; see for example \cite[Lemma 2.4]{FKSS20} for a self-contained summary.
\begin{proposition}
	\label{Wicksthm}
	Let $\varphi$ be as in \eqref{random_classical_initial}. Given $g \in H^{-\frac{1}{2} + \eps}$ for $\eps>0$, we let $\varphi(g) := \langle g, \varphi \rangle$ and $\overline{\varphi}(g) := \langle \varphi,g \rangle$. Furthermore, we let  $(\varphi)^*(g)$ denote either $\varphi(g)$ or $\overline{\varphi}(g)$. Then, given $n \in \N^*$ and $g_1,\ldots,g_n \in H^{-\frac{1}{2}+\eps}$, we have
	\begin{equation}
	\label{Wickthmsum}
	\E_{\mu}\left[\prod_{i=1}^n \left(\varphi(g_i)\right)^{*_i}\right] = \sum_{\Pi \in \mathcal{M}(n)} \prod_{(i,j) \in \Pi} \E_{\mu}\left[\left(\varphi(g_i)\right)^{*_i}\left(\varphi(g_j)\right)^{*_j}\right],
	\end{equation}
	where the sum is taken over all complete pairings of $\{1,\ldots,n\}$, and where edges of $\Pi$ are denoted by $(i,j)$ with $i < j$.
\end{proposition}
We note that, by gauge invariance, for all $(i,j) \in \Pi$
	\begin{equation*}
	\E_{\mu}\left[\varphi(g_i)\varphi(g_j)\right]=\E_{\mu}\left[\overline{\varphi}(g_i)\,\overline{\varphi}(g_j)\right]=0\,.
	\end{equation*}
Therefore, each non-zero factor arising on the right-hand side of \eqref{Wickthmsum} can be computed using
\begin{equation*}
\int d\mu \, \overline{\varphi}(\tilde g) \varphi(g) = \langle g, h^{-1}\tilde g \rangle,
\end{equation*}
for $g,\tilde g \in H^{-\frac{1}{2} + \eps}$. Here, the Green function $h^{-1}$ is the covariance of $\mu$.
We note that, under a suitable pushforward, we can identify $\mu$ with a probability measure on $H^s$; see e.g.\ \cite[Remark 1.3]{FKSS17}. As in \cite{FKSS17}, we work directly with the measure $\mu$ as above and do not use this identification.

Given $p \in \N^*$, the {\it p-particle space} $\mathfrak{h}^{(p)}$ is defined as the symmetric subspace of $\mathfrak{h}^{\otimes p}$, i.e.\ $u \in \mfh^{(p)}$ if and only if for any permutation $\pi$,
\begin{equation*}
	u(x_{\pi(1)},\ldots,x_{\pi (p)}) = u(x_1,\dots,x_p)\,.
\end{equation*}
For $\xi$ a closed linear operator on $\mathfrak{h}^{(p)}$, we can associate $\xi$ with its Schwartz integral kernel, which we denote by $\xi(x_1,\ldots,x_p;y_1,\ldots,y_p)$; see \cite[Corollary V.4.4]{RS80}.  For such a $\xi$ and for $\varphi$ as in \eqref{random_classical_initial}, we define the random variable
\begin{multline}
	\label{classical_Theta_definition}
\Theta(\xi) :=
\int dx_1 \ldots dx_p \, dy_1 \ldots dy_p \,
\xi(x_1,\ldots,x_p;y_1,\ldots,y_p) \\
\times \overline{\varphi}(x_1) \ldots \overline{\varphi}(x_p)\varphi(y_1)\ldots\varphi(y_p)\,.
\end{multline}
We denote by $\mathcal{L}(\mathcal{H})$ the set of all bounded operators on a Hilbert Space $\mathcal{H}$. 
If $\xi \in \mathcal{L}(\mathfrak{h}^{(p)})$, then $\Theta(\xi)$ defined in  \eqref{classical_Theta_definition} is almost surely well-defined, since $\varphi \in \mathfrak{h}$ almost surely.

Given $w$ as in Assumption \ref{w_definition}, we define the {\it classical interaction} as
\begin{equation}
\label{classical_interaction_definiton}
\mcW := \frac{1}{2} \int dx \, dy \, |\varphi(x)|^2 \, w(x-y) |\varphi(y)|^2\,.
\end{equation}
The {\it free classical Hamiltonian} is given by
\begin{equation}
H_0 := \Theta(h) = \int dx \, dy \, \overline{\varphi}(x) h(x;y) \varphi(y)\,.
\end{equation}
The {\it interacting classical Hamiltonian} is given by
\begin{equation}
\label{interacting_classical_Hamiltonian}
H:= H_0 +\mcW\,.
\end{equation}
The {\it mass} is defined as
\begin{equation}
\label{mass}
	\mcN := \int dx \, |\varphi(x)|^2\,.
\end{equation}

At this stage, we have to introduce the cut-off $f$ that appears in \eqref{Gibbs_formal_truncated}. We now state the precise assumptions on $f$ that we use in the sequel.

\begin{assumption}
\label{support_f}
Throughout the paper, we fix $f \in C_c^{\infty}(\R)$, which is not identically equal to zero such that $0 \leq f \leq 1$ and 
\begin{equation}
\label{support_f_2}
\mathrm{supp}(f) \subset [-K,K]\,,
\end{equation}
for some $K > 0$. 
\end{assumption}
All of our estimates depend on $K$ in \eqref{support_f_2}, but we do not track this dependence explicitly.

We define the {\it classical state} $\rho^f(\cdot) \equiv \rho(\cdot)$ by
\begin{equation}
\label{classical_state}
\rho(X) := \frac{\int d \mu \, X e^{-\mcW} f(\mcN)}{\int d\mu \, e^{-\mcW} f(\mcN)} \equiv \mathbb{E}_{\mathbb{P}_{\mathrm{Gibbs}}^{f}}(X)\,,
\end{equation}
where $X$ is a random variable. Let the {\it classical partition function} $z \equiv z_{\mathrm{Gibbs}}^{f}$ be defined as
\begin{equation}
\label{classical_partition_function_defn}
z := \int d\mu \, e^{-\mcW} f(\mcN)\,.
\end{equation}
Note that both $\rho$ and $z$ are well defined by Lemma \ref{main} and Corollary \ref{main_corollary} below. We characterise $\rho(\cdot)$ through its moments. Namely, we define the {\it classical p-particle correlation function} $\gamma_p \equiv \gamma_p^f$, which acts on $\mathfrak{h}^{(p)}$ through its kernel
\begin{equation}
\label{classical_p_particle_correlation_fn}
\gamma_p(x_1,\ldots,x_p;y_1,\ldots,y_p) := \rho(\overline{\varphi}(y_1)\,\ldots\, \overline{\varphi}(y_p)\,\varphi(x_1)\,\ldots \,\varphi(x_p))\,.
\end{equation}

\subsection{The Quantum Problem}
\label{The Quantum Problem}

We use the same conventions as in \cite[Section 1.4]{FKSS17}. For more details and motivation, we refer the reader to the aforementioned work. In the quantum setting, we work on the {\it bosonic Fock space}, which is defined as
\begin{equation*}
	\mathcal{F} \equiv \mathcal{F}(\mathfrak{h}):= \bigoplus_{p \in \N} \mathfrak{h}^{(p)}.
\end{equation*}
Let us denote vectors of $\mathcal{F}$ by $\Psi=(\Psi^{(p)})_{p \in \N}$. For $g \in \mathfrak{h}$, let $b^*(g)$ and $b(g)$ denote the bosonic creation and annihilation operators, defined respectively as
\begin{align*}
	\left(b^*(g)\Psi\right)^{(p)}(x_1,\ldots,x_p) &:= \frac{1}{\sqrt{p}}\sum_{i=1}^p g(x_i)\Psi^{(p-1)}(x_1,\ldots,x_{i-1},x_{i+1},\ldots,x_p)\,, \\
	\left(b(g)\Psi\right)^{(p)} (x_1,\ldots,x_p) &:= \sqrt{p+1} \int dx \, \overline{g(x)} \Psi^{(p+1)}(x,x_1,\ldots,x_p)\,.
\end{align*}
These are closed, densely-defined operators which are each other's adjoints.
The creation and annihilation operators satisfy the canonical commutation relations, i.e.\
\begin{equation}
\label{commutation_relations}
[b(g_1),b^*(g_2)] = \langle g_1,g_2\rangle_{\mathfrak{h}}, \quad [b(g_1),b(g_2)] = [b^*(g_1),b^*(g_2)] = 0\,,
\end{equation}
for all $g_1,g_2 \in \mathfrak{h}$.
Furthermore, we define the rescaled creation and annihilation operators 
\begin{equation}
\label{phi_tau}
\varphi^*_\tau(g) := \tau^{-1/2}\,b^*(g)\,, \quad \varphi_\tau(g) := \tau^{-1/2}\,b(g)\,,
\end{equation}
for $g \in \mathfrak{h}$.
Here, we think of $\varphi_\tau^*$ and $\varphi_\tau$ as operator valued distributions, and we denote their distribution kernels by $\varphi^*_\tau(x)$ and $\varphi_\tau(x)$, respectively. 
Formally, $\varphi^*_\tau(x)$ and $\varphi_\tau(x)$ correspond to taking $g=\delta_x$ (the Dirac delta function centred at $x$) in \eqref{phi_tau}. In analogy to \eqref{random_classical_initial}, we call $\varphi_\tau$ the \emph{quantum field}.

As before, let $\xi$ be a closed linear operator on $\mathfrak{h}^{(p)}$. The {\it lift} of $\xi$ to $\mathcal{F}$ is defined by
\begin{multline}
	\label{quantum_lift}
\Theta_\tau(\xi) :=
\int dx_1\ldots dx_p \, dy_1 \ldots dy_p \, \xi(x_1,\ldots,x_p;y_1,\ldots,y_p) \\ 
\times \varphi^*_\tau(x_1)\ldots\varphi^*_\tau(x_p)\varphi_\tau(y_1)\ldots\varphi_\tau(y_p)\,.
\end{multline}
For $w \in L^{\infty}(\T)$ real-valued and even, we define the {\it quantum interaction} as\footnote{In principle, we could consider $w$ as in Assumption \ref{w_definition} in the quantum setting at the level of the definition. In practice, we take the interaction potential to be bounded; see Section \ref{Statement of the results} below for the precise statements.}
\begin{equation}
	\label{quantum_interaction_defn}
	\mcW_\tau := \frac{1}{2}\,\Theta_\tau(W) = \frac{1}{2} \int dx\,dy\,\varphi^*_\tau(x)\varphi^*_\tau(y)w(x-y)\varphi_\tau(x)\varphi_\tau(y)\,.
\end{equation}
Here $W$ is the two particle operator on $\mathfrak{h}^{(2)}$ which acts by multiplication by $w(x_1-x_2)$ for $w \in L^\infty$. We define the {\it free quantum Hamiltonian} as
\begin{equation}
	\label{quantum_free_ham}
	H_{\tau,0} := \Theta_\tau(h) = \int dx\,dy\,\varphi_\tau^*(x)h(x;y)\varphi_\tau(y)\,,
\end{equation}
where $h$ is as in \eqref{one_body_Hamiltonian}. We define the {\it interacting quantum Hamiltonian} as
\begin{equation*}
	\label{quantum_interacting_ham}
	H_\tau := H_{\tau,0} + \mcWt\,.
\end{equation*}
We also define the rescaled particle number as
\begin{equation}
\label{rescaled_particle_number_defn}
	\mcNt := \int dx \, \varphi^*_\tau(x)\varphi_\tau(x)\,.
\end{equation}
On the $p{\mathrm{th}}$ sector of Fock space, $\mcNt$ acts as multiplication by $\frac{p}{\tau}$.
 
The {\it (untruncated) grand canonical ensemble} is defined as 
\begin{equation}
\label{P_tau_definition}
P_\tau := e^{-H_\tau}
\end{equation}
and the {\it (truncated) quantum state} $\rho^f_\tau(\cdot) \equiv \rho_\tau(\cdot)$ is defined as
\begin{equation}
\label{quantum_state_defn}
	\rho_\tau(\mathcal{A}) := \frac{\mathrm{Tr}(\mathcal{A} P_\tau f(\mcNt))}{\mathrm{Tr}(P_\tau f(\mcNt))}\,,
\end{equation}
where the traces are taken over Fock space.
Let the {\it quantum partition function} and the {\it free quantum partition function}, $Z_\tau \equiv Z_\tau^f$, $Z_{\tau,0}$ be defined respectively as
\begin{equation}
\label{quantum_partition_function_defn}
Z_\tau := \mathrm{Tr}(e^{-H_{\tau}}f(\mcNt)), \quad Z_{\tau,0} := \mathrm{Tr}(e^{-H_{\tau,0}})\,.
\end{equation}
With $Z_\tau,Z_{\tau,0}$ as in \eqref{quantum_partition_function_defn}, we define the \emph{relative quantum partition function} $\mathcal{Z}_{\tau} \equiv \mathcal{Z}_{\tau}^f$ by
\begin{equation}
\label{relative_quantum_partition_function_defn}
\mathcal{Z}_{\tau}:=\frac{Z_{\tau}}{Z_{\tau,0}}\,.
\end{equation}

In analogy to \eqref{classical_p_particle_correlation_fn}, we characterise the quantum state through its correlation functions. Namely, for $p \in \N^*$, we define the {\it quantum p-particle correlation function} $\gamma_{\tau,p}^f \equiv \gamma_{\tau,p}$, which acts on $\mathfrak{h}^{(p)}$ through its kernel
\begin{equation}
\label{quantum_p_particle_correlation_fn}
\gamma_{\tau,p} (x_1,\ldots,x_p;y_1,\ldots,y_p) := \rho_{\tau}(\varphi^*_\tau(y_1)\ldots\varphi_\tau^*(y_p)\varphi_\tau(x_1)\ldots\varphi_\tau(x_p))\,.
\end{equation}
Throughout the sequel, for a given quantity $Y = \rho, \mcN, H , \ldots \,$, we will use the abbreviation $Y_\#$ to denote either $Y_\tau$ or $Y$.

\subsection{Statement of the results}
\label{Statement of the results}
We can now state our main results. In Section \ref{The time-independent problem}, we state the time-independent results. In Section \ref{The time-dependent problem}, we state their time-dependent generalisations. In all of the results, we will consider the limit $\tau \rightarrow \infty$, which we interpret as being the \emph{mean-field} or \emph{semiclassical} limit, with semiclassical parameter $1/\tau \rightarrow 0$. Physically, this corresponds to taking a \emph{high-density} limit, where we let the \emph{mass} of the bosonic particles or the \emph{temperature} tend to infinity. For a precise justification of this terminology and the choice of parameters, we refer the reader to \cite[Section 1.1]{FKSS20} for a detailed discussion.
\subsubsection{The time-independent problem}
The first result that we prove concerns bounded interaction potentials.
\label{The time-independent problem}
\begin{theorem}[Convergence for $w \in L^{\infty}(\T)$]
	\label{bounded_trace_class_final_convergence_theorem}
Let $w \in L^\infty(\T)$ be real-valued and even. Given $p \in \N^*$, we recall the quantities $\gamma_{\tau,p}$ and $\gamma_p$ defined in \eqref{quantum_p_particle_correlation_fn} and \eqref{classical_p_particle_correlation_fn} respectively. We then have
\begin{equation}
	\label{bounded_trace_class_convergence_final}
\lim_{\tau \to \infty}  \|\gamma_{\tau,p} - \gamma_p\|_{\mathfrak{S}^1(\mathfrak{h}^{(p)})}=0\,.
\end{equation}
Moreover, recalling \eqref{classical_partition_function_defn}  and  \eqref{relative_quantum_partition_function_defn}, we have 
\begin{equation}
\label{bounded_trace_class_convergence_final_2}
\lim_{\tau \to \infty}  \mathcal{Z}_{\tau}= z\,.
\end{equation}
\end{theorem}
By applying an approximation argument, we prove results for $w$ as in Assumption \ref{w_definition}. Throughout the sequel, any object with a superscript $\eps$ is the corresponding object defined by taking the interaction potential to be $w^\eps$, which will be a suitable bounded approximation of $w$. In what follows, we always assume that all the approximating interaction potentials $w^{\eps}$ are \emph{real-valued and even}, without mentioning this explicitly. We can now state the result for $L^1(\T)$ interaction potentials.
\begin{theorem}[Convergence for $w \in L^1(\T)$]
\label{unbounded_correlation_convergence_thm}
Let $w$ be as in Assumption \ref{w_definition} (i). 
Suppose that $(w^\eps)$ is a sequence of interaction potentials which are in $L^\infty(\T)$ such that $w^\eps \to w$ in $L^1(\T)$. Then there exists a sequence $(\eps_\tau)$ satisfying $\eps_\tau \to 0$ as $\tau \to \infty$ such that for any $p \in \N^*$
\begin{equation}
	\label{unbounded_trace_class_convergence_final}
\lim_{\tau \to \infty} \|\gamma^{\eps_{\tau}}_{\tau,p} - \gamma_p\|_{\mathfrak{S}^1(\mathfrak{h}^{(p)})} = 0
\end{equation}
and such that
	\begin{equation}
	\label{Unbounded_potential_state_theorem_2}
	\lim_{\tau \to \infty}\mathcal{Z}^{\eps_\tau}_{\tau}=z\,.
	\end{equation}
\end{theorem}

Before considering $w=-\delta$ as in Assumption \ref{w_definition} (ii), we need to define the sequence more $w^{\eps}$ precisely.
We fix $U: \R \rightarrow \R$ to be a continuous even function, with $\mathrm{supp} \,U \subset \T$ satisfying 
\begin{equation}
\label{int_U}
\int_{\mathbb{R}} dx\,U(x)=\int_{\T} dx \, U(x)=-1\,.
\end{equation}
For $\eps \in (0,1)$, we define
\begin{equation}
\label{w_eps_delta_function}
w^\eps := \frac{1}{\eps} \,U\left(\frac{[x]}{\eps}\right),
\end{equation}
where $[x]$ is defined to be the unique element in $(x+\Z) \cap \T$. 
In particular, $w^{\eps} \in L^{\infty}(\T)$ and  $w^\eps$ converges to $-\delta$ weakly, with respect to continuous functions. 
\begin{theorem}[Convergence for $w=-\delta$]
	\label{delta_correlation_convergence_thm}
With notation as in \eqref{w_eps_delta_function}, there exists a sequence $(\eps_\tau)$ satisfying $\eps_\tau \to 0$ as $\tau \to \infty$ such that for any $p \in \N^*$
\begin{equation}
\label{delta_trace_class_convergence_final}
\lim_{\tau \to \infty} \|\gamma^{\eps_{\tau}}_{\tau,p} - \gamma_p\|_{\mathfrak{S}^1(\mathfrak{h}^{(p)})} = 0
\end{equation}
and such that
	\begin{equation}
	\label{delta_potential_state_theorem_2}
	\lim_{\tau \to \infty}\mathcal{Z}^{\eps_\tau}_{\tau}=z\,.
	\end{equation}
\end{theorem}
\begin{remark}
	\label{Remark_AppendixB}
We make the following observations about Theorems \ref{bounded_trace_class_final_convergence_theorem}, \ref{unbounded_correlation_convergence_thm}, and \ref{delta_correlation_convergence_thm}.
\begin{enumerate}
\item For a pointwise almost everywhere non-negative, bounded, even interaction potential $w$, Theorem \ref{bounded_trace_class_final_convergence_theorem} holds without the need for a cut-off function $f$. This is the content of \cite[Theorem 1.8]{FKSS17}. Moreover, by working with the non-normal ordered quantum interaction $\mcW'_\tau$ defined in \eqref{nonnormal_quantum_interaction}, for a bounded, real-valued, even interaction potential $w$ of positive type (i.e.\ $\hat{w}$ pointwise almost everywhere non-negative), the same proof as \cite[Theorem 1.8]{FKSS17} again shows that Theorem \ref{bounded_trace_class_final_convergence_theorem} holds without the need for a cut-off function $f$. We include the details of the proof of this claim in Appendix \ref{Appendix B1}.
\label{Remark_Appendix_B_1}
\item We conjecture that the results hold for $f$ a characteristic function of an interval. The method that we apply in Lemma \ref{bounded_explicit_term_convergence_lemma} of Section \ref{Convergence of the Explicit Terms} requires suitable smoothness assumptions on $f$. This is a technical assumption.
\item For an individual $w \in L^\infty$, Theorem \ref{bounded_trace_class_final_convergence_theorem} holds with a cut-off function of the form $f(x)=e^{-cx^2}$, for $c>0$ sufficiently large depending on $\|w\|_{L^\infty}$. This is also proved by working with a non-normal ordered quantum interaction. The details are given in Appendix \ref{Appendix B2}. We note this $c$ cannot be chosen uniformly in the $L^\infty$ norm of the interaction potential. So we cannot treat the unbounded interactions as in Theorems \ref{unbounded_correlation_convergence_thm} and \ref{delta_correlation_convergence_thm} using this kind of truncation.
\label{Remark_Appendix_B_2}
\item 
One could consider the questions from Theorems \ref{bounded_trace_class_final_convergence_theorem} and \ref{time_dependent_Linf_case} 
in the non-periodic setting when the spatial domain is $\R$ for the one-body Hamiltonian $h=-\Delta + \kappa + v$, where $v: \R \to [0, \infty)$ is a positive {\it one-body potential} such that $h$ has compact resolvent and $\mathrm{Tr} \,h^{-1}<\infty$ holds (as in \eqref{trace_hamiltonian}). The analysis that we present in the periodic setting would carry through to this case, provided that we know that the time evolution $S_t$ given in \eqref{flow_map} is well-defined on the support of the Gibbs measure. We do not address this question further in our paper.
\item By following the duality arguments in \cite[Section 3.3]{FKSS17}, we can get the equivalents of equations \eqref{bounded_trace_class_convergence_final}, \eqref{unbounded_trace_class_convergence_final}, and \eqref{delta_trace_class_convergence_final} in terms of $\rho_\tau$ and $\rho$. For more details when $w \in L^{\infty}$, see Corollary \ref{bounded_state_final_convergence_corollary}, Lemma \ref{Unbounded_potential_state_theorem}, and Lemma \ref{delta_function_potential_theorem} below. For the time-independent problem, we state the convergence as above in the trace class. For the time-dependent problem, we need to use the alternative formulation, which can be seen as a generalisation of the time-independent analysis. For more details, see Remark \ref{generalisation_remark} below.
\item
Our method works for more general interaction potentials. In particular, we can consider linear combinations of interaction potentials as in Assumption \ref{w_definition} (i) and (ii) with the same arguments.
\end{enumerate}
\end{remark}
\subsubsection{The time-dependent problem}
\label{The time-dependent problem}

We also prove time-dependent generalisations of the results in Section \ref{The time-independent problem}. In order to precisely state the results, we first introduce some notation.
\begin{definition}
	\label{classical_evolution}
Let $p \in \N^*$ and $\xi \in \mathcal{L}(\mathfrak{h}^{(p)})$ be given. For $t \in \R$, we define the random variable
\begin{multline}
\Psi^t \Theta(\xi) := \int dx_1 \ldots dx_p \, dy_1 \ldots dy_p \, \xi(x_1, \ldots, x_p; y_1 \ldots y_p) \\
\times \overline{S_t \varphi}(x_1) \ldots \overline{S_t \varphi}(x_p) S_t \varphi(x_p) \ldots S_t \varphi(y_p),
\end{multline}
where $S_t$ is the flow map defined in \eqref{flow_map}. This is well defined since $\varphi \in \mathfrak{h}$ almost surely and $S_t$ is norm preserving on $\mfh$.
\end{definition}
\begin{definition}
	\label{quantum_evolution}
Suppose $\mcA : \mcF \to \mcF$. Define the quantum time evolution of $\mcA$ as
\begin{equation*}
\Psi_{\tau}^t \mcA := e^{it\tau H_\tau}\, \mcA \,e^{-it\tau H_\tau}\,.
\end{equation*}
\end{definition}
We also recall the quantities $\rho_\tau$ and $\rho$ defined as in \eqref{quantum_state_defn} and \eqref{classical_state} respectively. 
\begin{theorem}[Convergence for $w \in L^{\infty}(\T)$]
	\label{time_dependent_Linf_case}
Let $w$ be as in Theorem \ref{bounded_trace_class_final_convergence_theorem}.
Given $m \in \N^*$, $p_i \in \N^*$, $\xi^{i} \in \mathcal{L}(\mfh^{(p_i)})$, and $t_i \in \R$, we have
\begin{equation*}
\lim_{\tau\to\infty} \rho_\tau\left(\Psi^{t_1}_\tau \Theta_\tau(\xi^{1})\ldots\Psi^{t_m}_\tau \Theta_\tau(\xi^{m})\right) = \rho\left(\Psi^{t_1} \Theta(\xi^{1})\ldots\Psi^{t_m} \Theta(\xi^{m})\right)\,.
\end{equation*}
\end{theorem}
\begin{theorem}[Convergence for $w \in L^1(\T)$]
	\label{time_dependent_L1_case}
Let $w,w^{\eps}$ be as in the assumptions of Theorem \ref{unbounded_correlation_convergence_thm}.
Then, there exists a sequence $(\eps_\tau)$ satisfying $\eps_\tau \to 0$ as $\tau \to \infty$ such that, given  $m \in \N^*$, $p_i \in \N^*$, $\xi^{i} \in \mathcal{L}(\mfh^{(p_i)})$, and $t_i \in \R$, we have
\begin{equation*}
\lim_{\tau\to\infty} \rho^{\eps_\tau}_\tau\left(\Psi^{t_1}_\tau \Theta_\tau(\xi^{1})\ldots\Psi^{t_m}_\tau \Theta_\tau(\xi^{m})\right) = \rho\left(\Psi^{t_1} \Theta(\xi^{1})\ldots\Psi^{t_m} \Theta(\xi^{m})\right).
\end{equation*}
\end{theorem}
\begin{theorem}[Convergence for $w=-\delta$]
	\label{time_dependent_delta_case}
Let $w,w^{\eps}$ be as in the assumptions of Theorem \ref{delta_correlation_convergence_thm}.
Then, there exists a sequence $(\eps_\tau)$ satisfying $\eps_\tau \to 0$ as $\tau \to \infty$ such that, given $m \in \N^*$, $p_i \in \N^*$, $\xi^{i} \in \mathcal{L}(\mfh^{(p_i)})$, and $t_i \in \R$, we have
\begin{equation*}
\lim_{\tau\to\infty} \rho^{\eps_\tau}_\tau\left(\Psi^{t_1}_\tau \Theta_\tau(\xi^{1})\ldots\Psi^{t_m}_\tau \Theta_\tau(\xi^{m})\right) = \rho\left(\Psi^{t_1} \Theta(\xi^{1})\ldots\Psi^{t_m} \Theta(\xi^{m})\right)\,.
\end{equation*}
\end{theorem}

\begin{remark}
\label{generalisation_remark}
Theorems \ref{time_dependent_Linf_case}--\ref{time_dependent_delta_case} can indeed be seen as generalisations of the results given in Theorems \ref{bounded_trace_class_convergence_final}--\ref{delta_correlation_convergence_thm} respectively (the latter of which correspond to setting $m=1$ and $t_1=0$). Namely, we use Remark \ref{Remark_AppendixB} (3) above and noting that the proofs show that the convergence is uniform in $\|\xi^1\| \leq 1$.
\end{remark}
\subsection{Previously known results}
\label{Previously known results}
In the context of the NLS, Gibbs measures \eqref{Gibbs_formal}--\eqref{Gibbs_formal_truncated} were relevant to study a substitute for a conservation law at low regularity. Namely, one can show that they are invariant under the flow and that they are supported on Sobolev spaces of low regularity. Consequently, it is possible to construct global solutions for random rough initial data. This was first rigorously obtained in the work of Bourgain \cite{Bou94,Bou96,Bou97}. Some preliminary results were previously known by Zhidkov \cite{Zhidkov}. This is an active area of research in nonlinear dispersive PDEs. We refer the reader to the expository works \cite{Burq_Thomann_Tzvetkov,Nahmod_Staffilani,Oh_Thomann} for further explanations and background. For more recent developments, we refer the reader to \cite{Bringmann1,Bringmann2,Bringmann_Deng_Nahmod_Yue, Deng_Nahmod_Yue, Gunaratnam_Oh_Tzvetkov_Weber, Oh_Sosoe_Tolomeo} and the references therein.

The focusing problem is more challenging. In one dimension, it was addressed in the earlier works \cite{Bou94,LRS}. The one-dimensional problem was revisited recently in \cite{Ammari_Sohinger_2021,Carlen_Frohlich_Lebowitz,Oh_Sosoe_Tolomeo}. For recent results on the fractional NLS, see \cite{Liang_Wang}.

It also makes sense to consider the higher-dimensional problem, i.e.\ when the spatial domain is $\mathbb{T}^d$, with $d=2,3$. Here, one needs to renormalise the interaction by means of \emph{Wick ordering}. Formally, this refers to replacing \eqref{classical_interaction_definiton} by
\begin{equation}
\label{Wick_ordering_interaction}
\mathcal{W}^{\mathrm{Wick}}:=\frac{1}{2}\int dx\,dy\, \bigl(|\varphi(x)|^2-\mathbb{E}_{\mu}[|\varphi(x)|^2]\bigr)\,w(x-y)\,\bigl(|\varphi(y)|^2-\mathbb{E}_{\mu}[|\varphi(x)|^2]\bigr)\,,
\end{equation}
One rigorously constructs \eqref{Wick_ordering_interaction} by means of a frequency truncation, see e.g. \cite[Lemma 1.5]{FKSS17} for a pedagogical overview. In this context, the notion of \emph{defocusing} refers to $w$ being of positive type (i.e.\ $\hat{w}$ being pointwise nonnegative almost everywhere), in which case the quantity \eqref{Wick_ordering_interaction} is formally nonnegative. When this assumption is relaxed, one needs to consider a truncation in the Gibbs measure, similarly as in \eqref{Gibbs_formal_truncated}. Since the $L^2$ norm is almost surely infinite, one needs to modify \eqref{Gibbs_formal_truncated} and consider
\begin{equation}
\label{Gibbs_formal_truncated_2}
d \tilde{\mathbb{P}}_{\mathrm{Gibbs}}^{f}(\varphi):=\frac{1}{\tilde{z}_{\mathrm{Gibbs}}^{f}}\,e^{-H^{\mathrm{Wick}}(\varphi)}\,f\bigl(:\|\varphi\|_{\mathfrak{h}}^2:\bigr)\,d\varphi\,.
\end{equation}
In \eqref{Gibbs_formal_truncated_2}, $H^{\mathrm{Wick}}$ denotes the Hamiltonian obtained by replacing \eqref{classical_interaction_definiton} by \eqref{Wick_ordering_interaction} in \eqref{interacting_classical_Hamiltonian}. Moreover,
$:\|\varphi\|_{\mathfrak{h}}^2:$ denotes the Wick-ordering of the mass \eqref{mass} formally given by
\begin{equation*}
:\|\varphi\|_{\mathfrak{h}}^2: \equiv \int dx \, \bigl(|\varphi(x)|^2-\mathbb{E}_{\mu}[|\varphi(x)^2|]\bigr)\,.
\end{equation*}
Finally, $\tilde{z}_{\mathrm{Gibbs}}^{f}$ denotes a normalisation constant.
For precise definitions of these objects, we refer the reader to \cite[(12)]{Bou97}.
The invariance of \eqref{Gibbs_formal_truncated_2} under the corresponding (Wick-ordered, focusing) NLS flow was first shown in \cite{Bou97} for $w$ satisfying appropriate decay conditions on its Fourier coefficients (or under appropriate integrability conditions on $w$)\,. It was noted in \cite{Brydges_Slade} that, when $d=2$, Wick ordering and truncation as in \eqref{Gibbs_formal_truncated_2} do not yield a well-defined probability measure when $w=-\delta$.
Gibbs measures for the focusing NLS and related models were also studied in \cite{Bourgain_Bulut,Oh_Okamoto_Tolomeo_1,Oh_Okamoto_Tolomeo_2, Xian}.

The first result showing how Gibbs measures for the NLS arise as limits of many-body quantum Gibbs states was proved by Lewin, Nam, and Rougerie \cite{LNR_1}. More precisely, the authors show that the quantum Gibbs state (as in \eqref{quantum_state_defn}, with $f=1$, which we henceforth take throughout this subsection) converges 
to the classical Gibbs state (as in \eqref{classical_state}) in the sense of partition functions and correlation functions as $\tau \rightarrow \infty$, as in Section \ref{Statement of the results}. In \cite{LNR_1}, the authors studied the full defocusing problem in one dimension, as well as systems in $d=2,3$ with suitably chosen non translation-invariant interactions (which do not require Wick ordering)\,. Their method is based on the Gibbs variational principle and the quantum de Finetti theorem. The techniques from \cite{LNR_1} were later applied to the regime of one-dimensional sub-harmonic traps in \cite{LNR_2}. 

In \cite{FKSS17}, Fr\"{o}hlich, Knowles, Schlein, and the second author developed an alternative approach based on a series expansion of the classical and quantum state in terms of the interaction, combined by a comparison of the explicit terms of the obtained series, and a Borel resummation. In doing so, they could give an alternative proof of the one-dimensional result obtained in \cite{LNR_1} and consider (Wick-ordered) Gibbs measures obtained from translation-invariant interaction potentials for $d=2,3$, under a suitable modification of the quantum Gibbs state.
The results for $d=2,3$ in \cite{FKSS17} (under the same modification of the quantum Gibbs state) were originally stated for interaction potentials $w \in L^{\infty}(\mathbb{T}^d)$ of positive type. 
In \cite{FKSS18}, the results from \cite{FKSS17} were used to study \emph{time-dependent correlations} for $d=1$. 
Moreover, the methods from \cite{FKSS17} were later extended to $w \in L^q(\mathbb{T}^d), d=2,3$ with optimal $q$ in \cite{Sohinger_2019}. The optimal range of $q$ was observed in \cite{Bou97}.

In \cite{LNR_3}, and in \cite{LNR_5}, Lewin, Nam, and Rougerie obtained the derivation of the (Wick-ordered) Gibbs measures obtained from translation-invariant Gibbs measures when $d=2,3$ without the modification of the quantum Gibbs state from \cite{FKSS17}. Their methods are based on a non-trivial extension of the ideas from \cite{LNR_1}. An expository summary of the results of Lewin, Nam, and Rougerie can be found in \cite{LNR_4}.

Independently, and simultaneously with \cite{LNR_5}, Fr\"{o}hlich, Knowles, Schlein, and the second author \cite{FKSS20} obtained a derivation of the (Wick-ordered) Gibbs measure when $d=2,3$ based on a \emph{functional integral representation}, and an infinite-dimensional saddle-point argument. The fundamental tool for setting up the functional integral representation in \cite{FKSS20} is the \emph{Hubbard-Stratonovich transformation}. In \cite{FKSS20}, convergence in the $L^{\infty}$ norm of Wick-ordered correlation functions was shown. 

The result of \cite{FKSS20} was shown for continuous interaction potentials of positive type. In recent work \cite{FKSS22}, the same group of authors obtained  the result with $w=\delta$ when $d=2$. Here, one takes a limit in which the range of the interaction potential varies appropriately. The limiting object corresponds to the \emph{(complex) Euclidean $\Phi^4_2$ theory}. The proof in \cite{FKSS22} is based on the combination quantitative analysis of the infinite-dimensional saddle point argument from \cite{FKSS20} and a 
Nelson-type estimate for a general nonlocal theory in two dimensions (analogous to \cite{Nelson})\,.

Related results were proved for systems on the lattice \cite{Kno09,Salmhofer_2020,FKSS20_3}. We refer to \cite{FKSS20_2} for an expository summary of some of the aforementioned results. We emphasise that all of the results mentioned in this subsection are proved in the \emph{defocusing} (or \emph{thermodynamically stable}) regime. We also refer the interested reader to subsequent applications of the methods in this paper to the three-body setting \cite{Rout_Sohinger_2023}.

\subsection{Main ideas of the proofs}
The starting point of our analysis of the time-independent problem with $w \in L^{\infty}$ (i.e.\ of the proof of Theorem \ref{bounded_trace_class_final_convergence_theorem}) is the perturbative expansion of the interaction $e^{-H_{\#}}$ in the interaction, similarly as in \cite[Section 2.2]{FKSS17} for the quantum and \cite[Section 3.2]{FKSS17} for the classical setting. Due to the presence of the truncation $f(\mathcal{N}_{\#})$ in \eqref{classical_state} and \eqref{quantum_state_defn}, the resulting series have infinite radius of convergence; see Propositions \ref{quantum_function_analytic_prop} and \ref{classical_function_analytic_prop} below. Thus, we avoid the need to apply Borel resummation techniques as in \cite{FKSS17}.

When analysing the remainder term in the quantum setting, we apply the Feynman-Kac formula and use the truncation property from Assumption \ref{support_f}. This analysis is possible since we are not Wick-ordering the interaction in one dimension; see Lemma \ref{quantum_bounded_remainder_bound_lemma}. The truncation is likewise crucially used in the analysis of the classical remainder term; see Lemma \ref{Classical_bounded_remainder_bound_lemma}.

When studying the convergence of the explicit terms of the obtained series, we use complex analytic methods as in \cite[Section 3.1]{FKSS18} to perform an expansion of the truncation $f(\mathcal{N}_{\#})$ and thus reduce to the study of the problem with a shifted chemical potential, but without a truncation. It is important that at this step, where we no longer have the control coming from the truncation, the analysis does not depend on the sign of the interaction. The details of this step are given in Lemmas \ref{nu_hamiltonian_proposition} and \ref{bounded_explicit_term_convergence_lemma} below.

The proofs of Theorems \ref{unbounded_correlation_convergence_thm} and \ref{delta_correlation_convergence_thm} are based on the application of Theorem \ref{bounded_trace_class_final_convergence_theorem} for appropriate $w^{\eps}$ and on a diagonal argument. At this step, we have to crucially use \cite[Lemma 3.10]{Bou94}, which is recalled in Lemma \ref{main} below. Even when we are working with $L^1$ interaction potentials, it is important that we apply the local version of this result (instead of Corollary \ref{main_corollary}). For details, see the proof of Lemma \ref{Unbounded_potential_state_theorem}, in particular see steps \eqref{uniform_epsilon_1}--\eqref{uniform_epsilon_2}.


For the time-dependent problem, we apply a \emph{Schwinger-Dyson expansion}, similarly as in \cite[Sections 3.2--3.3]{FKSS18}. For the precise statements, see Lemmas \ref{quantum_schwinger_dyson_prop}--\ref{classical_schwinger_dyson_prop} below. Note that, due to the presence of the truncation, we do not need to consider the large particle number regime as in \cite[Section 4]{FKSS18} (whose analysis, in turn, relies crucially on the defocusing assumption). With this setup, we can easily deduce Theorem \ref{time_dependent_Linf_case} from Theorem \ref{bounded_trace_class_final_convergence_theorem}.

In order to prove Theorems \ref{time_dependent_L1_case} and \ref{time_dependent_delta_case}, we need to apply an approximation argument. In particular, we want to estimate the difference of the flow map of the NLS with interaction potential $w$ and of the NLS with interaction potential $w^{\eps}$. For the precise statement, see Lemma \ref{L^1_Cauchy_approximation_theorem} when $w \in L^1$ and Lemma \ref{delta_Cauchy_approximation_theorem} when $w=-\delta$. We prove these results by working in $X^{s,b}$ spaces; see Definition \ref{X^{s,b}}.

\subsection{Organisation of the paper}
In Section \ref{Notation and auxiliary results}, we set up some more notation and recall several auxiliary results from analysis and probability theory. Section \ref{Section 3} is devoted to the analysis of the time-independent problem with bounded interaction potential. Here, we prove Theorem \ref{bounded_trace_class_final_convergence_theorem}. In Section \ref{Section 4}, we study the time-independent problem with unbounded interaction potential and prove Theorems \ref{unbounded_correlation_convergence_thm} and \ref{delta_correlation_convergence_thm}. Section \ref{Section 5} is devoted to the time-dependent problem and the proofs of Theorems \ref{time_dependent_Linf_case}--\ref{time_dependent_delta_case}. In Appendix \ref{Appendix_A}, we recall the proof of Lemma \ref{main}, which was originally given in \cite[Lemma 3.10]{Bou94}. In Appendix \ref{Appendix_cutoff_function}, we give a detailed proof of the comments on the cut-off $f$ given in Remark \ref{Remark_AppendixB} (1) and (2) above.

\section{Notation and auxiliary results}
\label{Notation and auxiliary results}
\subsection{Notation}
Throughout the paper, we use $C>0$ to denote a generic positive constant that can change from line to line. If $C$ depends on a finite set of parameters $\alpha_1,\ldots,\alpha_n$, we indicate this dependence by writing $C(\alpha_1, \ldots, \alpha_n)$. Sometimes, we also write $a \leq C b$ as $a \lesssim b$. We denote by $\N=\{0,1,2,\ldots,\}$ the set of nonnegative integers and by $\N^*=\{1,2,3,\ldots\}$ the set of positive integers.

We write $\mathbf{1}$ to denote the identity operator on a Hilbert space. For a separable Hilbert space $\mathcal{H}$ and $q \in [1,\infty]$, we define the Schatten space $\mathfrak{S}^q(\mathcal{H})$ to be the set of $\mathcal{A} \in 
\mathcal{L}(\mathcal{H})$ satisfying $\|\mathcal{A}\|_{\mathfrak{S}^q(\mathcal{H})}$, where
\begin{equation}
\label{Schatten_space}
\|\mathcal{A}\|_{\mathfrak{S}^q(\mathcal{H})} :=	
\begin{cases}
\left(\mathrm{Tr}\,|\mathcal{A}|^q\right)^{1/q} & \text{if }  q< \infty \\
\sup \mathrm{spec} \, |\mathcal{A}| & \text{if } q=\infty\,,
\end{cases}
\end{equation}
and $|\mathcal{A}| = \sqrt{\mathcal{A}^*\mathcal{A}}$. We usually omit the argument $\mathcal{H}$ where there is no confusion. We also have the following notation
\begin{equation}
\label{B_p}
\mathfrak{B}_p:= \{\xi \in \mathfrak{S}^2(\mathfrak{h}^{(p)}) : \|\xi\|_{\mathfrak{S}^2(\mathfrak{h}^{(p)})} \leq 1 \}\,.
\end{equation}

\subsection{Auxiliary Results}
We recall several auxiliary results that we use in the paper.
\subsubsection*{Gibbs measures for the focusing local NLS}
When analysing Gibbs measures for the focusing cubic NLS with $w \in L^{\infty}(\T)$, it is straightforward to make rigorous sense of \eqref{Gibbs_formal_truncated} due to the presence of the truncation as in Assumption \ref{support_f}; see Lemma \ref{classical_W_bound} (1) below.

For unbounded potentials, we will need to make use of the following result of Bourgain, found in \cite[Lemma 3.10]{Bou94}, whose proof is recalled in Appendix \ref{Appendix_A}. 
\begin{lemma}
\label{main} 
Let $(\C^\N,\mathcal{G},\mu)$ be the probability space defined in \eqref{Wiener_measure_defn}. Let $c>0$ be fixed. For $\varphi \equiv \varphi^\omega$, the quantity
\begin{equation}
e^{c \|\varphi\|_{L^p}^p} \chi_{\{\|\varphi\|_{L^2} \leq B\}} \label{func}
\end{equation} 
is in $L^1(d \mu)$ for $p\in [4,6)$ for $B>0$ arbitrary and $p=6$ for $B>0$ sufficiently small (chosen in terms of $c$). 
\end{lemma}
\begin{remark}
When $p=6$, the optimal value of $B$ in Lemma \ref{main} was recently determined in \cite[Theorem 1.1 (ii)]{Oh_Sosoe_Tolomeo}. We do not need to use this precise result since we work with $p=4$ in the remainder of the paper.
\end{remark}

\begin{remark}
When $p=6$, an upper bound for the choice of $B$ is determined by the constant $c>0$. For details, see \eqref{G_L^1} below.
\end{remark}

\begin{corollary}
	\label{main_corollary}
Let $(\C^\N,\mathcal{G},\mu)$ be the probability space defined in \eqref{Wiener_measure_defn}, and let $w \in L^1(\T)$. For $\varphi \equiv \varphi^\omega$,
\begin{equation*}
e^{-\frac{1}{2} \int dx\,dy\, |\varphi(x)|^2\, w(x-y)\,|\varphi(y)|^2} \chi_{\{\|\varphi\|_{L^2} \leq B\}}
\end{equation*}
is in $L^1(d \mu)$ for $B>0$ arbitrary.
\end{corollary}
We note that Corollary \ref{main_corollary} follows from Lemma \ref{main} with $p=4$ by the same argument as estimate \eqref{classical_W_bound_2_equation} below.

\subsubsection*{H\"{o}lder's inequality for Schatten spaces}
We have the following version of H\"older's inequality for Schatten spaces \eqref{Schatten_space}, found in \cite{Sim05}.
\begin{lemma}[H\"older's Inequality]
	\label{Holder's_inequality_Schatten}
Given $p_1,p_2 \in [1,\infty]$ with $\frac{1}{p_1} + \frac{1}{p_2} = \frac{1}{p}$ and $\mathcal{A}_j \in \mathfrak{S}^{p_j}(\mathcal{F})$ we have
\begin{equation*}
\|\mathcal{A}_1\mathcal{A}_2\|_{{\mathfrak{S}}^p(\mathcal{F})} \leq \|\mathcal{A}_1\|_{{\mathfrak{S}}^{p_1}(\mcF)} \|\mathcal{A}_2\|_{{\mathfrak{S}}^{p_2}(\mcF)}\,.
\end{equation*}
\end{lemma}
\subsubsection*{The Feynman-Kac Formula}
In our analysis, we make use of the Feynman-Kac formula. To this end, let $\tau >0$ and let $\Omega^{\tau}$ denote the space of continuous paths $\omega: [0,\tau] \to \T$. Given $x,\tilde{x} \in \T$, we let $\Omega^{\tau}_{x,\tilde{x}}$ denote the set of all elements of $\Omega^{\tau}$ such that $\omega(0)=\tilde{x}$ and $\omega(\tau)=x$.
Given $t>0$, we define 
\begin{equation}
\psi^t(y) := e^{t\Delta}(y) = \sum_{n \in \Z^d} (4\pi t)^{-1/2}e^{-|y-n|^2/4t}
\end{equation}
to be the periodic heat kernel on $\T$. For $x, \tilde{x} \in \T$, we characterise the Wiener measure $\mathbb{W}_{x,\tilde{x}}^\tau$ on $\Omega^{\tau}_{x,\tilde{x}}$ by its finite-dimensional distribution. Namely for $0 < t_1 < \ldots < t_n < \tau$ and $f:\T^n \to \R$ continuous
\begin{align*}
	\int \mathbb{W}^{\tau}_{x,\tilde{x}}(d \omega) f(\omega(t_1),\ldots,\omega(t_n))& \\
	= \int dx_1 \, \ldots \, dx_n &\, \psi^{t_1}(x_1-\tilde{x}) \psi^{t_2-t_1}(x_2-x_1) \ldots\\ & \times \psi^{t_{n}-t_{n-1}}(x_n-x_{n-1}) \psi^{\tau-t_{n}}(x-x_n) f(x_1 , \ldots, x_n)\,.
\end{align*}
Then we have the following result, see for example \cite[Theorem X.68]{RS75}.
\begin{proposition}[Feynman-Kac Formula]
	\label{Feynman-Kac_formula}
Let $V: \T \to \C$ be continuous and bounded below. For $t > 0$
	\begin{equation*}
		e^{t(\Delta-V)}(x;\tilde{x}) = \int \mathbb{W}^{t}_{x,\tilde{x}} (d \omega) e^{-\int^{t}_0 ds \, V(\omega(s))}\,.
	\end{equation*}
\end{proposition}

\section{The time-independent problem with bounded interaction potential. Proof of Theorem \ref{bounded_trace_class_final_convergence_theorem}.}
\label{Section 3}

In this section, we study the time-independent problem with bounded interaction potential. In Section \ref{Basic Estimates}, we state some basic estimates which will be used throughout the rest of the paper. In Section \ref{Duhamel Expansion}, we set up the Duhamel expansion in the quantum setting. For this expansion, bounds on the explicit term are shown in Section \ref{Bounds on the explicit terms} and bounds on the remainder term are shown in Section \ref{Bounds on the remainder term}. The analogous expansion in the classical setting is analysed in Section \ref{The classical setting}. In Section \ref{The classical setting}, we prove convergence of the explicit terms. The proof of Theorem \ref{bounded_trace_class_final_convergence_theorem} is given in Section \ref{correlation_fn_argument}.
\subsection{Basic Estimates}

\label{Basic Estimates}
Let us first note the following bound on the classical interaction.
\begin{lemma} Suppose that $\mcW=\frac{1}{2} \int dx \, dy \, |\varphi(x)|^2 \, w(x-y) |\varphi(y)|^2$ is defined as in \eqref{classical_interaction_definiton}. The following estimates hold.
	\label{classical_W_bound}
\begin{enumerate}
\item \label{classical_W_bound_1}
For $w \in L^\infty(\T)$, we have
\begin{equation}
\left|\mcW\right| \leq \frac{1}{2}\|w\|_{L^\infty} \|\varphi\|_{L^2}^4.
\end{equation}
\item For $w \in L^1(\T)$, we have
\begin{equation}
\label{classical_W_bound_2_equation}
\left|\mcW\right| \leq \frac{1}{2}\|w\|_{L^1} \|\varphi\|_{L^4}^4.
\end{equation}
\label{classical_W_bound_2}
\end{enumerate}
\end{lemma}
\begin{proof}
For \eqref{classical_W_bound_1}, we note that
\begin{align*}
\left|\mcW\right| &= \frac{1}{2}\left|\int dx\,dy\, |\varphi(x)|^2w(x-y)|\varphi(y)|^2\right| \\
&\leq \frac{1}{2} \|w\|_{L^\infty} \int dx\,dy \, |\varphi(x)|^2|\varphi(y)|^2 = \frac{1}{2} \|w\|_{L^\infty} \|\varphi\|_{L^2}^4.
\end{align*}
For \eqref{classical_W_bound_2}, we apply Cauchy-Schwarz and Young's inequality to get \eqref{classical_W_bound_2_equation}.
\end{proof}
For the remainder of this section, we fix $p \in \N^*$. Unless otherwise specified, we consider $\xi \in \mathcal{L}(\mfh^{(p)})$. Moreover, $\|\cdot\|$ denotes the operator norm.
The following lemma follows from the definition of $\Theta(\xi)$ in \eqref{classical_Theta_definition}.
\begin{lemma}
	\label{Theta_classical_bound}
We have
\begin{equation*}
\left| \Theta(\xi) \right| \leq \|\varphi\|_{\mfh}^{2p} \|\xi\|\,.
\end{equation*}
\end{lemma}
Let us note that with $\Theta_\tau$ as in \eqref{quantum_lift}, we have
\begin{equation}
\label{quantum_lift_on_hn}
\Theta_\tau(\xi)\big|_{\mathfrak{h}^{(n)}} =
\begin{cases}
\frac{p!}{\tau^p}{n \choose p} P_+\left(\xi \otimes \mathbf{1}^{(n-p)} \right)P_+ & \text{if } n \geq p\\
0 & \text{otherwise}\,,
\end{cases}
\end{equation}
where $\mathbf{1}^{(q)}$ denotes the identity map on $\mathfrak{h}^{(q)}$ and $P_+$ is the orthogonal projection onto the subspace of symmetric tensors. More details of the above equality can be found in \cite[(3.88)]{Kno09}.
We also have the quantum analogue of Lemma \ref{Theta_classical_bound}, which follows from \eqref{quantum_lift_on_hn}.
\begin{lemma}
	\label{Theta_quantum_bound}
For all $n \in \N^*$, we have
\begin{equation*}
\left\|\Theta_\tau(\xi)\big|_{\mathfrak{h}^{(n)}}\right\| \leq \left(\frac{n}{\tau}\right)^{p} \|\xi\|.
\end{equation*}
\end{lemma}

\subsection{Duhamel Expansion}
\label{Duhamel Expansion}
Throughout this section, we take $w \in L^\infty(\T)$. Note that with $\rho_\tau$ defined as in \eqref{quantum_state_defn}, we have
\begin{equation}
\label{quantum_state_identity}
\rho_\tau(\Theta_\tau(\xi)) = \frac{\tilde{\rho}_{\tau,1}(\Theta_\tau(\xi))}{\tilde{\rho}_{\tau,1}(
\mathbf{1})},
\end{equation}
where
\begin{equation}
\label{quantum_state_identity_2}
\tilde{\rho}_{\tau,\zeta}(\mathcal{A}) := \frac{1}{Z_{\tau,0}} \mathrm{Tr}\left(\mathcal{A}e^{-H_{\tau,0} - \zeta \mcWt}f(\mcNt)\right),
\end{equation}
and $\mathbf{1}$ denotes the identity operator on $\mathcal{F}$. Here, we recall the definition \eqref{quantum_partition_function_defn} of $Z_{\tau,0}$. With notation as above, we define
\begin{align*}
A^{\xi}_\tau(\zeta) &:= \tilde{\rho}_{\tau,\zeta}(\Theta_\tau(\xi))\,.
\end{align*}
Performing a Duhamel expansion by up to order $M \in \N$ by iterating the identity $e^{X+\zeta Y} = e^X + \zeta \int_{0}^{1} dt \, e^{(1-t)X}Ye^{t(X+\zeta Y)}$ yields the following result.
\begin{lemma}
For $M \in \N$, we have $A^\xi_\tau(\zeta) = \sum_{m=0}^{M-1} a_{\tau,m}^\xi \zeta^m + R^\xi_{\tau,M}(\zeta)$, where
\begin{align}
\notag
a^\xi_{\tau,m}:=&\frac{(-1)^m}{Z_{\tau,0}}\mathrm{Tr}\bigg(\int_0^{1} dt_1 \int_{0}^{t_1} dt_2 \ldots \int^{t_{m-1}}_0 dt_m \, \Theta_{\tau} (\xi) e^{-(1-t_1)H_{\tau,0}} \, \mcWt \\
&
\label{explicit_term_quantum}
\times e^{-(t_1-t_2)H_{\tau,0}}\, \mcWt \, e^{-(t_2-t_3)H_{\tau,0}} \ldots e^{-(t_{m-1}-t_m)H_{\tau,0}} \, \mcWt \, e^{-t_m H_{\tau,0}}f(\mcN_\tau) \bigg)\,
\end{align}
and
\begin{align*}
R^\xi_{\tau,M}(\zeta) := \frac{(-1)^M\zeta^M}{Z_{\tau,0}}\mathrm{Tr}\bigg(&\int_0^{1} dt_1 \int_{0}^{t_1} dt_2 \ldots \int^{t_{M-1}}_0 dt_M \, \Theta_\tau(\xi)e^{-(1-t_1)H_{\tau,0}} \, \mcWt \, \\ 
&\times e^{-(t_1-t_2)H_{\tau,0}} \ldots \mcWt \, e^{-(t_{M-1}-t_M)H_{\tau,0}} \, \mcWt \\
&\times e^{-t_M(H_{\tau,0} + \zeta\mcWt)} f(\mcNt)\bigg)\,.
\end{align*}
\end{lemma}
We also define
\begin{equation}
\label{mfA_definition}
\mathfrak{A} := \{\mbt \in \R^{m} : 0 < t_m < t_{m-1} \ldots < t_{1} < 1\}\,.
\end{equation}

\subsection{Bounds on the explicit terms}
\label{Bounds on the explicit terms}
Throughout the following proofs, we will use without mention that for any function $g: \C \to \C$, $g(\mcNt)$ commutes with all operators on $\mcF$ that commute with $\mcNt$, which is clear from the definition of $g(\mcNt)$.
Namely, $g(\mcNt)$ acts on the $n^{\mathrm{th}}$ sector of Fock space as multiplication by $g(n/\tau)$.
In particular, all of the operators appearing in the integrands of $a^\xi_{\tau,m}$ and $R^\xi_{\tau,M}$ commute with $g(\mcNt)$.
\begin{lemma}
	\label{bounded_explicit_term_estimate_lemma}
For $m \in \N$, we have
\begin{equation}
	\label{explicit_estimate_wanted}
\left|a^\xi_{\tau,m}\right| \leq  \frac{K^p\|\xi\|\left(K^2\|w\|_{L^\infty}\right)^m}{2^m m!}.
\end{equation}
\end{lemma}
\begin{proof}
Lemma \ref{Holder's_inequality_Schatten} implies
\begin{align}
	\nonumber
\left|a^\xi_{\tau,m}\right| \leq \frac{1}{Z_{\tau,0}}&\int_0^{1} dt_1 \int_{0}^{t_1} dt_2 \ldots \int^{t_{m-1}}_0 dt_m \left\|\Theta_\tau(\xi)f^{\frac{1}{m+1}}(\mcNt)\right\|_{\mathfrak{S}^{\infty}}\\
\nonumber
&\times \left\|e^{-(1-t_1)H_{\tau,0}}\right\|_{\mathfrak{S}^{\frac{1}{1-t_1}}} \left\|\mcWt f^{\frac{1}{m+1}}(\mcNt)\right\|_{\mathfrak{S}^{\infty}} \left\|e^{-(t_1-t_2)H_{\tau,0}}\right\|_{\mathfrak{S}^{\frac{1}{t_1-t_2}}} \\
\label{quantum_explicit_bound_1}
&\times \ldots \left\|\mcWt f^{\frac{1}{m+1}}(\mcNt)\right\|_{\mathfrak{S}^{\infty}} \left\|e^{-t_m H_{\tau,0}}\right\|_{\mathfrak{S}^{\frac{1}{t_m}}}.
\end{align}
Since $e^{-sH_{\tau,0}}$ is a positive operator for $s \in [0,1]$, we have $\|e^{-sH_{\tau,0}}\|_{\mathfrak{S}^{1/s}} = \left(Z_{\tau,0}\right)^s$. So it follows from \eqref{quantum_explicit_bound_1} that
\begin{equation}
\label{quantum_explicit_bound_2}
\left|a^{\xi}_{\tau,m}\right| \leq \frac{Z_{\tau,0}}{Z_{\tau,0}} \frac{1}{m!} \left\|\Theta_\tau(\xi)f^{\frac{1}{m+1}}(\mcNt)\right\|_{\mathfrak{S}^{\infty}} \left\|\mcWt f^{\frac{1}{m+1}}(\mcNt)\right\|^m_{\mathfrak{S}^{\infty}}.
\end{equation}
From Lemma \ref{Theta_quantum_bound}, for fixed $n$ we have
\begin{equation}
	\label{quantum_lift_nth_bound}
\left\|\Theta_\tau(\xi)f^{\frac{1}{m+1}}(\mcNt)\big|_{\mfh^{(n)}}\right\|_{\mathfrak{S}^{\infty}} \leq \left(\frac{n}{\tau}\right)^p\left|f^{\frac{1}{m+1}}\left(\frac{n}{\tau}\right)\right|\|\xi\| \leq K^p\|\xi\|,
\end{equation}
where the final inequality follows from Assumption \ref{support_f}.
It follows from \eqref{quantum_lift_nth_bound} that, when viewed as an operator on $\mcF$
\begin{equation}
	\label{quantum_explicit_bound_3}
\left\|\Theta_\tau(\xi)f^{\frac{1}{m+1}}(\mcNt)\right\|_{\mathfrak{S}^{\infty}} \leq K^p \|\xi\|.
\end{equation}
To bound $\left\|\mcWt f^{\frac{1}{m+1}}(\mcNt)\right\|_{\mathfrak{S}^{\infty}}$ we note that $\mcWt$ acts on $\mfh^{(n)}$ as multiplication by
\begin{equation}
\label{mcWt_hn}
\frac{1}{\tau^2} \sum_{1 \leq i < j \leq n} w(x_i-x_j)\,.
\end{equation}
In particular, arguing as in \eqref{quantum_explicit_bound_3}, it follows that
\begin{equation}
\label{quantum_W_nth_bound}
\left\|\mcWt f^{\frac{1}{m+1}}(\mcNt)\right\|_{\mathfrak{S}^{\infty}} \leq \frac{1}{2}K^2\|w\|_{L^\infty}.
\end{equation}
Combining \eqref{quantum_explicit_bound_2} with \eqref{quantum_explicit_bound_3} and \eqref{quantum_W_nth_bound}, we have \eqref{explicit_estimate_wanted}.
\end{proof}

\subsection{Bounds on the remainder term}
\label{Bounds on the remainder term}
The following bound holds on the remainder term.
\begin{lemma}
	\label{quantum_bounded_remainder_bound_lemma}
	Let $M \in \N$, and $\mbt \in \mathfrak{A}$ (as in \eqref{mfA_definition}) be given. Define
	\begin{align*}
		\mathcal{R}^\xi_{\tau,M}(\mbt,\zeta) := \Theta&_\tau(\xi)e^{-(1-t_1)H_{\tau,0}}\,\mcWt\, e^{-(t_1-t_2)H_{\tau,0}} \ldots \\
		&\times \mcWt \, e^{-(t_{M-1}-t_M)H_{\tau,0}} \, \mcWt \, e^{-t_M(H_{\tau,0} + \zeta\mcWt)} f(\mcNt)\,.
	\end{align*}
	Then for any $\zeta \in \C$,
	\begin{equation}
	\label{quantum_remainder_term_bound}
	\frac{1}{Z_{\tau,0}}\left|\mathrm{Tr}\left(\mathcal{R}^{\xi}_{\tau,M}(\mbt,\zeta)\right)\right| \leq e^{\left|\mathrm{Re}(\zeta)\right| K^2 \|w\|_{L^\infty}}  \frac{K^p\|\xi\|\left(K^2\|w\|_{L^\infty}\right)^M}{2^M}.
	\end{equation}
\end{lemma}
\begin{proof}
	Define
	\begin{equation*}
		\mathcal{S}(\mbt) := \Theta_\tau(\xi)e^{-(1-t_1)H_{\tau,0}}\mcWt  e^{-(t_1-t_2)H_{\tau,0}} \ldots \mcWt \, e^{-(t_{M-1}-t_M)H_{\tau,0}} \mcWt.
	\end{equation*}
	Then
	\begin{equation}
	\label{Trace_curlyR_bound}
	\mathrm{Tr}\left(\mathcal{R}^\xi_{\tau,M}(\mbt,\zeta)\right) = \sum_{n \geq 0} \mathrm{Tr}\left(\left[\mathcal{S}(\mbt)f^{\frac{1}{2}}(\mcNt)\right]\left[e^{-t_M(H_{\tau,0}+\zeta\mcWt)}f^{\frac{1}{2}}(\mcNt)\right]\right)^{(n)},
	\end{equation}
where the trace on the left hand side of \eqref{Trace_curlyR_bound} is taken over Fock space, whereas on the right hand side for each term it is taken over the $n{\mathrm{th}}$ sector of Fock space. For $n \in N$, we have
	\begin{align*}
		\mathrm{Tr}&\left(\left[\mathcal{S}(\mbt)f^{\frac{1}{2}}(\mcNt)\right]\left[e^{-t_M(H_{\tau,0}+\zeta\mcWt)}f^{\frac{1}{2}}(\mcNt)\right]\right)^{(n)} \\ 
		&= \int_{\T^n} d\mbx \int_{\T^n} d\mby \left(\mcS(\mbt)f^{\frac{1}{2}}(\mcNt)\right)^{(n)}(\mby;\mbx)\left(e^{-t_M(H_{\tau,0}+\zeta\mcWt)}f^{\frac{1}{2}}(\mcNt)\right)^{(n)}(\mbx;\mby)\,.
	\end{align*}
	
We now rewrite $\left(e^{-t_M(H_{\tau,0}+\zeta\mcWt)}f^{\frac{1}{2}}(\mcNt)\right)^{(n)}(\mbx;\mby)$ using Proposition \ref{Feynman-Kac_formula}.
	\begin{align*}
\bigg(&e^{-t_M(H_{\tau,0}+\zeta\mcWt)} f^{\frac{1}{2}}(\mcNt)\bigg)^{(n)}(\mbx;\mby) \\
& =\int \mathbb{W}^{t_M}_{\mbx,\mby}(d\bm{\omega}) e^{-\frac{\kappa n}{\tau}t_M} e^{-\int^{t_M}_0 ds \, \zeta\left(\frac{1}{\tau^2}\sum_{1 \leq i < j \leq n} w_{ij}( \omega(s))\right)}
f^{\frac{1}{2}}\left(\frac{n}{\tau}\right),
	\end{align*}
where $\mathbb{W}^{t}_{\mbx,\mby}(d\bm{\omega}) := \prod_{i=1}^n \mathbb{W}_{x_i,y_i}^{t}(d\omega_i)$. Here we used that $$\left(\mcWt\right)^{(n)} (\mathbf{u};\mathbf{v}) = \frac{1}{\tau^2}\sum_{1 \leq i < j \leq n} w(u_i-u_j) \prod_{k=1}^n \delta(u_k-v_k)$$ and defined $w_{ij}(\mathbf{u}) := w(u_i-u_j)$ for $\mathbf{u} = (u_1,\ldots,u_n) \in \T^n$. Then
	\begin{align}
\nonumber
&\left|\left(e^{-t_M(H_{\tau,0}+\zeta\mcWt)}f^{\frac{1}{2}}(\mcNt)\right)^{(n)}(\mbx;\mby)\right| \\
		\nonumber
&\leq \int \mathbb{W}^{t_M}_{\mbx,\mby}(d\bm{\omega}) e^{-\frac{\kappa n}{\tau}t_M} \big| e^{-\int^{t_M}_0 ds \, \zeta\left(\frac{1}{\tau^2}\sum_{1 \leq i < j \leq n} w_{ij}( \bm{\omega}(s))\right)} \times f^{\frac{1}{2}}\left(\frac{n}{\tau}\right) \big| \\
		\label{remainder_nth_kernel_estimate}
&\leq \sup_{\bm{\omega}}\left| e^{-\int^{t_M}_0 ds \, \zeta\left(\frac{1}{\tau^2}\sum_{1 \leq i < j \leq n} w_{ij}( \bm{\omega}(s))\right)} f^{\frac{1}{2}}\left(\frac{n}{\tau}\right) \right| \left( e^{-t_M H_{\tau,0}} \right)^{(n)}(\mbx; \mby),
	\end{align}
where we have used Proposition \ref{Feynman-Kac_formula} in the second line. We have
	\begin{equation}
	\label{remainder_nth_kernel_estimate2}
	\sup_{\bm{\omega}} \left| e^{-\int^{t_M}_0 ds \, \zeta\left(\frac{1}{\tau^2}\sum_{1 \leq i < j \leq n} w_{ij}( \bm{\omega}(s))\right)} f^{\frac{1}{2}}\left(\frac{n}{\tau}\right) \right| \leq e^{|\mathrm{Re}(\zeta)|t_M \left(\frac{n}{\tau}\right)^2\|w\|_{L^\infty}} \left|f^{\frac{1}{2}}\left(\frac{n}{\tau}\right)\right|.
	\end{equation}
	It follows from \eqref{remainder_nth_kernel_estimate2} that
	\begin{equation}
	\label{remainder_nth_kernel_estimate3}
	\sup_{\bm{\omega}}\left| e^{-\int^{t_M}_0 ds \, \zeta\left(\frac{1}{\tau^2}\sum_{1 \leq i < j \leq n} w_{ij}( \bm{\omega}(s))\right)} f^{\frac{1}{2}}\left(\frac{n}{\tau}\right) \right| \leq e^{|\mathrm{Re}(\zeta)|K^2\|w\|_{L^\infty}}.
	\end{equation}
	Combining \eqref{remainder_nth_kernel_estimate} with \eqref{remainder_nth_kernel_estimate3} and the triangle inequality, we have shown
	\begin{equation}
	\label{remainder_term_FK_bound}
	\left|\left(e^{-t_M(H_{\tau,0}+\zeta\mcWt)}f^{\frac{1}{2}}(\mcNt)\right)^{(n)}(\mbx;\mby)\right| \leq e^{|\mathrm{Re}(\zeta)|K^2\|w\|_{L^\infty}} \left( e^{-t_M H_{\tau,0}} \right)^{(n)}(\mbx; \mby)\,.
	\end{equation}
	Combining \eqref{Trace_curlyR_bound} with \eqref{remainder_term_FK_bound}, it follows that
	\begin{multline*}
		\big|\mathrm{Tr}\left(\mathcal{R}^\xi(\mbt,\zeta)\right)\big| \leq e^{|\mathrm{Re}(\zeta)|K^2\|w\|_{L^\infty}} \mathrm{Tr}\bigg(\int_{0}^{1}dt_1 \int_{0}^{t_1}dt_2 \ldots \int_{0}^{t_{M-1}} dt_M \, \Theta_\tau(\tilde{\xi})\,
		\\
		\times e^{-(1-t_1)H_{\tau,0}} \, \widetilde{\mcWt}e^{-(t_1-t_2)H_{\tau,0}} \widetilde{\mcWt}e^{-(t_2-t_3)H_{\tau,0}} \ldots
		\widetilde{\mcWt}e^{-t_MH_{\tau,0}} f^{\frac{1}{2}}(\mcNt) \bigg)\,,
	\end{multline*}
	where $\tilde{\xi}$ is the operator with kernel $|\xi|$ and
	\begin{align}
	\label{mcWt_tilde}	
	\widetilde{\mcWt} := \frac{1}{2} \int dx\,dy\, \varphi_\tau^*(x)\varphi_\tau^*(y) |w(x-y)| \varphi_\tau(x)\varphi_\tau(y)\,.
	\end{align}
	Then \eqref{quantum_remainder_term_bound} follows by arguing as in the proof of Lemma \ref{bounded_explicit_term_estimate_lemma}.
\end{proof}
Integrating \eqref{quantum_remainder_term_bound} in the variables $\mbt \in \mathfrak{A}$, as defined in \eqref{mfA_definition}, implies
\begin{equation}
\label{quantum_final_bound_remainder}
\left|R^\xi_{\tau,M}(\zeta)\right| \leq e^{|\mathrm{Re}(\zeta)|K^2\|w\|_{L^\infty}}  \frac{K^p\|\xi\|\left(K^2\|w\|_{L^\infty}\right)^M}{2^M M!}\,|\zeta|^M.
\end{equation}
We note that this converges to $0$ as $M \rightarrow \infty$ for any fixed $\zeta \in \C$. Moreover, since the radius of convergence of $a_{\tau,m}^\xi$ is infinite by Lemma \ref{bounded_explicit_term_estimate_lemma}, we conclude the following proposition.
\begin{proposition}
	\label{quantum_function_analytic_prop}
	The function $A^\xi_{\tau}(\zeta) = \sum_{m=0}^\infty a^\xi_{\tau,m} \zeta^m$ is analytic on $\C$.
\end{proposition}

\subsection{The classical setting}
\label{The classical setting}
We now analyse the analogous expansion in the classical setting.
Let us note that
\begin{equation}
\label{classical_state_identity}
\rho(\Theta(\xi)) = \frac{\tilde{\rho}_{1}(\Theta(\xi))}{\tilde{\rho}_{1}(\mathbf{1})}\,,
\end{equation}
where
\begin{equation}
\label{tilde_rho_zeta}
\tilde{\rho}_\zeta(X) := \int d\mu \, X e^{-\zeta\mcW}f(\mcN)\,.
\end{equation}
Define
\begin{align*}
	A^\xi(\zeta) &:= \tilde{\rho}_{\zeta}(\Theta(\xi))\,.
\end{align*}
Then, for $M \in \N$
\begin{equation*}
	A^\xi(\zeta) = \sum_{m=0}^{M-1} a_{m}^\xi \zeta^m + R^\xi_{M}(\zeta)\,,
\end{equation*}
where
\begin{align}
	\label{explicit_term_classical} 
	a^\xi_m &:= \frac{(-1)^m}{m!} \int d\mu \, \Theta(\xi)\mcW^m f(\mcN)\,\\
	\label{classical_remainder}
	R^\xi_{M}(\zeta) &= \frac{(-1)^M \zeta^M}{M!} \int d\mu \,  \Theta(\xi)\mcW^M f(\mcN) e^{-\tilde{\zeta}\mcW} \quad \text{for some} \quad \tilde{\zeta} \in [0,\zeta].
\end{align}
\begin{lemma}
	\label{Classical_bounded_explicit_bound_lemma}
	For each $m \in \N$, we have
	\begin{equation}
	\label{Classical_bounded_explicit_bound}
	\left|a^\xi_m\right| \leq \frac{K^p\|\xi\|\left(K^2\|w\|_{L^\infty}\right)^m}{2^m m!}
	\end{equation}
\end{lemma}
\begin{proof}
	We have
	\begin{align}
		\label{classical_explicit_bound1}
		\left|a^\xi_m\right| \leq \frac{1}{m!} \int d\mu \,\left|\Theta(\xi)f^{\frac{1}{m+1}}(\mcN)\right| \left|\mcW f^{\frac{1}{m+1}}(\mcN)\right|^m.
	\end{align}
	From Lemma \ref{Theta_classical_bound} and Assumption \ref{support_f}, we have
	\begin{equation}
	\label{classical_lift_bound}
	\left|\Theta(\xi)f^{\frac{1}{m+1}}(\mcN)\right| \leq \left\|f^{\frac{1}{m+1}}\right\|_{L^\infty} K^p \|\xi\|.
	\end{equation}
	Moreover, Lemma \ref{classical_W_bound} \eqref{classical_W_bound_1} and Assumption \ref{support_f} imply
	\begin{equation}
	\label{classical_potential_bound}
	\left|\mcW f^{\frac{1}{m+1}}(\mcN)\right| \leq \frac{1}{2} \|w\|_{L^\infty} \left\|f^{\frac{1}{m+1}}\right\|_{L^\infty} K^2.
	\end{equation}
	Recalling $\|f\|_{L^\infty} \leq 1$, \eqref{Classical_bounded_explicit_bound} follows from \eqref{classical_explicit_bound1} combined with \eqref{classical_lift_bound} and \eqref{classical_potential_bound}.
\end{proof}
Note that Lemma \ref{classical_W_bound} implies that
\begin{equation*}
	\left|e^{-\tilde{\zeta}\mcW}f^{\frac{1}{M+2}}(\mcN)\right| \leq e^{\frac{1}{2} |\mathrm{Re}(\zeta)| K^2 \|w\|_{L^\infty}}
\end{equation*}
for $\tilde{\zeta} \in [0,\zeta]$. Applying the same arguments as the proof of Lemma \ref{Classical_bounded_explicit_bound_lemma}, we have the following lemma.
\begin{lemma}
	\label{Classical_bounded_remainder_bound_lemma}
	For any $M \in \N$, we have
	\begin{equation}
	\label{Classical_bounded_remainder_bound}
	\left|R^\xi_M(\zeta)\right| \leq e^{\frac{1}{2} |\mathrm{Re}(\zeta)| K^2 \|w\|_{L^\infty}} \frac{K^p\|\xi\|\left(K^2\|w\|_{L^\infty}\right)^M}{M! \,2^M}|\zeta|^M.
	\end{equation}
\end{lemma}
Like in the quantum case, for each $\zeta \in \C$, $R^\xi_M(\zeta)$ converges to $0$ as $M\to \infty$ and $a^\xi_m$ has infinite radius of convergence, so we have the following result.
\begin{proposition}
	\label{classical_function_analytic_prop}
	The function $A^\xi(\zeta) = \sum_{m=0}^\infty a_m^\xi \zeta^m$ is analytic in $\C$.
\end{proposition}

\subsection{Convergence of the Explicit Terms}
\label{Convergence of the Explicit Terms}
When analysing the convergence of the explicit terms, we argue similarly as in \cite[Section 3.1]{FKSS18} and rewrite $f(\mathcal{N}_{\#})$ as an integral of the form
\begin{equation}
\label{f(N_sharp)_1}
f(\mathcal{N}_{\#})=\int_\C d\zeta \, \frac{\psi(\zeta)}{\mcN_\# - \zeta}\,,
\end{equation}
for suitable $\psi \in C_c^{\infty}(\C)$.
For the precise setup, see \eqref{psi_inequality}--\eqref{f(N_sharp)} below. 
Using \eqref{f(N_sharp)_1}, we use that 
\begin{equation}
\label{1/{f(N)-zeta}}
\frac{1}{\mcN_\# - \zeta} = \int_{0}^{\infty} d\nu \, e^{-\nu(\mcN_\# - \zeta)}\,,
\end{equation}
for $\mathrm{Re} \, \zeta < 0$, which leads us to analyse analogues of \eqref{explicit_term_quantum} and \eqref{explicit_term_classical} without the truncation $f(\mathcal{N})$ and with chemical potential shifted by $\nu>0$. More precisely, we note the following boundedness and convergence result. We recall that in this section, we are considering $w \in L^{\infty}$.

\begin{lemma}
	\label{nu_hamiltonian_proposition}
	Fix $\nu > 0$. We recall $\mathcal{B}_p$ given by \eqref{B_p} and consider $\xi \in \mathcal{C}_p$, where
	\begin{equation}
	\label{C_p}
	\mathcal{C}_p := \mathfrak{B}_p \cup \{\mathbf{1}_p\}\,. 
	\end{equation}
	Let
	\begin{align*}
		b^{\xi,\nu}_{\tau,m} := \frac{(-1)^m}{Z_{\tau,0}}&\mathrm{Tr}\bigg( \int_0^{1} dt_1 \int_{0}^{t_1} dt_2 \ldots \int^{t_{m-1}}_0 dt_m \, \Theta_{\tau} (\xi) e^{-(1-t_1)(H_{\tau,0}  + \nu \mcNt)} \\
		& \times \mcWt e^{-(t_1-t_2)(H_{\tau,0}  + \nu \mcNt)} \mcWt e^{-(t_2-t_3)(H_{\tau,0} + \nu \mcNt)} \ldots \\
		& \times e^{-(t_{m-1}-t_m)(H_{\tau,0}  + \nu \mcNt)} \mcWt
		e^{-t_m (H_{\tau,0}  + \nu \mcNt)}\bigg) \\
		b^{\xi,\nu}_m := \frac{(-1)^m}{m!}&\int d\mu \, \Theta(\xi)\mcW^me^{-\nu \mcN}.
	\end{align*}
	Then, the following results hold
	\begin{enumerate}
		\item $\left| b^{\xi,\nu}_{\#,m}\right| \leq C(m,p,\nu)$.
		\item $b^{\xi,\nu}_{\tau,m} \to b^{\xi,\nu}_m$ as $\tau \to \infty$ uniformly in $\xi \in \mathcal{C}_p$.
	\end{enumerate}
\end{lemma}
\begin{proof}
	Let us first consider the case when $\xi \in \mathfrak{B}_p$. We define
	\begin{equation*}
		h^{\nu} := h+\nu = \sum_{k \in \N} (\lambda_k + \nu)u_ku^*_k\,.
	\end{equation*}
Then the deformed classical state defined by
	\begin{align}
		\widetilde{\rho}_{0}^{\nu}(X) &:= \frac{\int d\mu \, Xe^{-\nu \mcN}}{\int d\mu \, e^{-\nu \mcN}}
	\end{align}
	satisfies a Wick theorem with Green function given by $G^\nu := \frac{1}{h^{\nu}}$. This follows directly from Proposition \ref{Wicksthm}, since all we have done is shift the chemical potential by $\nu$.

	Moreover, the deformed quasi-free state  defined by
	\begin{align}
		\widetilde{\rho}_{\tau,0}^{\nu}(\mathcal{A}) &:= \frac{\mathrm{Tr}\left(\mathcal{A} \, e^{-H_{\tau,0}-\nu\mcNt}\right)}{\mathrm{Tr}\left(e^{-H_{\tau,0} - \nu\mcNt}\right)}
	\end{align}
	satisfies a quantum Wick theorem similar to \cite[Lemma B.1]{FKSS17} with quantum Green function $G_\tau = \frac{1}{\tau(e^{h/\tau} - 1)}$ replaced by
	\begin{equation*}
		G^{\nu}_\tau := \frac{1}{\tau(e^{h^\nu/\tau} -1)}\,.
	\end{equation*}
	In particular, we have that $\|G_{\#}^\nu\|_{\mathfrak{S}^2} \leq \|G_\#\|_{\mathfrak{S}^2}<\infty$. Let us define
	\begin{align*}
		\tilde{b}^{\xi,\nu}_{\tau,m} := \frac{(-1)^m}{\mathrm{Tr}\left(e^{-H_{\tau,0} - \nu\mcNt}\right)}\mathrm{Tr}\bigg( \int_0^{1} dt_1 \int_{0}^{t_1} dt_2 \ldots \int^{t_{m-1}}_0 dt_m \, \Theta_{\tau} (\xi) e^{-(1-t_1)(H_{\tau,0}  + \nu \mcNt)} \\
		 \times \mcWt e^{-(t_1-t_2)(H_{\tau,0}  + \nu \mcNt)} \mcWt e^{-(t_2-t_3)(H_{\tau,0} + \nu \mcNt)} \ldots \\
		 \times e^{-(t_{m-1}-t_m)(H_{\tau,0}  + \nu \mcNt)} \mcWt
		e^{-t_m (H_{\tau,0}  + \nu \mcNt)}\bigg)
		\end{align*}
and
\begin{equation*}		
\tilde{b}^{\xi,\nu}_m := \frac{1}{m!}\,\frac{(-1)^m}{\int d\mu \, e^{-\nu \mcN}}\,\int d\mu \, \Theta(\xi)\mcW^me^{-\nu \mcN}\,.
\end{equation*}

	Noting that noting that the arguments in \cite[Sections 2.3--2.6]{FKSS17} concerning explicit terms do not use any positivity properties of $w$, we hence obtain that the following properties hold.
	\begin{enumerate}
		\item[(1')] $\left| \tilde{b}^{\xi,\nu}_{\#,m}\right| \leq C(m,p,\nu)$.
		\item[(2')] $\tilde{b}^{\xi,\nu}_{\tau,m} \to \tilde{b}^{\xi,\nu}_m$ as $\tau \to \infty$ uniformly in $\xi \in \mathcal{B}_p$.
	\end{enumerate}
More precisely, $(1')$ and $(2')$ correspond to the 1 dimensional versions\footnote{Throughout the paper, when referring to \cite[Corollary 2.21, Proposition 2.26]{FKSS17}, we mean these 1 dimensional versions.} of \cite[Corollary 2.21, Proposition 2.26]{FKSS17} proved in \cite[Section 4.1]{FKSS17}, as well as \cite[Lemma 3.1]{FKSS17}. 

When $\xi \in \mathfrak{B}_p$, we deduce the claim from (1') and (2') by noting that by \cite[Lemma 3.4]{FKSS18}, we have
\begin{equation*}
\lim_{\tau \rightarrow \infty} \frac{\mathrm{Tr}\left(\mathcal{A} \, e^{-H_{\tau,0}-\nu\mcNt}\right)}{\mathrm{Tr}\left(\mathcal{A} \, e^{-H_{\tau,0}}\right)}=\int d\mu \, e^{-\nu \mcN}\,.
\end{equation*}
It remains to consider the case when $\xi = \mathbf{1}_p$ is the identity operator on $\mfh^{(p)}$. We then have
	\begin{equation}
	\label{identity_kernel}
	\xi(x_1,\ldots,x_p;y_1,\ldots,y_p) = \prod_{j=1}^p \delta(x_j-y_j)\,. 
	\end{equation} 
	Since $\widetilde{\rho}_{\tau,0}$ satisfies the quantum Wick theorem, we can argue analogously as in \cite[Section 4.2]{FKSS17} to get the required bounds and convergence as before. We omit the details.
\end{proof}

We also need the following result.
\begin{lemma}
	\label{operator_projection_bound_lemma}
	Let $\mathcal{A}: \mathcal{F} \to \mathcal{F}$ and $g \in L^\infty(\R)$. Then $|\mathrm{Tr}(\mathcal{A}\,g(\mcNt))| \leq \|g\|_{L^\infty} \, \mathrm{Tr}(\hat{\mathcal{A}})$, where $\hat{\mathcal{A}}^{(n)}$ has kernel $\left|\mathcal{A}^{(n)}(x;y)\right|$.
\end{lemma}

\begin{proof}
	For an operator $\mathcal{A}: \mathcal{F} \to \mathcal{F}$, we define $\mathcal{A}^{(n)} := P^{(n)}\,\mathcal{A}\,P^{(n)}$, where $P^{(n)}$	is the projection of an operator on Fock space to the $n{\mathrm{th}}$ component of Fock space. We also define $\hat{\mathcal{A}} := \oplus_{n \geq 0} \hat{\mathcal{A}}^{(n)}$. We have
	\begin{align*}
		|\mathrm{Tr}(\mathcal{A}\,g(\mcNt))| &= \left|\sum_{n\geq0} \int_{\T^n} d\mathbf{x} \, \mathcal{A}^{(n)}(\mathbf{x};\mathbf{x}) g\left(\frac{n}{\tau}\right)\right| \\
		&\leq \sup_{n\geq0} \left\{\left|g\left(\frac{n}{\tau}\right) \right| \right\} \sum_{n \geq 0} \int_{\T^n} d\mathbf{x} \, |A^{(n)}(\mathbf{x};\mathbf{x})| \\
		&\leq \|g\|_{L^\infty} \mathrm{Tr}(\hat{\mathcal{A}})\,.
	\end{align*}
\end{proof}
\begin{lemma}
	\label{bounded_explicit_term_convergence_lemma}
	We recall the definitions \eqref{explicit_term_quantum} and \eqref{explicit_term_classical}. For each $m \in \N$, we have
	\begin{equation}
	\label{explicit_convergence}
	\lim_{\tau \to \infty}a^\xi_{\tau,m} = a^\xi_m
	\end{equation}
	uniformly in $\xi \in\mathcal{C}_p$, defined in \eqref{C_p}.
\end{lemma}
\begin{proof}
	For $\zeta \in \C \backslash [0,\infty)$, we define
	\begin{align*}
		\alpha^{\xi}_{\tau,m} (\zeta) := \frac{1}{Z_{\tau,0}} \mathrm{Tr}\bigg( &\int_0^{1} dt_1 \int_{0}^{t_1} dt_2 \ldots \int^{t_{m-1}}_0 dt_m \, \Theta_{\tau} (\xi) e^{-(1-t_1)H_{\tau,0}}\mcWt e^{-(t_1-t_2)H_{\tau,0}} \\
		& \times \mcWt \ldots e^{-(t_{m-1}-t_m)H_{\tau,0}} \mcWt e^{-t_m H_{\tau,0}}\frac{1}{\mcNt-\zeta} \bigg)\,
	\end{align*}
	and
	\begin{equation}
	\label{classical_alpha}
	\alpha^\xi_m (\zeta) := \frac{1}{m!}\int d\mu \, \Theta(\xi) \mathcal{W}^m \frac{1}{\mcN-\zeta}\,.
	\end{equation}
	We prove that $\alpha_{\tau,m}^\xi$ and $\alpha^\xi_m$ are analytic in $\zeta \in \C \backslash [0,\infty)$. We first deal with $\alpha^\xi_m$. Note that
	\begin{equation*}
		\left|\alpha^\xi_m(\zeta)\right| \leq \frac{1}{m!}\int d\mu \left|\Theta(\xi)\mathcal{W}^m\right|\left|\frac{1}{\mcN - \zeta}\right|.
	\end{equation*}
	Using Lemma \ref{Theta_classical_bound}, Lemma \ref{classical_W_bound} \eqref{classical_W_bound_1}, and that $\int d\mu \, \|\varphi\|_{\mathfrak{h}}^{2p} \leq C(p)$ by Remark \ref{finite_phi_integral_remark}, we have 
	\begin{equation}
	\label{classical_alpha_estimate}
	|\alpha^\xi_m(\zeta)| \leq \frac{C(m,p)}{\mathrm{max}\{-\mathrm{Re} \, \zeta,|\mathrm{Im} \, \zeta|\}} .
	\end{equation}
	Arguing similarly to \eqref{classical_alpha_estimate}, it follows that
	\begin{equation*}
		\frac{1}{m!}\int d\mu \left|\Theta(\xi)\mathcal{W}^m\right|\left|\frac{1}{\left(\mcN - \zeta \right)^2}\right| \leq \frac{C(m,p)}{\mathrm{max}\{-\mathrm{Re} \, \zeta,|\mathrm{Im} \, \zeta|\}^2}\,,
	\end{equation*}
	so by the dominated convergence theorem, we can differentiate under the integral sign in \eqref{classical_alpha} and conclude that $\alpha^\xi_m$ is analytic in $\C \backslash [0,\infty)$.
	
	To show $\alpha_{\tau,m}^\xi$ is analytic in $\C \backslash [0,\infty)$, we first note that $\frac{1}{\mcNt-\zeta}$ acts as multiplication by $\frac{1}{(n/\tau)-\zeta}$ on the $n{\mathrm{th}}$ sector of Fock space. By using Lemma \ref{operator_projection_bound_lemma} we get
	\begin{align*}
		|\alpha_{\tau,m}(\zeta)| \leq \frac{1}{Z_{\tau,0}} \mathrm{Tr}\bigg( \bigg[ &\int_0^{1} dt_1 \int_{0}^{t_1} dt_2 \ldots \int^{t_{m-1}}_0 dt_m \, \Theta_{\tau} (\xi) e^{-(1-t_1)H_{\tau,0}}\mcWt e^{-(t_1-t_2)H_{\tau,0}} \\
		& \times \mcWt \ldots e^{-(t_{m-1}-t_m)H_{\tau,0}} \mcWt e^{-t_m H_{\tau,0}} \bigg] \,\,^{\widehat{ }} \,\, \bigg) \frac{1}{\mathrm{max}\{-\mathrm{Re} \, \zeta,|\mathrm{Im} \, \zeta|\}} \\
		\leq \frac{1}{Z_{\tau,0}} \mathrm{Tr}\bigg( &\int_0^{1} dt_1 \int_{0}^{t_1} dt_2 \ldots \int^{t_{m-1}}_0 dt_m \, \Theta_{\tau} (\tilde{\xi}) e^{-(1-t_1)H_{\tau,0}}\widetilde{\mcWt} e^{-(t_1-t_2)H_{\tau,0}} \\
		& \times \widetilde{\mcWt} \ldots e^{-(t_{m-1}-t_m)H_{\tau,0}} \widetilde{\mcWt} e^{-t_m H_{\tau,0}}\bigg) \frac{1}{\mathrm{max}\{-\mathrm{Re} \, \zeta,|\mathrm{Im} \, \zeta|\}},
	\end{align*}
	where we recall $\tilde{\xi}$ is the operator with kernel $|\xi|$, and $\widetilde{\mcWt}$ is as in \eqref{mcWt_tilde}. Applying \cite[Corollary 2.21]{FKSS17}, we have
	\begin{equation}
	\label{analytic_alpha_tau_estimate}
	|\alpha^\xi_{\tau,m}(\zeta)| \leq \frac{C(m,p)}{\mathrm{max}\{-\mathrm{Re} \, \zeta,|\mathrm{Im} \, \zeta|\}}.
	\end{equation}
	Define
	\begin{equation}
	\label{restricted_fock_space_defn}
		\mfh^{(\leq p)} := \bigoplus_{n = 0}^p \mfh^{(n)}\,,
	\end{equation}
	and
	\begin{equation*}
		P^{(\leq p)} : \mathcal{F} \to \mfh^{(\leq p)}
	\end{equation*}
	as the orthogonal projection. Define
	\begin{align*}
		\alpha_{\tau,m,n}(\zeta) := \frac{1}{Z_{\tau,0}} \,\mathrm{Tr}\bigg( &\int_0^{1} dt_1 \int_{0}^{t_1} dt_2 \ldots \int^{t_{m-1}}_0 dt_m \, P^{(\leq n)} \, \Theta_{\tau} (\xi) \,e^{-(1-t_1)H_{\tau,0}}\,\mcWt \\
		& \times e^{-(t_1-t_2)H_{\tau,0}} \,\mcWt \ldots e^{-(t_{m-1}-t_m)H_{\tau,0}}\, \mcWt e^{-t_m H_{\tau,0}} \frac{1}{\mcN_\tau - \zeta} \bigg)\,.
	\end{align*} 
	Since $P^{(\leq n)}$ commutes with $\Theta_\tau(\xi)$, $H_{\tau,0}$, and $\mcWt$, it follows that $\alpha_{\tau,m,n}$ is analytic in $\C \backslash [0,\infty)$. By construction we have $\lim_{n \to \infty} \alpha_{\tau,m,n}(\zeta) = \alpha_{\tau,m}(\zeta)$ for all $\zeta \in \C \backslash [0,\infty)$ and by the same argument as \eqref{analytic_alpha_tau_estimate}, we have
	\begin{equation}
	\label{analytic_truncated_alpha_tau_estimate}
	|\alpha^\xi_{\tau,m,n}(\zeta)| \leq \frac{C(m,p)}{\mathrm{max}\{-\mathrm{Re} \, \zeta,|\mathrm{Im} \, \zeta|\}}\,.
	\end{equation}
	The pointwise convergence, \eqref{analytic_truncated_alpha_tau_estimate} and the dominated convergence theorem imply that
	\begin{equation*}
		\lim_{n \to \infty} \int_{\partial T} d \zeta \, \alpha_{\tau,m,n}(\zeta) = \int_{\partial T} d\zeta \, \alpha_{\tau,m}(\zeta)
	\end{equation*}
	for any triangle $T$ contained in $\C \backslash [0,\infty)$. Morera's theorem implies that $\alpha_{\tau,m}$ is analytic in $\C \backslash [0,\infty)$.
	
	We now prove that $\alpha^\xi_{\tau,m}(\zeta) \to \alpha^\xi_m(\zeta)$ as $\tau \to \infty$ for all $\zeta \in \C \backslash [0,\infty)$. First, for $\mathrm{Re} \, \zeta <0$, we recall \eqref{1/{f(N)-zeta}}.
Therefore, 
	\begin{align}
		\nonumber
		\left|\alpha_{\tau,m}^\xi(\zeta) -\alpha^\xi_m(\zeta)\right| = \bigg|&\frac{1}{Z_{\tau,0}} \mathrm{Tr}\bigg( \int_0^{1} dt_1 \int_{0}^{t_1} dt_2 \ldots \int^{t_{m-1}}_0 dt_m \, \Theta_{\tau} (\xi) e^{-(1-t_1)H_{\tau,0}}\mcWt \\
		\nonumber
		& \times e^{-(t_1-t_2)H_{\tau,0}} \mcWt \ldots e^{-(t_{m-1}-t_m)H_{\tau,0}} \mcWt e^{-t_m H_{\tau,0}}\frac{1}{\mcNt-\zeta} \bigg) \\
		\nonumber 
		& - \frac{1}{m!} \int d\mu \, \mcW^m \frac{1}{\mcN-\zeta} \bigg| \\
		\nonumber
		\leq \int_{0}^\infty d\zeta \, e^{\nu\zeta} \, \bigg| \frac{1}{Z_{\tau,0}}&\mathrm{Tr}\bigg( \int_0^{1} dt_1 \int_{0}^{t_1} dt_2 \ldots \int^{t_{m-1}}_0 dt_m \, \Theta_{\tau} (\xi) e^{-(1-t_1)(H_{\tau,0}  + \nu \mcNt)} \\
		\nonumber
		& \times \mcWt e^{-(t_1-t_2)(H_{\tau,0}  + \nu \mcNt)} \mcWt \ldots e^{-(t_{m-1}-t_m)(H_{\tau,0}  + \nu \mcNt)} \mcWt \\
		\label{alpha_difference_estimate}
		& \times e^{-t_m (H_{\tau,0}  + \nu \mcNt)}\bigg)  - \frac{1}{m!} \int d\mu \, e^{-\nu \mcN}\mcW^m\bigg|\,,
	\end{align}
	where we have used part (1) of Lemma \ref{nu_hamiltonian_proposition} and $\mathrm{Re} \, \zeta < 0$ to apply Fubini's theorem. Lemma \ref{nu_hamiltonian_proposition}, \eqref{alpha_difference_estimate}, and the dominated convergence theorem give 
	\begin{equation}
	\label{quantum_alpha_convergence}
	\lim_{\tau \to \infty} \alpha_{\tau,m}^\xi(\zeta) = \alpha_m^\xi(\zeta)
	\end{equation}
	uniformly in $\xi \in \mathcal{C}_p$ for $\mathrm{Re} \, \zeta < 0$.
	
	We define $\beta_{\tau,m}^\xi:=\alpha_{\tau,m}^\xi - \alpha_m^\xi$. We follow the argument in \cite[Proposition 3.3]{FKSS18} to prove
	\begin{equation}
	\label{uniform_convergence_beta}
	\lim_{\tau \to \infty} \sup_{\xi \in \mathcal{C}_p} |\beta_{\tau,m}^\xi(\zeta)| = 0 \, \text{ for all $\zeta \in \C \backslash [0,\infty)$}\,.
	\end{equation}
	From the analyticity of $\alpha_{\#,m}^\xi$ on $\C \backslash [0,\infty)$, \eqref{classical_alpha_estimate} and \eqref{analytic_alpha_tau_estimate}, and \eqref{quantum_alpha_convergence}, we know that $\beta^\xi_{\tau,m}$ satisfy the following properties.
	\begin{enumerate}
		\item $\beta^\xi_{\tau,m}$ is analytic on $\C \backslash [0,\infty)$.
		\label{beta_property1}
		\item $\lim_{\tau \to \infty} \sup_{\xi \in \mathcal{C}_p} |\beta_{\tau,m}^\xi(\zeta)| = 0$ for all $\mathrm{Re \, \zeta} < 0$.
		\label{beta_property2}
		\item $\sup_{\xi \in \mathcal{C}_p} |\beta_{\tau,m}^\xi(\zeta)| \leq \frac{C(m,p)}{|\mathrm{Im}\,\zeta|}$ for all $\zeta \in \C \backslash [0,\infty)$.
		\label{beta_property3}
	\end{enumerate}
	Given $\varepsilon > 0$, define
	\begin{equation*}
		\mathcal{D}_{\varepsilon} := \{\zeta : \mathrm{Im} \, \zeta > \varepsilon\}
	\end{equation*}
	and 
	\begin{equation*}
		\mathcal{T}_{\varepsilon} := \{\zeta_0 \in \mathcal{D}_\varepsilon : \lim_{\tau \to \infty} \sup_{\xi \in \mathcal{C}_p} \left|\partial_\zeta^n \beta_{\tau,m}^\xi(\zeta_0) = 0\right| \,  \text{for all  $n \in \N$}\,.
	\end{equation*}
	So $\mcTe$ is the set of points in  $\mcDe$ at which all $\zeta$-derivatives of $\beta^\xi_{\tau,m}$ converge to $0$ as $\tau \to \infty$ uniformly in $\xi \in \mathcal{C}_p$. Using properties \eqref{beta_property1}--\eqref{beta_property3} of $\beta$, Cauchy's integral formula, and the dominated convergence theorem, we have $\mcDe \cap \{\zeta: \mathrm{Re} \, \zeta < 0 \} \subset \mcTe$. In particular, $\mcTe$ is not empty.
	
	So to prove \eqref{uniform_convergence_beta} on $\mcDe$, it suffices to show that $\mcTe = \mcDe$. Since $\mcDe$ is connected, the latter claim follows from showing that $\mcTe$ is both open and closed in $\mcDe$. We first show that $\mcTe$ is open in $\mcDe$. Given $\zeta_0 \in \mcTe$, note that $B_{\varepsilon/2}(\zeta_0) \subset \mcD_{\varepsilon/2}$. So by property \eqref{beta_property3}, $\left|\beta_{\tau,m}^\zeta\right| \leq C(\varepsilon)$ on $B_{\varepsilon/2}(\zeta_0)$. Analyticity and Cauchy's integral formula imply that the Taylor series of $\beta_{\tau,m}^\xi$ at $\zeta_0$ converges on $B_{\varepsilon/2}(\zeta_0)$. So we can differentiate term by term and use the dominated convergence theorem and $\zeta_0 \in \mcTe$ to get that $B_{\delta}(\zeta_0) \subset \mcTe$ for $\delta \in (0,\varepsilon/2)$ sufficiently small such that $B_{\delta}(\zeta_0) \subset \mcDe$. So $\mcTe$ is open in $\mcDe$.
	
	To show that $\mcTe$ is closed in $\mcDe$, let $(\zeta_n)$ be a sequence in $\mcTe$ which converges to some $\zeta \in
	\mcDe$. Since $\zeta \in \mcDe$ which is open, there is $\varepsilon' \in (0,\varepsilon/2)$ such that $B_{\varepsilon'}(\zeta) \subset \mcDe$. Since $(\zeta_n) \to \zeta$, for $n$ sufficiently large, $\zeta \in B_{\varepsilon'/2}(\zeta_n)$. Since $B_{\varepsilon'/2}(\zeta_n) \subset B_{\varepsilon'}(\zeta) \subset \mcDe$, the argument that $\mcTe$ is open in $\mcDe$ implies that $B_{\varepsilon'/2}(\zeta_n) \subset \mcTe$. In particular, $\zeta \in \mcTe$, so $\mcTe$ is closed in $\mcDe$. By symmetry, the same argument shows that \eqref{uniform_convergence_beta} holds on $\widetilde{\mcD}_{\varepsilon} := \{\zeta : \mathrm{Im} \, \zeta < -\varepsilon\}$. Then \eqref{uniform_convergence_beta} holds on $\C \backslash [0,\infty)$ by letting $\varepsilon \to 0$ and recalling that \eqref{uniform_convergence_beta} holds for $\zeta < 0$ by property \eqref{beta_property2} above.
	
	Applying the Helffer-Sj\"ostrand formula and arguing as in \cite[(3.29)-(3.33)]{FKSS18}, we can find $\psi \in C^{\infty}_c(\C)$ satisfying
	\begin{equation}
	\label{psi_inequality}
	|\psi(\zeta)| \leq C|\mathrm{Im}\, \zeta|
	\end{equation} 
	such that
	\begin{equation}
	\label{f(N_sharp)}
		f(\mcN_\#) = \int_\C d\zeta \, \frac{\psi(\zeta)}{\mcN_\# - \zeta}\,.
	\end{equation}
	Then \eqref{classical_alpha_estimate}, \eqref{analytic_alpha_tau_estimate}, \eqref{psi_inequality}, and $\psi \in C^{\infty}_c(\C)$ imply that
	\begin{equation}
	\label{alpha_psi_estimate}
	|\alpha^{\xi}_{\#,m}(\zeta)\psi(\zeta)| \leq F(\zeta)
	\end{equation}
	for some $F \in L^1(\C)$. By \eqref{alpha_psi_estimate}, we can use Fubini's theorem to write
	\begin{equation*}
		|a^\xi_{\tau,m} - a^\xi_m| \leq \int_\C |\psi(\zeta)|\left|\alpha^\xi_{\tau,m}(\zeta) - \alpha^\xi_{m}(\zeta)\right|.
	\end{equation*}
	Using $\beta_{\tau,m}^\xi \to 0$ as $\tau \to \infty$ almost everywhere in $\C$ uniformly in $\xi$ and \eqref{alpha_psi_estimate}, the dominated convergence theorem implies \eqref{explicit_convergence}.
\end{proof}
\begin{remark}
	\label{finite_phi_integral_remark}
In the proof of Lemma \ref{bounded_explicit_term_convergence_lemma}, we used that $\int d\mu \, \|\varphi\|_{\mathfrak{h}}^{2p} \leq C(p) < \infty$. To see this, recall \eqref{random_classical_initial} implies
\begin{align*}
\int d\mu \, \|\varphi\|_{\mfh}^{2p} &= \E_\omega \left[\left(\sum_{n \in \N} \frac{|\omega_n|^2}{\lambda_n}\right)^p\right] \\
&= \E_{\omega} \left[\sum_{n_i \in \N} \frac{\omega_{n_1}\overline{\omega_{n_1}} \ldots \omega_{n_p}\overline{\omega_{n_p}}}{\lambda_{n_1} \ldots \lambda_{n_p}}\right] \\
& \leq C(p) \left(\sum_n \frac{1}{\lambda_n}\right)^p \leq C(p) < \infty\,.
\end{align*}
The final line follows from Proposition \ref{Wicksthm}.
\end{remark}
\subsection{Convergence of correlation functions. Proof of Theorem \ref{bounded_trace_class_final_convergence_theorem}}
\label{correlation_fn_argument}
We recall the class $\mathcal{C}_p$, defined in \eqref{C_p}. The following convergence result holds.
\begin{lemma}
	\label{bounded_power_series_convergence_lemma}
	$A^\xi_{\tau}(\zeta) \to A^\xi(\zeta)$ as $\tau \to \infty$ uniformly in $\xi \in \mathcal{C}_p$.
\end{lemma}
\begin{proof}
	Since $A^\xi_\#$ are analytic in $\C$, for all $\zeta \in \C$
	\begin{equation*}
		\sup_{\xi \in \mathcal{C}_p}\left|A^\xi_\tau(\zeta) - A^\xi(\zeta)\right| \leq \sum_m \sup_{\xi \in \mathcal{C}_p}\left| a^\xi_{\tau,m} - a^\xi_m\right| |\zeta|^m \to 0
	\end{equation*}
	as $\tau \to \infty$. Here we have used Lemma \ref{bounded_explicit_term_convergence_lemma}, Lemma \ref{bounded_explicit_term_estimate_lemma}, and the dominated convergence theorem. We also recall the notation \eqref{B_p}.
\end{proof}
Recalling \eqref{quantum_state_identity} and \eqref{classical_state_identity} and taking $\zeta = 1$, we have the following result.
\begin{corollary}
	\label{bounded_state_final_convergence_corollary}
	$\rho_\tau(\Theta_\tau(\xi)) \to \rho(\Theta(\xi))$ as $\tau \to \infty$ uniformly in $\xi \in \mathcal{C}_p$.
\end{corollary}
Before proceeding to the proof of Theorem \ref{bounded_trace_class_final_convergence_theorem}, we first need to prove the following technical lemma.
\begin{lemma}
\label{state_technical_lemma}
Recalling \eqref{classical_p_particle_correlation_fn} and \eqref{quantum_p_particle_correlation_fn}, we have $\gamma_{\#,p} \geq 0$ in the sense of operators.
\end{lemma}
\begin{proof}
	For $\eta \in \mathfrak{h}^{(p)}$, define the orthogonal projection $\Pi_\eta(\cdot) := \langle \eta,\cdot \rangle \eta$. Let us first note that
\begin{equation}
\label{state_computation}
\langle \eta,\gamma_{\#,p} \eta \rangle_{\mathfrak{h}^{(p)}}=\rho_\# (\Theta_\#(\Pi_\eta))\,.
\end{equation}
In the quantum setting, we use \eqref{quantum_p_particle_correlation_fn} and linearity to compute
\begin{multline}
\label{state_computation_1}
\langle \eta,\gamma_{\tau,p} \eta \rangle_{\mathfrak{h}^{(p)}}=\int dx_1\cdots dx_p\,dy_1\cdots dy_p \,\overline{\eta}(x_1,\ldots,x_p)\,\eta(y_1,\ldots,y_p)\,
\\
\times \rho_\tau \bigl(\varphi_{\tau}^*(y_1)\cdots \varphi^*_{\tau}(y_p) \varphi_\tau(x_1)\,\cdots \varphi_{\tau}(x_p)\bigr)
\\
=\rho_{\tau} \biggl(\int dx_1\cdots dx_p\,dy_1\cdots dy_p \,\overline{\eta}(x_1,\ldots,x_p)\,\eta(y_1,\ldots,y_p)\,
\\
\times
\varphi_{\tau}^*(y_1)\cdots \varphi^*_{\tau}(y_p) \varphi_\tau(x_1)\,\cdots \varphi_{\tau}(x_p) \biggr)\,.
\end{multline}
By \eqref{quantum_lift} and the definition of $\Pi_{\eta}$ we deduce that that the expression in \eqref{state_computation_1} equals $\rho_{\tau} (\Theta_{\tau}(\Pi_\eta))$, thus showing \eqref{state_computation} in the quantum setting. 
Similarly in the classical setting, we use \eqref{classical_p_particle_correlation_fn} and \eqref{classical_Theta_definition} to compute 
\begin{multline*}
\langle \eta,\gamma_p \eta \rangle_{\mathfrak{h}^{(p)}}=\rho \biggl(\int dx_1\cdots dx_p\,dy_1\cdots dy_p \,\overline{\eta}(x_1,\ldots,x_p)\,\eta(y_1,\ldots,y_p)\,
\\
\times
\overline{\varphi} (y_1)\cdots \overline{\varphi} (y_p) \varphi (x_1)\,\cdots \varphi (x_p) \biggr)=\rho (\Theta(\Pi_\eta))\,,
\end{multline*}
as was claimed.

We now show that the expression on the right-hand side of \eqref{state_computation} is non-negative. Let us first show this in the quantum setting. By \eqref{quantum_lift}, we note that  $\Theta_\tau(\Pi_\eta)$ is a positive operator.  Furthermore $f(\mcNt)$ is a positive operator which commutes with $\Theta_\tau(\Pi_\eta)$. In particular, their composition is a positive operator. Recalling \eqref{P_tau_definition}, we know that 
\begin{equation*}
\mathcal{A} \mapsto \frac{\mathrm{Tr}(\mathcal{A} P_\tau)}{\mathrm{Tr} (P_\tau)}
\end{equation*}
is a quantum state. In particular, when applied to positive operators it is nonnegative, so we obtain that
\begin{equation}
\label{state_computation_2}
\frac{\mathrm{Tr}(\Theta_\tau(\Pi_\eta) f(\mcNt) P_\tau)}{\mathrm{Tr} (P_\tau)} \geq 0\,.
\end{equation}
Since $P_\tau$ and $f(\mcNt)$ commute, by using \eqref{state_computation_2}, and recalling \eqref{P_tau_definition} as well as Assumption \ref{support_f}, it follows that
\begin{equation}
\label{state_computation_3}
\rho_\tau(\Theta_\tau(\Pi_\eta)) = \frac{\mathrm{Tr}(\Theta_\tau(\Pi_\eta) f(\mcNt) P_\tau )}{\mathrm{Tr}(P_\tau f(\mcNt))} \geq 0\,.
\end{equation}
We deduce the claim in the quantum setting from \eqref{state_computation} and \eqref{state_computation_3}.

In the classical setting, we use \eqref{classical_Theta_definition} to write 
\begin{multline}
\label{state_computation_4}
\rho (\Theta(\Pi_\eta))=\rho\biggl(\int dx_1\cdots dx_p\,dy_1\cdots dy_p \overline{\eta}(x_1,\ldots,x_p)\,\eta(y_1,\ldots,y_p)\,
\\
\times
\varphi(x_1) \cdots \varphi(x_p)\,\overline{\varphi}(y_1)\cdots \overline{\varphi}(y_p) \biggr)
\\
= \rho \biggl( \biggl| \int dx_1 \cdots dx_p \overline{\eta}(x_1,\ldots,x_p)\,\varphi(x_1)\cdots \varphi(x_p) \biggr|^2 \biggr) \geq 0\,.
\end{multline}
For the last inequality in \eqref{state_computation_4}, we recalled \eqref{classical_state}. We deduce the claim in the classical setting from \eqref{state_computation} and \eqref{state_computation_4}.
\end{proof}
\begin{remark}
By following the same duality argument as \cite[Proposition 3.3 $(ii)$]{FKSS18}, we can deduce from Lemma \ref{state_technical_lemma} that Corollary \ref{bounded_state_final_convergence_corollary} holds for all $\xi \in \mathcal{L}(\mathfrak{h}^{(p)})$.
\end{remark}

To prove Theorem \ref{bounded_trace_class_final_convergence_theorem}, we argue similarly as in \cite[Sections 4.2-4.3]{FKSS17}, and use the following result.

\begin{lemma}
	\label{duality_lemma}
Let $p \in \N^*$ be fixed. Suppose that for all $\tau > 0$, $\gamma_{\tau,p} \in \mathfrak{S}^1(\mfh^{(p)})$ are positive and that $\gamma_p \in \mathfrak{S}^1(\mfh^{(p)})$ is positive (both in the sense of operators). Furthermore, suppose that
\begin{equation}
	\label{duality_conditions}
\lim_{\tau \to \infty} \|\gamma_{\tau,p} -\gamma_p\|_{\mathfrak{S}^2(\mfh^{(p)})} = 0
\,, \qquad  \quad \lim_{\tau \to \infty} \mathrm{Tr}\,\gamma_{\tau,p} = \mathrm{Tr}\, \gamma_p.
\end{equation}
Then $\lim_{\tau \to \infty} \|\gamma_{\tau,p} - \gamma_p\|_{\mathfrak{S}^1(\mfh^{(p)})} = 0$.
\end{lemma}

The result of Lemma \ref{duality_lemma} is based on \cite[Lemma 2.20]{Sim05}, and proved in this form in \cite[Lemma 4.10]{FKSS17}. We refer the reader to the latter reference for the details of the proof. 

\begin{proof}[Proof of Theorem \ref{bounded_trace_class_final_convergence_theorem}]
We first prove \eqref{bounded_trace_class_convergence_final}. Let $p \in \N^*$ be given. We verify the conditions of Lemma \ref{duality_lemma}. Using the fact that $\mathfrak{S}^2(\mfh^{(p)}) \cong \mathfrak{S}^2(\mfh^{(p)})^*$ and recalling \eqref{B_p}, we have
\begin{equation}
\label{duality_lemma_condition_1_proof}
\|\gamma_{\tau,p}-\gamma_p\|_{\mathfrak{S}^2(\mfh^{(p)})} = \sup_{\xi \in \mathfrak{B}_p} \left|\mathrm{Tr}\left(\gamma_{\tau,p}\,\xi - \gamma_p \,\xi\right)\right| = \sup_{\xi \in \mathfrak{B}_p} \left|\rho_\tau(\Theta_{\tau}(\xi)) - \rho(\Theta(\xi)) \right| \to 0
\end{equation}
as $\tau \to \infty$ by Corollary \ref{bounded_state_final_convergence_corollary}. For the second equality in \eqref{duality_lemma_condition_1_proof}, we used the identity
\begin{equation}
\label{duality_lemma_condition_1_proof_B}
\mathrm{Tr} (\gamma_{\#,p}\,\xi) = \rho_{\#}(\Theta_{\#}(\xi))\,,
\end{equation}
for all $\xi \in \mathcal{L}(\mathfrak{h}^{(p)})$. One directly verifies \eqref{duality_lemma_condition_1_proof_B} from \eqref{quantum_lift}, \eqref{quantum_state_defn}, \eqref{quantum_p_particle_correlation_fn} in the quantum setting, and from \eqref{classical_Theta_definition}, \eqref{classical_state}, \eqref{classical_p_particle_correlation_fn} in the classical setting.

Setting $\xi=\mathbf{1}$ in \eqref{duality_lemma_condition_1_proof_B}, we obtain that
\begin{equation}
\label{duality_lemma_condition_1_proof_C}
\mathrm{Tr} \,\gamma_{\#,p} = \rho_{\#}(\Theta_{\#}(\mathbf{1}))\,.
\end{equation}
Corollary \ref{bounded_state_final_convergence_corollary} and \eqref{duality_lemma_condition_1_proof_C} hence imply that
\begin{equation}
\label{duality_lemma_condition_2_proof}
\lim_{\tau \to \infty} \mathrm{Tr} \,\gamma_{\tau,p} = \mathrm{Tr}\, \gamma_{p}.
\end{equation}
We now deduce \eqref{bounded_trace_class_convergence_final} from Lemma \ref{state_technical_lemma}, \eqref{duality_lemma_condition_1_proof}, \eqref{duality_lemma_condition_2_proof}, and Lemma \ref{duality_lemma}.

The proof of  \eqref{bounded_trace_class_convergence_final_2} is similar. Namely we start from \eqref{quantum_state_identity_2} with $\mathcal{A}=\mathbf{1}$ and repeat the previous argument (in which we formally set $p=0$). We note that in this case, we do not need to use Lemma \ref{duality_lemma} above.
\end{proof}

\section{The time-independent problem with unbounded interaction potentials. Proofs of Theorems \ref{unbounded_correlation_convergence_thm} and \ref{delta_correlation_convergence_thm}.}
\label{Section 4}
In this section, we analyse the time-dependent problem for general $w$ as in Assumption \ref{w_definition}. In particular, we no longer assume that $w$ is bounded, as in Section \ref{Section 3}. In Section \ref{Section 4 A}, we consider $w$ satisfying Assumption \ref{w_definition} (i) and prove Theorem \ref{unbounded_correlation_convergence_thm}. In Section \ref{Section 4 B}, we consider $w$ satisfying Assumption \ref{w_definition} (ii) and prove Theorem \ref{delta_correlation_convergence_thm}. As before, we fix $p \in \N^*$ throughout the section.

\subsection{$L^1$ interaction potentials. Proof of Theorem \ref{unbounded_correlation_convergence_thm}.}
\label{Section 4 A}
We first consider the case where $w$ satisfies Assumption \ref{w_definition} (i), i.e. when it is taken to be an even and real-valued function in $L^1(\T)$. To do this, we approximate $w$ with bounded potentials $w^\eps$, which are even and real-valued. For instance, we can take $w^\eps := w \chi_{\{|w| \leq 1/\eps\}}$. We then use the results of the previous section combined with a diagonal argument. 


Let us first note the following result.
\begin{lemma}
	\label{Unbounded_potential_state_theorem}
	Let $w$ be as in Assumption \ref{w_definition} (i), and suppose $w^\eps \in L^\infty$ is a sequence of even, real-valued interaction potentials satisfying $w^\eps \to w$ in $L^1(\T)$ as $\eps \to 0$. Then there exists a sequence $(\eps_\tau)$ converging to $0$ as $\tau \to \infty$ such that for all $p \in \N^*$ 
	\begin{equation}
	\label{Unbounded_potential_state_theorem_1}
	\lim_{\tau \to \infty}\rho_\tau^{\eps_\tau}\left(\Theta_\tau(\xi)\right) = \rho\left(\Theta\left(\xi\right)\right)\,,
	\end{equation}
	uniformly in $\xi \in \mathcal{C}_p$. We recall that $\mathcal{C}_p$ is given by \eqref{C_p}.
\end{lemma}

\begin{proof}
	Using a standard diagonal argument, it suffices to prove that for each fixed $\eps >0$
	\begin{equation}
	\label{unbounded_potential_tau_state_convergence}
	\lim_{\tau \to \infty}\rho^\eps_\tau\left(\Theta_\tau(\xi)\right) \to \rho^\eps\left(\Theta\left(\xi\right)\right)
	\end{equation}
	uniformly in $\xi \in \mathcal{C}_p$, and 
	\begin{equation}
	\label{unbounded_potential_epsilon_state_convergence}
	\lim_{\eps \to 0}\rho^\eps\left(\Theta\left(\xi\right)\right) \to \rho\left(\Theta\left(\xi\right)\right)
	\end{equation} 
	uniformly in $\xi \in \mathcal{C}_p$. The convergence in \eqref{unbounded_potential_tau_state_convergence} holds by Corollary \ref{bounded_state_final_convergence_corollary} because $w^\eps \in L^\infty(\T)$. To show \eqref{unbounded_potential_epsilon_state_convergence}, we first note that by Lemma \ref{classical_W_bound} \eqref{classical_W_bound_2} and the Sobolev embedding theorem
	\begin{align}
		\label{mcWeps_convergence_mcW_L1}
		\left|\mcW^\eps - \mcW\right| &\lesssim \|w^\eps-w\|_{L^1} \|\varphi\|_{H^{\frac{1}{2}-}}^4\,.
	\end{align}
	Since $\varphi \in H^{\frac{1}{2}-}$ almost surely, it follows that  
	\begin{equation}
		\label{W_eps_L1_convergence_almost_surely}
	\lim_{\eps \rightarrow 0} \mcW^\eps = \mcW
	\end{equation}
	almost surely. Continuity of the exponential implies that
	\begin{equation*}
	\lim_{\eps \rightarrow 0} e^{-\mcW^\eps} = e^{-\mcW}
	\end{equation*}
	almost surely.
	By Lemma \ref{classical_W_bound} \eqref{classical_W_bound_2}, we have
	\begin{align*}
		|\mcW| &\leq \frac{1}{2} \|w\|_{L^1} \|\varphi\|_{L^4}^4\,,
	\end{align*}
	and for $\eps$ sufficiently small
	\begin{equation*}
		|\mcW^\eps| \leq \frac{1}{2}\|w^\eps\|_{L^1} \|\varphi\|_{L^4}^4 \leq \|w\|_{L^1} \|\varphi\|_{L^4}^4\,.
	\end{equation*}
	It follows that 
	\begin{equation}
	\label{uniform_epsilon_1}
	\left|e^{-\mcW^\eps} - e^{-\mcW}\right| \leq 2e^{\|w\|_{L^1}\|\varphi\|^4_{L^4}}\,.
	\end{equation}
	By Lemma \ref{main} and Assumption \ref{support_f}, we know that 
	\begin{equation}
	\label{uniform_epsilon_2}
	e^{\|w\|_{L^1}\|\varphi\|^4_{L^4}}f^{\frac{1}{2}}(\mcN) \in L^1(d \mu)\,.
	\end{equation}
	By Lemma \ref{Theta_classical_bound}, we have that 
	\begin{equation}
	\label{uniform_epsilon_3}
	\Theta(\xi)f^{\frac{1}{2}}(\mcN) \in L^\infty(d \mu)\,.
	\end{equation}
	Using \eqref{uniform_epsilon_1}--\eqref{uniform_epsilon_3} and the dominated convergence theorem, it follows that
	\begin{equation}
	\label{unbounded_epsilon_convergence_intermdediate_1}
	\lim_{\eps \rightarrow 0} \int d\mu \, \left| \Theta(\xi) \right|\,\left|e^{-\mcW^\eps} - e^{-\mcW}\right| f(\mcN) = 0\,.
	\end{equation}
        The same argument implies
	\begin{equation}
	\label{unbounded_classical_partition_epsilon_convergence}
	\lim_{\eps \rightarrow 0} z^\eps = z\,.
	\end{equation}
	Noting that
	\begin{equation*}
		\rho^\eps\left(\Theta\left(\xi\right)\right) - \rho\left(\Theta\left(\xi\right)\right) = \frac{1}{z}\int d\mu \, \Theta(\xi)f(\mcN) \left(\frac{z}{z^\eps}e^{-\mcW^\eps} - e^{-\mcW}\right),
	\end{equation*}
	\eqref{unbounded_potential_epsilon_state_convergence} follows from \eqref{unbounded_epsilon_convergence_intermdediate_1} and \eqref{unbounded_classical_partition_epsilon_convergence}.
\end{proof}
We can now prove Theorem \ref{unbounded_correlation_convergence_thm}.
\begin{proof}[Proof of Theorem \ref{unbounded_correlation_convergence_thm}]
We deduce \eqref{unbounded_trace_class_convergence_final} from Lemma \ref{Unbounded_potential_state_theorem} by arguing analogously as in the proof of \eqref{bounded_trace_class_convergence_final}. The proof of \eqref{Unbounded_potential_state_theorem_2} is similar to that of \eqref{unbounded_trace_class_convergence_final}. Instead of \eqref{unbounded_potential_tau_state_convergence}, we use 
\begin{equation*}
	\lim_{\tau \to \infty}\mathcal{Z}_\tau^{\eps} =z^{\eps}\,,
\end{equation*}
for fixed $\eps>0$, which follows from \eqref{bounded_trace_class_convergence_final_2}.
Instead of \eqref{unbounded_potential_epsilon_state_convergence}, we use \eqref{unbounded_classical_partition_epsilon_convergence}.
\end{proof}

\subsection{The Delta Function. Proof of Theorem \ref{delta_correlation_convergence_thm}}
\label{Section 4 B}
We now deal with the case $w=-\delta$. Let us first recall the definition \eqref{w_eps_delta_function} of $w^{\eps}$. Let us note that since $U$ is even, it is not necessary to take $U$ to be non-positive, since we can argue as in \cite[(5.33)]{FKSS18} using $|U|$ (note that in \cite{FKSS18}, one writes $\widetilde{w}$ for $U$). In what follows, we again denote objects corresponding to the interaction potential $w^\eps$ by using a superscript $\eps$. Again, by following Section \ref{correlation_fn_argument}, to prove Theorem \ref{delta_correlation_convergence_thm}, it suffices to prove the following proposition.
\begin{lemma}
	\label{delta_function_potential_theorem}
	Let $w := -\delta$, and let $w^\eps$ be defined as in \eqref{w_eps_delta_function}. Then there is a sequence $(\eps_\tau)$ satisfying $\eps_\tau$ converging to $0$ as $\tau \to \infty$ such that
	\begin{equation}
	\label{delta_function_potential_equation}
	\lim_{\tau \to \infty} \rho_\tau^{\eps_\tau} \left(\Theta_\tau(\xi)\right) = \rho\left(\Theta(\xi)\right)\,,
	\end{equation}
	uniformly in $\xi \in \mathcal{C}_p$, where $\mathcal{C}_p$ is given by \eqref{C_p}.
\end{lemma}
\begin{proof}
	As in the proof of Lemma \ref{Unbounded_potential_state_theorem}, it suffices to prove for fixed $\eps$ that
	\begin{equation}
	\label{delta_potential_tau_state_convergence}
	\lim_{\tau \to \infty}\rho^\eps_\tau\left(\Theta_\tau(\xi)\right) \to \rho^\eps\left(\Theta\left(\xi\right)\right)\,,
	\end{equation}
	uniformly in $\xi \in \mathcal{C}_p$, and 
	\begin{equation}
	\label{delta_potential_epsilon_state_convergence}
	\lim_{\eps \to 0}\rho^\eps\left(\Theta\left(\xi\right)\right) \to \rho\left(\Theta\left(\xi\right)\right)
	\end{equation}
	uniformly in $\xi \in \mathcal{C}_p$. Since $w^\eps \in L^{\infty}(\T)$, \eqref{delta_potential_tau_state_convergence} follows from Lemma \ref{Unbounded_potential_state_theorem}. To prove \eqref{delta_potential_epsilon_state_convergence}, we note that Lemma \ref{classical_W_bound} \eqref{classical_W_bound_2} now implies
	\begin{align}
	\notag
	\left|\mcW\right| &\leq \frac{1}{2}\|U\|_{L^1}\|\varphi\|_{L^4}^4\,, \\
	\label{W^{eps}_bound}
	\left|\mcW^\eps\right| &\leq \frac{1}{2}\|U\|_{L^1} \|\varphi\|_{L^4}^4\,.
	\end{align}
	Since $\int dx \, U = -1$ and $U$ is even,
	\begin{equation*}
		\mcW^\eps-\mcW = \frac{1}{2} \int dx \, dy \, w^\eps(x-y) \left(|\varphi(x)|^2|\varphi(y)|^2-|\varphi(x)|^4\right)\,.
	\end{equation*}
	So
	\begin{equation}
	\label{mcWeps_convergence_mcW_delta}
		\left|\mcW^\eps-\mcW\right| \leq \frac{1}{2}\int dx\,dy\, |w^\eps(x-y)| |\varphi(x)|^2 |\varphi(x)-\varphi(y)|\left(|\varphi(x)| + |\varphi(y)| \right).
	\end{equation}
	We can then follow the argument in \cite[(5.49) in the proof of Theorem 1.6.]{FKSS18} to conclude that
	\begin{equation} 
	\label{mcWeps_convergence_mcW_delta_B}
	\mcW^\eps \to \mcW\,.
	\end{equation} 
We omit the details.  Arguing as in the proof of Lemma \ref{Unbounded_potential_state_theorem}, we obtain \eqref{delta_potential_epsilon_state_convergence}, and thus \eqref{delta_function_potential_equation}. We emphasise that, in order to apply the dominated convergence theorem as in the proof of Lemma \ref{Unbounded_potential_state_theorem}, it is important that the upper bound \eqref{W^{eps}_bound} is uniform in $\eps$.
\end{proof}

\begin{proof}[Proof of Theorem \ref{delta_correlation_convergence_thm}]
We obtain \eqref{delta_trace_class_convergence_final} by arguing analogously as for \eqref{unbounded_trace_class_convergence_final}. Here, instead of Lemma \ref{Unbounded_potential_state_theorem}, we use Lemma \ref{delta_function_potential_theorem}. The proof of \eqref{delta_potential_state_theorem_2} is analogous to that of \eqref{Unbounded_potential_state_theorem_2}.
\end{proof}
\section{The time-dependent problem}
\label{Section 5}
In this section, we consider the time-dependent problem.  The analysis for bounded $w$ and the proof of Theorem \ref{time_dependent_Linf_case} are given in Section \ref{Section 5 A}. The case when $w$ is unbounded is analysed in Section \ref{Section 5 B}. Here, we prove Theorems \ref{time_dependent_L1_case} and \ref{time_dependent_delta_case}.
Throughout the section, we fix $p \in \N^*$ and $\xi \in \mathcal{L}(\mathfrak{h}^{(p)})$. In particular, we have the following two lemmas.
\subsection{Bounded interaction potentials. Proof of Theorem \ref{time_dependent_Linf_case}.}
\label{Section 5 A}
In order to deal with bounded interaction potentials, we recall the \emph{Schwinger-Dyson expansion} outlined in \cite[Sections 3.2 and 3.3]{FKSS18}.
\begin{lemma}
	\label{quantum_schwinger_dyson_prop}
	Given $\mathcal{K} > 0$, $\eps > 0$, and $t \in \R$, there exists $L = L(\mathcal{K}, \eps, t, \|\xi\|, p) \in \N$, a finite sequence $(e^l)_{l=0}^L$, with $e^l = e^l(\xi,t) \in \mathcal{L}(\mathfrak{h}^{(p)})$ and $\tau_0 = \tau_0(\mathcal{K}, \eps, t, \|\xi\|) > 0$ such that
	\begin{equation*}
		\left\| \left(\Psi^t_\tau \Theta_\tau(\xi) - \sum_{l=0}^L \Theta_\tau(e^l)\right)\Bigg|_{\mathfrak{h}^{(\leq\mathcal{K}\tau)}}\right\| < \eps\,,
	\end{equation*}
	for all $\tau \geq \tau_0$. Here we recall the definition of $\mathfrak{h}^{(\leq p)}$ from \eqref{restricted_fock_space_defn}.
\end{lemma}
In other words, for large $\tau$ and restricted numbers of particles, we can approximate the evolution of the lift of an arbitrary operator with finitely many unevolved lifts. We also have the corresponding classical result.
\begin{lemma}
	\label{classical_schwinger_dyson_prop}
	Given $\mathcal{K} > 0$, $\eps > 0$, and $t \in \R$, then there exist $L = L(\mathcal{K}, \eps, t, \|\xi\|, p) \in \N$, $\tau_0 = \tau_0(\mathcal{K}, \eps, t, \|\xi\|) > 0$ both possibly larger than in Lemma \ref{quantum_schwinger_dyson_prop}, and for the same choice of $e^{l} = e^l(\xi,t)$ as in Lemma \ref{quantum_schwinger_dyson_prop}, we have
	\begin{equation*}
		\left| \left(\Psi^t \Theta(\xi) - \sum_{l=0}^L \Theta(e^l)\right)\chi_{\{\mcN \leq \mathcal{K}\}}\right| < \eps\,,
	\end{equation*}
	for all $\tau \geq \tau_0$.
\end{lemma}
We note that the proofs of Lemmas \ref{quantum_schwinger_dyson_prop} and \ref{classical_schwinger_dyson_prop}, respectively \cite[Lemmas 3.9 and 3.12]{FKSS18}, do not use the sign of the interaction potential, so still hold in our case. The proofs of both results also require a compactly supported cut-off function, demonstrating the cut-off function of the form $f(x)=e^{-cx^2}$ discussed in Remark \ref{Remark_AppendixB} \eqref{Remark_Appendix_B_2} would not suffice here.
\begin{proof}[Proof of Theorem \ref{time_dependent_Linf_case}]
By using Theorem \ref{bounded_trace_class_final_convergence_theorem}, Lemmas \ref{quantum_schwinger_dyson_prop}--\ref{classical_schwinger_dyson_prop}, and following the proof of \cite[Proposition 2.1]{FKSS18}, we obtain Theorem \ref{time_dependent_Linf_case}.
\end{proof}
\begin{remark}
\label{uniform_convergence_remark}
Recalling the proof of \cite[Proposition 2.1]{FKSS18}, it follows that the convergence in Theorem \ref{time_dependent_Linf_case} is uniform on the set of parameters 
\begin{equation*}
w \in L^{\infty},\, m \in \N,\, t_i \in \R,\, p_i \in \N^*,\,\xi^i \in \mathcal{L}(\mathfrak{h}^{(p_i)});\, i=1,\ldots,m\,,
\end{equation*}
satisfying
\begin{equation*}
\mathrm{max}\bigl\{\|w\|_{L^{\infty}}, m,|t_1|,\ldots,|t_m|,p_1,\ldots,p_m,\|\xi^1\|,\ldots,\|\xi^m\|\bigr\} \leq M\,,
\end{equation*}
 for any fixed choice of $M>0$.
\end{remark}
\subsection{Unbounded interaction potentials. Proofs of Theorems \ref{time_dependent_L1_case} and \ref{time_dependent_delta_case}}
\label{Section 5 B}
Before proceeding, we need to prove a technical result concerning the flow of the NLS.
\begin{lemma}
	\label{L^1_Cauchy_approximation_theorem}
	Let $w \in L^1(\T)$ and $s \geq \frac{3}{8}$ be given, and suppose $\varphi \in H^s$. Consider the Cauchy problem on $\T$ given by
	\begin{equation}
	\label{L^1_cauchy_problem}
	\begin{cases}
	i\partial_t u + (\Delta - \kappa)u = \left(w * |u|^2\right)u \\
	u_0 = \varphi.
	\end{cases}
	\end{equation}
	In addition, given $\eps > 0$ and letting $w^\eps \in L^\infty$ be a sequence satisfying $w^\eps \to w$ in $L^1$, we consider
	\begin{equation}
	\label{L^1_approximate_cauchy_problem}
	\begin{cases}
	i\partial_t u^\eps + (\Delta - \kappa)u^\eps = \left(w^\eps * |u^\eps|^2\right)u^\eps \\
	u^\eps_0 = \varphi.
	\end{cases}
	\end{equation}
	Since $s > 3/8 \geq 0$, the flow map defined in \eqref{flow_map} is globally well defined. Denote by $u$ and $u^\eps$ the solutions of \eqref{L^1_cauchy_problem} and \eqref{L^1_approximate_cauchy_problem} respectively. Then for $T>0$
	\begin{equation*}
		\lim_{\eps \to 0}\|u-u^\eps\|_{L^\infty_{[-T,T]}\mathfrak{h}} = 0.
	\end{equation*}
\end{lemma}
We need to recall the {\it dispersive Sobolev $X^{s,b}$ spaces}.
\begin{definition}
\label{X^{s,b}}
	Given $f: \T \times \R \to \C$ and $s,b \in \R$ we define
	\begin{equation*}
		\|f\|_{X^{s,b}} := \left\|\left(1+|2\pi k|\right)^s \left(1+|\eta +2\pi k^2|\right)^b \tilde{f}\right\|_{L^2_\eta \ell^2_k},
	\end{equation*}
	where 
	\begin{equation*}
	\tilde{f}(k,\eta) := \int^{\infty}_{-\infty} dt \int_\T dx \, f(x,t) e^{-2\pi i kx - 2 \pi i \eta t}
	\end{equation*}
denotes the {\it spacetime Fourier transform} of $f$.
\end{definition}
In the following, we always take $b=\frac{1}{2} + \nu$, for $\nu > 0$ small.
\begin{proof}[Proof of Lemma \ref{L^1_Cauchy_approximation_theorem}]
We recall the details of proof of \cite[Proposition 5.1]{FKSS18}. Firstly, we can take $\kappa = 0$ by considering $\tilde{u}:=e^{i\kappa t}u$. So, we construct global mild solutions to \eqref{L^1_cauchy_problem} and \eqref{L^1_approximate_cauchy_problem} in the following way.
	
	Let $\zeta, \psi: \R \to \R$ be smooth functions with
	\begin{equation}
	\zeta(t) = 
	\begin{cases}
	1 & \text{if } \, |t| \leq 1 \\
	0 & \text{if } \, |t| > 2.
	\end{cases}
	\end{equation}
	\begin{equation}
	\psi(t) = 
	\begin{cases}
	1 & \mathrm{if } \, |t| \leq 2 \\
	0 & \mathrm{if } \, |t| > 4.
	\end{cases}
	\end{equation}
	We also define $\zeta_\delta(t) := \zeta(t/\delta)$ and $\psi_\delta(t) := \psi(t/\delta)$. We consider
	\begin{align}
		\label{Lnormal}
		(Lv)(\cdot, t) &:= \zeta_\delta(t) e^{it\Delta}\varphi_0 - i \zeta_\delta(t) \int^t_0 dt' \, e^{i(t-t')\Delta}\left(w*|v_\delta|^2\right)v_\delta(t'), \\
		\label{Leps}
		(L^\eps v)(\cdot, t) &:= \zeta_\delta(t) e^{it\Delta}\varphi_0 - i \zeta_\delta(t) \int^t_0 dt' \, e^{i(t-t')\Delta}\left(w^\eps*|v_\delta|^2\right)v_\delta(t'),
	\end{align}
	where $v_\delta(x,t) := \psi_\delta(t)v(x,t)$. By proving $L$ and $L^\eps$ are both contractions on appropriate function spaces for $\delta > 0$ sufficiently small, we are able to find local mild solutions to \eqref{L^1_cauchy_problem} and \eqref{L^1_approximate_cauchy_problem}. The arguments used to prove \eqref{Lnormal} and \eqref{Leps} are contractions in \cite[Proposition 5.1]{FKSS18} still hold if we can show  that
	\begin{equation}
	\label{convolution_xsb_bound}
	\left\|\left(w*|v_\delta|^2\right)v_\delta\right\|_{X^{0,b-1}} \lesssim \|w\|_{L^1} \|v_\delta\|^3_{X^{0,b}}.
	\end{equation}
	To show \eqref{convolution_xsb_bound}, we define $\mathcal{V}_\delta$ as the function satisfying $\widetilde{\mathcal{V}}_\delta =|\tilde{v}_\delta|$. Note that by construction, $\|\mathcal{V}_\delta\|_{X^{0,b}} = \|v_{\delta}\|_{X^{0,b}}$. Then
	\begin{multline*}
		\big|\left(\left(w*|v_\delta|^2\right)v_\delta\right)\,\widetilde{ }\,\,(k,\eta)\big| \\
		\leq \|\hat{w}\|_{\ell^\infty} \int dk_1\, dk_2 \, dk_3 \, d\eta_1 \, d\eta_2 \, d\eta_3 \, |\tilde{v}_\delta(k_1,\eta_1)||\tilde{v}_\delta(-k_2,-\eta_2)||\tilde{v}_\delta(k_3,\eta_3)| \\
		\times\delta(k_1+k_2+k_3-k) \delta(\eta_1+\eta_2+\eta_3-\eta) \\
		 = \|\hat{w}\|_{\ell^\infty} \left(|\mathcal{V}_\delta|^2\mathcal{V}_\delta\right) \, \widetilde{ } \,\,(k,\eta) \leq \|w\|_{L^1} \left(|\mathcal{V}_\delta|^2\mathcal{V}_\delta\right) \, \widetilde{ } \,\,(k,\eta)\,.
	\end{multline*}

To prove \eqref{convolution_xsb_bound}, it remains to show
\begin{equation}
\label{mcVd_x0b_bound}
\left\||\mathcal{V}_\delta|^2 \mcV \right\|_{X^{0,b-1}} \lesssim \|\mcVd\|_{X^{0,b}}^3 = \|v_\delta\|_{X^{0,b}}.
\end{equation}
To show \eqref{mcVd_x0b_bound} we argue as in \cite[(2.147)-(2.153)]{Soh11}, where similar bounds are proved for the quintic case, and use a duality argument. Choose $c: \Z \times \R \to \C$ such that $\sum_k \int d\eta\, |c(k,\eta)|^2 = 1$. We consider
\begin{equation*}
I:= \sum_k \int d\eta \, \left(1+|\eta+k^2|\right)^{b-1} \left(|\mcVd|^2\mcVd\right)\, \widetilde{ } \,\, (k,\eta)\,c(k,\eta)\,.
\end{equation*}
We have
\begin{align*}
|I| \leq \sum_k \sum_{k_1+k_2+k_3=k} \int_{\eta_1+\eta_2+\eta_3=\eta} d\eta_1\,&d\eta_2\,d\eta_3\,d\eta \, \frac{|c(k,\eta)|}{\left(1+|\eta+k^2|\right)^{1-b}} \\
&\times |\tilde{v}_\delta(k_1,\eta_1)||\tilde{v}_\delta(-k_2,-\eta_2)||\tilde{v}_\delta(k_3,\eta_3)|.
\end{align*}
Define
\begin{align*}
F(x,t) &:= \sum_k \int d\eta \, \frac{|c(k,\eta)|}{\left(1+|\eta+k^2|\right)^{1-b}} e^{2\pi ikx+2\pi it\eta}\,, \\
G(x,t)&:= \sum_k\int d\eta \, |\tilde{v}_\delta(k,\eta)|e^{2\pi i k x + 2 \pi it\eta}\,.
\end{align*}
Parseval's identity implies
\begin{align}
	\nonumber
I &\lesssim \int\int dx\,dt\, F \overline{G} G \overline{G} = \left|\int\int dx\,dt\, F \overline{G} G \overline{G} \right| \\
\label{Parseval_I_inequality}
&\leq \|F\|_{L^4_{t,x}} \|G\|_{L^4_{t,x}}^3\,.
\end{align}
Since $b>3/8$, we have the estimate $\|\phi\|_{L^4_{t,x}} \lesssim 
\|\phi\|_{X^{0,b}}$ (see \cite[Proposition 2.6]{Bou93}, \cite[Lemma 2.1 (i)]{Grun00}, and \cite[Proposition 2.13]{Tao06}). So
\begin{equation}
\label{F_X0b_bound}
\|F\|_{L^4_{t,x}} \lesssim \|F\|_{X^{0,3/8}} \leq \|F\|_{X^{0,b-1}} = \|c\|_{\ell^2_k L^2_\eta} = 1\,.
\end{equation}
Moreover
\begin{equation}
\label{G_X0b_bound}
\|G\|_{L^4_{t,x}} \lesssim \|v_\delta\|_{X^{0,3/8}} \lesssim \delta^{\theta}\|v_\delta\|_{X^{0,b}}\,,
\end{equation}
where $\theta > 0$. Here the final inequality follows from \cite[Lemma 5.3 (iv)]{FKSS18}. Combining \eqref{Parseval_I_inequality} with \eqref{F_X0b_bound} and \eqref{G_X0b_bound} yields \eqref{mcVd_x0b_bound}.

So, for a time of existence $\delta$ that depends only on the $L^2$ norm of the initial data, we are able to construct local mild solutions, $v_{(n)}$ and $v^\eps_{(n)}$ on $[n\delta, (n+1)\delta]$. We then piece these solutions together to create mild solutions $u$ and $u^\eps$ to \eqref{L^1_cauchy_problem} and \eqref{L^1_approximate_cauchy_problem} respectively. Using $v$ and $v^\eps$ to denote $v_{(0)}$ and $v^\eps_{(0)}$ respectively, we have
	\begin{align}
		\nonumber
		\|u-u^\eps\|_{L^\infty_{[0,\delta]}L^2_x} = & \|v-v^\eps\|_{L^\infty_{[0,\delta]}L^2_x} \\
		\nonumber
		\leq & \left\|\zeta_\delta(t) \int^t_0 dt' \, e^{i(t-t')\Delta} \left[\left(w-w^\eps\right)*|v_\delta(t')|^2\right]v_\delta(t')\right\|_{X^{0,b}} \\
		\nonumber
		+ & \left\|\zeta_\delta(t) \int^t_0 dt' \, e^{i(t-t')\Delta} \left[w^\eps*\left(|v_\delta(t')|^2-|v_\delta^\eps(t')|^2\right)\right]v_\delta(t')\right\|_{X^{0,b}} \\
		\label{cauchy_approximation_bounds1}
		+ & \left\|\zeta_\delta(t) \int^t_0 dt' \, e^{i(t-t')\Delta} \left[w^\eps * |v_\delta^\eps(t')|^2\right]\left(v_\delta(t') - v_\delta^\eps(t')\right)\right\|_{X^{0,b}}.
	\end{align}
	For the first term of \eqref{cauchy_approximation_bounds1}, we have
	\begin{align*}
		\bigg\|\zeta_\delta(t) \int^t_0 dt' \, e^{i(t-t')\Delta} \left[\left(w-w^\eps\right)*|v_\delta(t')|^2\right]&v_\delta(t')\bigg\|_{X^{0,b}}\\ \leq \, &C\delta^{\frac{1-2b}{2}}\left\|\left[\left(w-w^\eps\right)*|v_\delta|^2\right]v_\delta\right\|_{X^{0,b-1}},
	\end{align*}
	where the $\delta^\frac{1-2b}{2}$ comes from the estimates for local $X^{s,b}$ spaces proved in \cite{KPV93a} and \cite{KPV93b}. For a summary of these local $X^{s,b}$ spaces, we direct the reader to \cite[Appendix A]{FKSS18}.
	
	Arguing as in \eqref{convolution_xsb_bound}, we have
	\begin{align*}
		\left\|\left[\left(w-w^\eps\right)*|v_\delta|^2\right]v_\delta\right\|_{X^{0,b-1}} \leq \|w-w^\eps\|_{L^1} \|v_\delta\|^3_{X^{0,b}} \to 0.
	\end{align*}
	The bound on the second term in \eqref{cauchy_approximation_bounds1} follows by the same argument as in the proof of \cite[Proposition 5.1]{FKSS18}, although we note that since $\|w^\eps\|_{L^1}$ is only bounded rather than equal to $1$, we may get a larger constant times a positive power of $\eps$, which is not a problem. The third term in \eqref{cauchy_approximation_bounds1} then follows for the same reasons combined with \cite[Proposition 5.1]{FKSS18}.
	
	Following the remainder of the argument from \cite[Proposition 5.1]{FKSS18} and noting that there we gain no negative powers of $\eps$, we have
	\begin{equation*}
		\|u-u^\eps\|_{L^\infty_{[0,T]}\mfh} \to 0\,.
	\end{equation*}
The corresponding negative time estimates follow from an analogous argument. 
\end{proof}
We also have the corresponding result for the focusing local NLS.
\begin{lemma}
	\label{delta_Cauchy_approximation_theorem}
	Let $s \geq \frac{3}{8}$ be given, and suppose $\varphi \in H^s(\T)$. Consider the Cauchy problem on $\T$ given by
	\begin{equation}
	\label{delta_cauchy_problem}
	\begin{cases}
	i\partial_t u + (\Delta - \kappa)u = -|u|^2u \\
	u_0 = \varphi\,.
	\end{cases}
	\end{equation}
	In addition, given $\eps > 0$, let $w^\eps$ be as in \eqref{w_eps_delta_function}. We consider
	\begin{equation}
	\label{delta_approximate_cauchy_problem}
	\begin{cases}
	i\partial_t u^\eps + (\Delta - \kappa)u^\eps = \left(w^\eps * |u^\eps|^2\right)u^\eps \\
	u^\eps_0 = \varphi\,.
	\end{cases}
	\end{equation}
	Since $s>3/8 \geq 0$, the flow map defined in \eqref{flow_map} is globally well defined. Denote by $u$ and $u^\eps$ the solutions of \eqref{delta_cauchy_problem} and \eqref{delta_approximate_cauchy_problem} respectively. Then for $T>0$
	\begin{equation*}
		\lim_{\eps \to 0}\|u-u^\eps\|_{L^\infty_{[-T,T]}\mathfrak{h}} = 0\,.
	\end{equation*}
\end{lemma}
\begin{proof}
We can follow exactly the proof of \cite[Proposition 5.1]{FKSS18}, recalling \eqref{int_U}, and noting that the function 
$w^{\eps}$ defined in \eqref{w_eps_delta_function} is even and  to deduce 
\begin{equation*}
\bigl||v_{\delta}(x)|^2-(w^{\eps}*|v^{\delta}|^2)(x)\bigr| \leq \int dy\,|w^{\eps}(x-y)|\,|v_{\delta}(x)-v_{\delta}(y)| (|v_{\delta}(x)|+|v_{\delta}(y)|)\,,
\end{equation*}
similarly as in \cite[(5.27)]{FKSS18}. We also have the same point about $\|w^{\eps}\|_{L^1}$ not necessarily equal to $1$ as in the proof of Lemma \ref{L^1_Cauchy_approximation_theorem}, which does not affect the argument.
\end{proof}

Before proving Theorem \ref{time_dependent_L1_case}, we recall the following diagonalisation result, proved in \cite[Lemma 5.5]{FKSS18}.
\begin{proposition}
\label{diagonalisation_lemma}
Let $(Z_k)_{k \in \N}$ be an increasing sequence of sets in the sense that $Z_k \subset Z_{k+1}$. Let us define $Z:= \cup_{k \in \N} Z_k$. For $\eps, \tau > 0$, suppose that $g,g^\eps,g_\tau^\eps : Z \to \C$ are functions with the following properties.
\begin{enumerate}
\item For each fixed $k \in \N$ and $\eps > 0$, $\lim_{\tau \to \infty} g_\tau^\eps(\zeta) = g^\eps(\zeta)$ uniformly in $\zeta \in Z_k$.
\item For each fixed $k \in \N$, $\lim_{\eps \to 0} g^\eps(\zeta) = g(\zeta)$ uniformly in $\zeta \in Z_k$.
\end{enumerate}
Then there is a sequence $(\eps_\tau)$ such that $\lim_{\tau \to \infty} \eps_\tau = 0$ and
\begin{equation*}
\lim_{\tau \to \infty} g^{\eps_\tau}_\tau(\zeta) = g(\zeta) 
\end{equation*}
for any $\zeta \in Z$.
\end{proposition}

\begin{proof}[Proof of Theorem \ref{time_dependent_L1_case}]
Throughout this proof we use $X^\eps$ or $X$ to denote an object defined using $w^\eps$ or $w$ respectively. Define
\begin{align}
\label{Z_set_definition}
Z &:= \{(m,t_i,p_i,\xi^i) : m \in \N,\,  t_i \in \R,\, p_i \in \N^*,\, \xi^i \in \mathcal{L}(\mfh^{(p_i)})\} \\
\label{Z_k_set_definition}
Z_k &:= \{(m,t_i,p_i,\xi^i) : m \leq k,\, |t_i| \leq k,\,p_i \in \N^*\,, p_i \leq k,\, \|\xi^i\| \leq k\},
\end{align}
where in \eqref{Z_set_definition}--\eqref{Z_k_set_definition}, we take $i \in \{1,\ldots,m\}$. Let us also define
\begin{align}
\label{g^eps_sharp_zeta}
g^\eps_{\#}(\zeta) &:= \rho^\eps_\#\left(\Psi_\#^{t_1,\eps}(\Theta_\#(\xi^1)) \cdots \Psi_\#^{t_m,\eps}(\Theta_\#(\xi^m))\right), \\
\label{g_zeta}
g(\zeta) &:= \rho\left(\Psi^{t_1}(\Theta(\xi^1)) \cdots \Psi^{t_m}(\Theta(\xi^m))\right).
\end{align}
By Theorem \ref{time_dependent_Linf_case}, Remark \ref{uniform_convergence_remark}, and Proposition \ref{diagonalisation_lemma}, we note that it suffices to show that for fixed $k \in \N$, we have
\begin{equation}
	\label{rho_eps_time_convergence_1}
\lim_{\eps \to 0} \rho^\eps \left( \Psi^{t_1,\eps}\Theta(\xi^1) \ldots \Psi^{t_m,\eps}\Theta(\xi^m) \right) = \rho \left( \Psi^{t_1}\Theta(\xi^1) \ldots \Psi^{t_m}\Theta(\xi^m) \right),
\end{equation}
uniformly in $Z_k$. 
Using \eqref{W_eps_L1_convergence_almost_surely}--\eqref{uniform_epsilon_3} and the dominated convergence theorem, we have
\begin{equation}
	\label{rho_eps_tilde_identity_convergence}
\lim_{\eps \to 0} \tilde{\rho}^\eps_1(\mathbf{1}) = \tilde{\rho}_1(\mathbf{1}).
\end{equation}
Here we recalled the definition of $\tilde{\rho}_{\zeta}(\cdot)$ from \eqref{tilde_rho_zeta}. 
By \eqref{classical_state_identity} and \eqref{rho_eps_tilde_identity_convergence}, we note that \eqref{rho_eps_time_convergence_1} follows if we prove that 
\begin{equation}
\label{rho_eps_time_convergence_2}
\lim_{\eps \to 0} \tilde{\rho}_1^\eps\left( \Psi^{t_1,\eps}\Theta(\xi^1) \ldots \Psi^{t_m,\eps}\Theta(\xi^m) \right) = \tilde{\rho}_1\left( \Psi^{t_1}\Theta(\xi^1) \ldots \Psi^{t_m}\Theta(\xi^m) \right)
\end{equation}
uniformly in $Z_k$.

Let $S_t$ and $S^\eps_t$ denote the flow maps for \eqref{L^1_cauchy_problem} and \eqref{L^1_approximate_cauchy_problem} respectively. Let $\varphi_0 \in H^{\frac{1}{2}-} \subset \mfh$ be the classical free field defined in \eqref{random_classical_initial}. Recalling \eqref{classical_Theta_definition}, it follows that for $\xi \in \mathcal{L}(\mfh^{(k)})$, we have
\begin{equation}
\label{Psi_eps_convergence_1_C}
\Psi^{t,\eps}\Theta(\xi) = \left\langle \left(S_t^\eps \varphi_0\right)^{\otimes_k} , \xi \left(S_t^\eps \varphi_0\right)^{\otimes_k} \right\rangle_{\mfh^{\otimes_k}}, \quad 
\Psi^{t}\Theta(\xi) = \left\langle \left(S_t \varphi_0\right)^{\otimes_k} , \xi \left(S_t \varphi_0\right)^{\otimes_k} \right\rangle_{\mfh^{\otimes_k}}\,.
\end{equation}
We apply Lemma \ref{L^1_Cauchy_approximation_theorem} in \eqref{Psi_eps_convergence_1_C} to deduce that
\begin{equation}
\label{Psi_eps_convergence_1_B}
\left(S^\eps_t \varphi_0\right)^{\otimes_k} \to \left(S_t \varphi_0\right)^{\otimes_k}
\end{equation}
almost surely in $\mfh^{\otimes_k}$ as $\eps \to 0$. Moreover, since $\xi \in \mathcal{L}(\mfh^{(k)})$, we deduce from \eqref{Psi_eps_convergence_1_B} that
\begin{equation}
\label{Psi_eps_convergence_1}
\lim_{\eps \to 0} \Psi^{t,\eps} \Theta(\xi) = \Psi^t\Theta(\xi),
\end{equation}
almost surely.

By \eqref{W_eps_L1_convergence_almost_surely} and \eqref{Psi_eps_convergence_1}, we have
\begin{equation}
\label{Psi_eps_convergence_2}
\lim_{\eps \to 0} \Psi^{t_1,\eps} \Theta(\xi^1) \ldots \Psi^{t_m,\eps} \Theta(\xi^m) e^{-\mcW^\eps} = \Psi^{t_1} \Theta(\xi^1) \ldots \Psi^{t_m} \Theta(\xi^m) e^{-\mcW},
\end{equation}
almost surely. Using conservation of mass for \eqref{L^1_cauchy_problem}--\eqref{L^1_approximate_cauchy_problem}, as well as  Lemma \ref{classical_W_bound} \eqref{classical_W_bound_2} and Lemma \ref{Theta_classical_bound}, we have
\begin{align}
	\notag
\left|\Psi^{t_1,\eps} \Theta(\xi^1) \ldots \Psi^{t_m,\eps} \Theta(\xi^m) e^{-\mcW^\eps} f\left(\mcN\right)\right| &\leq \prod_{j=1}^m \|\xi^j\| \|\varphi_0\|^{2p_j}_{\mfh} e^{\frac{1}{2}\|w^\eps\|_{L^1} \|\varphi_0\|_{L^4}^4} f(\mcN), \\
\label{Psi_eps_bound}
\left|\Psi^{t_1} \Theta(\xi^1) \ldots \Psi^{t_m} \Theta(\xi^m) e^{-\mcW} f\left(\mcN\right)\right| &\leq \prod_{j=1}^m \|\xi^j\| \|\varphi_0\|^{2p_j}_{\mfh} e^{\frac{1}{2}\|w\|_{L^1} \|\varphi_0\|_{L^4}^4} f(\mcN).
\end{align}
Using Lemma \ref{main} and Assumption \ref{support_f}, both of the bounding functions in \eqref{Psi_eps_bound} are $L^1(d\mu)$. Furthermore, by construction of $w^{\eps}$, the quantity $\|w^{\eps}\|_{L^1}$ is bounded uniformly in $\eps$. 
Therefore, the first function is in $L^1(d\mu)$ uniformly in $\eps$.
Consequently, we deduce \eqref{rho_eps_time_convergence_2} follows from \eqref{Psi_eps_convergence_2}, \eqref{Psi_eps_bound}, and the dominated convergence theorem. The claim now follows.
\end{proof}

\begin{proof}[Proof of Theorem \ref{time_dependent_delta_case}]
We argue analogously as in the proof of Theorem \ref{time_dependent_L1_case}, with the same definitions of $Z,Z_k,g^\eps_\tau,g^\eps,g$ as in \eqref{Z_set_definition}--\eqref{g_zeta} above, except that now $w^{\eps}$ is chosen as in \eqref{w_eps_delta_function}. We recall that again $\|w^{\eps}\|_{L^1}$ is bounded uniformly in $\eps$. 
The proof is analogous to that of Theorem \ref{time_dependent_L1_case}. The only difference is that instead of \eqref{W_eps_L1_convergence_almost_surely} and Lemma \ref{L^1_Cauchy_approximation_theorem}, we use their local analogues \eqref{mcWeps_convergence_mcW_delta_B} and Lemma \ref{delta_Cauchy_approximation_theorem} respectively.
\end{proof}

\appendix
\section{Proof of Lemma \ref{main}}
\label{Appendix_A}
In this appendix, we prove Lemma \ref{main}, which was originally proved in \cite[Lemma 3.10]{Bou94}. For the convenience of the reader, we present the full details of the proof in a self-contained way. For an alternative summary, see also \cite[Section 2]{Oh_Sosoe_Tolomeo}. Before proceeding with the proof, we recall in Section \ref{Auxiliary Results} several auxiliary results concerning Fourier multipliers in the periodic setting and concentration inequalities. In Section \ref{Norming Sets}, we recall the notion of an \emph{norming set}, which we use to prove duality results in $L^p$ spaces.
The proof of Lemma \ref{main} is given in Section \ref{Lemma main proof}.

\subsection{Auxiliary Results}
\label{Auxiliary Results}

We will need the following result about Fourier multipliers on the torus, the full statement and proof of which can be found in full generality in \cite[VII, Theorem 3.8]{SW16}. We recall our convention \eqref{Fourier_transform} for the Fourier transform.
\begin{lemma}[Mikhlin Multiplier Theorem in the periodic setting]
	\label{MMT}
	Let $p \in [1,\infty]$ and $T \in (L^p(\R),L^p(\R))$ be a Fourier multiplier operator. Let $\hat{u}$ be the multiplier corresponding to $T$ and suppose that $\hat{u}$ is continuous at every point of  $\Z$. For $k \in \Z$, let $\lambda(k) := \hat{u}(k)$. Then there is a unique periodised lattice operator $\tilde{T}$ defined by
	\begin{equation*}
		\tilde{T}f(x) \sim \sum_{k \in \Z} \lambda(k)\hat{f}(k)e^{2 \pi i k x} 
	\end{equation*}
	such that $\tilde{T} \in (L^p(\T),L^p(\T))$ and $\|\tilde{T}\|_{L^p \to L^p} \leq \|T\|_{L^p \to L^p}$.
\end{lemma}
We also recall the definition of a sub-gaussian random variable.
\begin{definition}
Let $(\Omega, \mathcal{A}, \mathbb{P})$ be a probability space. We say a random variable $X$ is sub-gaussian if there exist constants $C,v > 0$ such that for all $t>0$ we have
	\begin{equation*}
		\mathbb{P}(|X|>t) \leq Ce^{-vt^2}.
	\end{equation*}
\end{definition}
We will use the following inequality about sub-gaussian random variables. For a proof, see \cite[Proposition 5.10]{Ver12}.
\begin{lemma}[Hoeffding's Inequality]
	\label{Hoeff}
	Suppose that $X_1, \ldots, X_N$ are all independent, centred sub-gaussian random variables. Let $Q:=\max_i \|X_i\|_{\psi_2}$ for 
	\begin{equation*}
	\|X\|_{\psi_2} := \sup_{p\geq 1} p^{-1/2}(\E|X|^p)^{1/p} 
	\end{equation*}
	and let $a \in \R^N$. Then, for any $t>0$, we have
	\begin{equation*}
		\mathbb{P}\left[\left|\sum_{i=1}^N a_iX_i\right| >t\right] \lesssim \exp \left(-\frac{ct^2}{Q^2\|a\|_{\ell^2}^2}\right)\,.
	\end{equation*}
\end{lemma}
\subsection{Norming Sets}
\label{Norming Sets}
To prove Lemma \ref{main}, we need the following result about duality in $L^p$ spaces. We emphasise that this is a known result, but whose proof we could not find in the literature, so we write out the proof for the convenience of the reader.
\begin{lemma}
	\label{norming}
	Suppose that $\mathcal{M} \subset \Z$ has cardinality $m$, and let
	\begin{equation*}
		S := \mathrm{Span}_\C\left\{e^{2\pi i kx} : k \in \mcM\right\}.
	\end{equation*}
	Then there is some subset $\Xi$ of the unit sphere of $L^{p'}$ satisfying the following properties.
	\begin{enumerate}
		\item $\max_{\varphi \in \Xi} |\langle g, \varphi\rangle| \geq \frac{1}{2}\|g\|_{L^p}$ for all $g \in S$.
		\item $\log |\Xi| \leq Cm$ for some universal constant $C>0$. \label{Xi_prop_2}
	\end{enumerate}
\end{lemma}
\begin{remark*}
	This result can be extended to finite dimensional subsets of normed vector spaces, but we do not need the result in full generality.
\end{remark*}

\subsubsection{Norming sets and $\eps$-nets}
Before proceeding, we introduce several notions in Banach spaces.
\begin{definition}
	Let $X$ be a Banach space, $Y \subset X$ a linear subspace, and $\theta \in (0,1]$.  We denote by $X^*$ the (continuous) dual space of $X$. We say that a set $F \subset X^*$ is \emph{$\theta$-norming over $Y$} if
	\begin{equation*}
		\sup_{g \in F \setminus \{0\}} \frac{g(y)}{\|g\|} \geq \theta \|y\|
	\end{equation*}
	for all $y \in Y$.
\end{definition}
\begin{definition}
	Let $X$ be a Banach space. Given $x \in X$ and $\eps>0$, we write $B_\eps(x) = \{y \in X : \|x-y\| < \eps\}$ for the ball in X of radius $\eps$ around $x$. Let $Y \subset X$ be a subset of $X$. Given $\eps>0$, we call $N_\eps \subset Y$ an \emph{$\eps$-net of $Y$} if $Y \subset \bigcup_{x \in N_\eps} B_{\eps}(x)$. 
\end{definition}
We write $S_X:= \{x \in X : \|x\| = 1\}$ for the unit sphere of $X$.

We want to relate norming sets to $\eps$-nets. To do this, we take inspiration from the following result, the proof of which comes from \cite[Section 17.2.4, Theorem 1]{Kad18}.
\begin{lemma}
	Suppose $X$ is a Banach space, $Y \subset X$ is a linear subspace, and $G \subset S_{X^*}$ a set that is $1$-norming over $Y$. Let $\varepsilon \in (0,1)$, and suppose that $N_\eps$ is an $\varepsilon $-net on the unit sphere of $Y$. For each element $x \in N_\eps$, fix a functional $g_x \in G$ such that $g_x(x)>1-\varepsilon$ (which we can do since $N_\eps \subset S_Y$ and $G$ is $1$-norming over $Y$). Then the set $F=\{g_x\}_{x \in N_{\varepsilon}}$ is $\theta$-norming over $S_Y$ for $\theta=1-2\varepsilon$.
\end{lemma}
\begin{proof}
	Let $y \in S_Y$. By definition, there is some $x_y \in N_\eps$ satisfying $\|y-x_y\| < \eps$. Then, by definition of $F$, linearity, the definition of $x_y$, and $G \subset S_{X^*}$, we have
	\begin{align*}
		\sup_{g  \in F} |g(y)| &= \sup_{x \in N_\eps} |g_x(y)| \geq |g_{x_y}(y)| = |g_{x_y}(x_y) - g_{x_y}(y-x_y)|\\
		&> 1 - \eps - \|y-x_y\| > 1-2\eps = \theta\,.
	\end{align*}
\end{proof}
\subsubsection{Conclusion of the proof of Lemma \ref{norming}}
We begin by bounding the size of an $\eps$-net of $\C^m$. For $\mcM \subset \Z$ with $|\mcM| = m$, we consider the following norm on $\C^m$
\begin{align*}
	\vertiii{(a)_{k \in \mathcal{M}}} := \left\|\sum_{k \in \mathcal{M}} a_k e^{2\pi i k x} \right\|_{L^{p'}}\,.
\end{align*}
We define $\Sigma := \{(a_k)_{k \in \mathcal{M}} : \vertiii{a} =1\}$. Notice that since $\C^m$ is finite dimensional, the unit ball with respect to any norm is compact. 
Let $N_{\eps}$ be a maximal subset of $\Sigma$ satisfying the property
\begin{equation}
\label{2.5}
x,y \in N_{\eps}, x \neq y \implies \vertiii{x-y} > \eps\,.
\end{equation}
In other words, any subset of $\Sigma$ strictly containing $N_\eps$ fails to have property \eqref{2.5}. Such a set exists and is finite by the compactness of $\Sigma$. Any such set must be an $\eps$-net of $\Sigma$ by maximality. We have the following bound, whose proof is an adaptation of \cite[Lemma 5.2]{Ver12}.
\begin{lemma}
\label{Lemma_A.9}
	For $N_\eps \subset \Sigma$ maximal satisfying \eqref{2.5}, we have
	\begin{equation*}
		|N_\eps| \leq \left(1+\frac{2}{\eps}\right)^m =: C_\eps^m\,.
	\end{equation*}
\end{lemma}
\begin{proof}
	The result follows from a volume bound. Since $N_\eps$ is $\eps$-separated, it follows that $\{B_{\eps/2}(x)\}_{x \in N_\eps}$ are pairwise disjoint. Moreover, since $x \in \Sigma$, it follows from the triangle inequality that all such balls lie inside the ball of radius $1+\eps/2$ centred at the origin. So
	\begin{equation}
	\label{2.7}
	\mathrm{vol}[B_{\eps/2}(x)] \cdot |N_\eps| \leq \mathrm{vol}[B_{1+\eps/2}(0)].
	\end{equation}
	We also have the following identity
	\begin{align*}
		\mathrm{vol}\bigl[cB_1(0)\bigr] &= \mathrm{vol}\Biggl[\biggl\{(ca_k)_{k \in \mcM} : \Bigl\|\sum_{k \in \mcM}a_k e^{2\pi i k x} \Bigr\|_{L^{p'}} \leq 1\biggr\}\Biggr] \\
		&= c^m \,\mathrm{vol} [B_1(0)]\,.
	\end{align*}
	Combining this with \eqref{2.7} (and using translation invariance), we have
	\begin{equation*}
		|N_\eps| \leq \left(\frac{1+\eps/2}{\eps/2}\right)^m = \left(1+\frac{2}{\eps}\right)^m.
	\end{equation*}
\end{proof}
We are now able to prove Lemma \ref{norming}.
\begin{proof}[Proof of Lemma \ref{norming}]
	We let $\Xi := N_{1/4} \subset \Sigma$ be obtained by setting $\eps = \frac{1}{4}$ in the construction above. Let $g = \sum_{k \in \mathcal{M}} a_k e^{2\pi i k x} $. By duality, there is some $\psi \in \Sigma$ with $|\langle g,\psi\rangle| \geq \frac{3}{4}\|g\|_{L^p}$. Moreover, since $N_{1/4}$ is a $\frac{1}{4}$ net of $\Sigma$, we can find $\varphi \in N_{1/4}\equiv \Xi$ with $\|\psi - \varphi\|_{L^{p'}} \leq \frac{1}{4}$. Hence, it follows that $|\langle g, \psi - \varphi \rangle | \leq \frac{1}{2}\|g\|_{L^p}$. Therefore, we obtain
	\begin{equation}
	\label{Lemma_A.4_claim}
		\frac{1}{2}\|g\|_{L^p} \leq |\langle g, \psi \rangle | - |\langle g, \psi - \varphi \rangle | \leq |\langle g, \psi \rangle  - \langle g, \psi - \varphi \rangle | = |\langle g, \varphi \rangle |\,,
	\end{equation}
	where in the second step above, we used the reverse triangle inequality. The result follows from \eqref{Lemma_A.4_claim} and Lemma \ref{Lemma_A.9}.
\end{proof}

\subsection{Proof of Lemma \ref{main}}
\label{Lemma main proof}

We now prove Lemma \ref{main}, which was originally proved in \cite[Lemma 3.10]{Bou94}. For the convenience of the reader, we present the full details of the proof in a self-contained way. Throughout, $(\C^\N,\mathcal{G},\mu)$ is the probability space defined in \eqref{Wiener_measure_defn} above.
\begin{proof}[Proof of Lemma \ref{main}]
	We show the following bound for large $\lambda$.
	\begin{align}
		\mu \left[\left\|\sum_{k \in \Z}\ \frac{\omega_k}{\sqrt{\lambda_k}}e^{2\pi i k x} \right\|_{L^p} > \lambda, \left( \sum_{k \in \Z} \frac{|\omega_k|^2}{\lambda_k}\right)^{1/2} \leq B    \right] \lesssim \exp(-cM_0^{1+2/p}\lambda^2), \label{eq1}
	\end{align}
	where
	\begin{equation}
	M_0 \sim \left(\frac{\lambda}{B}\right)^\frac{1}{1/2-1/p}. \label{M_0}
	\end{equation}
	Let us assume \eqref{eq1} and we show that it implies the claim. We write
	\begin{align*}
		F&:= \left\|\sum_{k \in \Z}\ \frac{\omega_k}{\sqrt{\lambda_k}}e^{2\pi i k x} \right\|_{L^p}^p \chi_{\left( \sum_{k\in \Z} \frac{|\omega_k|^2}{\lambda_k}\right)^{1/2} \leq B} \\
		G&:= e^{\frac{2}{p}\|\sum_{k \in \Z}\ \frac{\omega_k}{\sqrt{\lambda_k}}e^{2\pi i k x} \|_{L^p}^p}\,\chi_{\left( \sum_{k\in \Z} \frac{|\omega_k|^2}{\lambda_k}\right)^{1/2} \leq B}\,.
	\end{align*}
	Then
	\begin{align*}
		\|G\|_{L^1} &= \int_{y>0} dy \, y \, \mu\left(|G|>y\right) \, \\
		&\leq \int_{y>1} dy \, y \, \mu\left(|G|>y\right)  + 1,
	\end{align*}
	where the inequality follows because $\mu$ is a probability measure. Now defining $y:= \exp\left(\frac{2}{p}\lambda^p\right)$, we have
	\begin{align}
	\notag
		\|G\|_{L^1} &\leq \int_{\lambda>0} d\lambda \, 2\lambda^{p-1} e^{\frac{2}{p}\lambda^p} \mu\left(|F|>\lambda\right) + 1 \\
		\label{G_L^1}
		& \lesssim \int_{\lambda > 0} d\lambda \, \exp\left( \frac{2}{p} \lambda^p - cB^{\frac{-2p+4}{p-2}} \lambda^{\frac{4p}{p-2}}\right)  \lambda^{p-1} + 1.
	\end{align}
	Since $p < \frac{4p}{p-2}$ for $p \in [4,6)$ for $\|G\|_{L^1}$ to be finite, $B$ can be arbitrary. We have $p = \frac{4p}{p-2}$ for $p=6$, so in this case we have to take $B$ sufficiently small.
	
	We now prove \eqref{eq1}. Throughout, $M$ is a dyadic integer and $|k| \sim M$ means $\frac{3M}{4} \leq |k| < \frac{3M}{2}$. We make use of the following inequality.
	\begin{equation}
	\left\| \sum_{|k| \sim M} a_k e^{2\pi i k x}  \right\|_{L^p} \lesssim M^{1/2-1/p} \left\| \sum_{|k| \sim M} a_k e^{2\pi i k x}  \right\|_{L^2}. \label{Bern}
	\end{equation}
	For $p=2$, \eqref{Bern} is trivial, and for $p=\infty$, it follows from Cauchy-Schwarz and Plancherel's theorem. We then use the Riesz-Thorin interpolation theorem to deduce \eqref{Bern} for all $p \in (2, \infty)$.
	
	With $M_0$ as in \eqref{M_0}, we consider a sequence $(\sigma_M)_{M>M_0}$ of positive numbers with
	\begin{equation}
	\sum_{M>M_0} \sigma_M = \delta\,, \label{sum}
	\end{equation} 
	with $\delta > 0$ sufficiently small to be determined later.
	
	Consider $\omega \in \Om$ such that
	\begin{equation}
	\label{3.5}
	\left\|\sum_{k \in \Z} \frac{\omega_k}{\sqrt{\lambda_k}}e^{2\pi i k x} \right\|_{L^p} > \lambda , \quad \left(\sum_{k \in \Z} \frac{|\omega_k|^2}{\lambda_k}\right) \leq B\,.
	\end{equation}
	With $\omega$ as in \eqref{3.5}, we show that there is some $M>M_0$ such that
	\begin{equation}
	\left\| \sum_{|k| \sim M} \omega_ke^{2\pi i k x}   \right\|_{L^p} > \sigma_M M \lambda\,. \label{3.13}
	\end{equation}
	We argue by contradiction. First, we note that for $\omega$ as in \eqref{3.5}, we have
	\begin{equation}
	 \label{small}
	\sum_{M \leq M_0} \left\| \sum_{|k| \sim M} \frac{\omega_k}{\sqrt{\lambda_k}}e^{2\pi i k x}  \right\|_{L^p} \lesssim M_0^{1/2-1/p}\left(\sum_{k \in \Z} \frac{|\omega_k|^2}{\lambda_k}\right)^{1/2} \lesssim \lambda\,.
	\end{equation}
We used \eqref{Bern}, Plancherel's theorem, and summed a geometric sequence for the first inequality in \eqref{small}. For the second inequality in \eqref{small}, we used the $L^2$ bound in \eqref{3.5}, and \eqref{M_0}. By taking the implied constant in \eqref{M_0} to be sufficiently small, let us note that the proof of \eqref{small} implies
	\begin{equation}
	\sum_{M \leq M_0} \left\| \sum_{|k| \sim M} \frac{\omega_k}{\sqrt{\lambda_k}}e^{2\pi i k x}  \right\|_{L^p} \leq \frac{\lambda}{2}. \label{3.8}
	\end{equation}
We henceforth work with such a small implied constant in \eqref{M_0}.

	Suppose that \eqref{3.13} did not hold for any $M>M_0$. Then it would follow that, for an appropriate choice of $\delta $ in \eqref{sum}, we would have
	\begin{equation}
	\sum_{M>M_0}\left\| \sum_{|k| \sim M}\frac{\omega_k}{\sqrt{\lambda_k}} e^{2\pi i k x}  \right\|_{L^p} \leq \frac{\lambda}{2}\,. \label{big}
	\end{equation}
	We note that \eqref{big} combined with \eqref{small} would give us a contradiction with the first inequality in \eqref{3.5}. Let us explain how we have obtained \eqref{big}. First we note
	\begin{equation}
	\left\| \sum_{|k| \sim M} \frac{\omega_k}{\sqrt{\lambda_k}}e^{2\pi i k x}   \right\|_{L^p} \leq \frac{C}{M}\left\|\sum_{|k| \sim M} \omega_k e^{2\pi i k x} \right\|_{L^p}\,, \label{Mik}
	\end{equation}
which we justify as follows. Let $\Phi: \R \rightarrow \C $ be a smooth, compactly supported function which is equal to $1$ on $1/2 \leq |\xi| \leq 2$ and zero for $|\xi|\leq 1/4$ and $|\xi| \geq 4$. Consider the function
\begin{equation}
\label{Psi_M}
\Psi_M(\xi) := \frac{1}{\sqrt{\frac{4 \pi^2 |\xi|^2}{M^2}+\frac{\kappa}{M^2}}}\Phi\left(\frac{\xi}{M}\right)\,.
\end{equation} 
Since 
\begin{equation*}
\tilde{\Phi}(\xi):=
\frac{1}{\sqrt{4 \pi^2 |\xi|^2 + (\kappa/M^2)}}\Phi(\xi)
\end{equation*}
has bounded derivatives of all order (with bound depending on $\kappa$), 
the same holds for $\Psi_M=\tilde{\Phi}(\xi/M)$ given by \eqref{Psi_M} above. Hence, the Mikhlin multiplier theorem (on $\R$) implies that the map $T_M$ defined by $(T_Mf)\,\hat{ }\,(\xi) := \Psi_M(\xi)\hat{f}(\xi)$ is bounded as a map on $L^p(\R)$. Applying the support properties of $\Phi$ and using Lemma \ref{MMT}, we obtain
	\begin{align*}
		M \left\| \sum_{|k| \sim M} \frac{\omega_k}{\sqrt{\lambda_k}} e^{2\pi i k x}  \right\|_{L^p} &= M \left\| \sum_{|k| \sim M}\Phi_M(k) \frac{\omega_k}{\sqrt{\lambda_k}} e^{2\pi i k x}  \right\|_{L^p} \\ 
		&= \left\| \sum_{|k| \sim M} \Psi_M(k) \omega_k e^{2\pi i k x} \right\|_{L^p} \\
		&\lesssim \left\| \sum_{|k| \sim M} \omega_k e^{2\pi i k x}  \right\|_{L^p}\,.
	\end{align*}
	Here we use the fact that the Fourier coefficients are supported on $|k| \sim M$, for which $\Phi_M(k) = 1$. We hence deduce \eqref{Mik}. By summing in $M>M_0$ and applying \eqref{sum} with $\delta$ sufficiently small, we obtain \eqref{big}. Therefore \eqref{3.13} holds for some $M>M_0$.
	
	To estimate the contribution for each dyadic $M$, we consider the subspace $S$ of $L^p$ given by $\mathrm{Span}_\C\{e^{2\pi i k x}  : |k| \sim M\}$. We want to construct a $\frac{1}{2}$-norming set, $\Xi$, contained in the unit sphere of $L^{p'}$ with the following properties.
	\begin{enumerate}
		\item $\max_{\varphi \in \Xi} |\langle g, \varphi\rangle| \geq \frac{1}{2}\|g\|_{L^p}$ for all $g \in S$.
		\item $\|\varphi\|_{L^2} \lesssim M^{1/2-1/p}$ for any $\varphi \in \Xi$.
		\item $\log |\Xi| \lesssim M$.
	\end{enumerate}
	To find this set, we apply Lemma \ref{norming} and take the orthogonal projection of $\Xi$ onto $S$. We obtain the first and third properties from Lemma \ref{norming}, and the second follows from Plancherel's theorem, H\"{o}lder's inequality, and the Hausdorff-Young inequality (applied to $p'$). Namely, for $\varphi \in \Xi$, we have
	\begin{align*}
		\|\varphi\|_{L^2} = \|\hat{\varphi}\|_{\ell^2} &\lesssim M^{1/2-1/p}\|\hat{\varphi}\|_{\ell^{p}} \\
		&\lesssim M^{1/2-1/p} \|\varphi\|_{L^{p'}} = M^{1/2-1/p}\,.
	\end{align*}
	Having constructed the set, we now estimate the norm. We choose $M>M_0$ satisfying \eqref{3.13}. Then
	\begin{align*}
		\sigma_M M \lambda < \left\| \sum_{|k| \sim M} \omega_ke^{2\pi i k x} \right\|_{L^p}  &\leq 2 \max_{\varphi \in \Xi} \left| \sum_{|k| \sim M} \omega_k \hat{\varphi}(k) \right| \\
		&= 2 \max_{\varphi \in \Xi} \left| \sum_{|k|\sim M} \omega_k \frac{\hat{\varphi}(k)}{\|\varphi\|_{L^2}} \right| \|\varphi\|_{L^2} \\
		&\lesssim 2 M^{1/2-1/p} \max_{\varphi \in \Xi} \left| \sum_{|k|\sim M} \omega_k \frac{\hat{\varphi}(k)}{\|\varphi\|_{L^2}} \right|\,,
	\end{align*} 
	where the first line uses property (1) of $\Xi$ and the final inequality follows from property (2) of $\Xi$. So
	\begin{equation}
	\sigma_M M^{1/2+1/p} \lambda \lesssim \max_{\varphi \in \Xi} \left| \sum_{|k|\sim M} \omega_k \frac{\hat{\varphi}(k)}{\|\varphi\|_{L^2}} \right|. \label{final}
	\end{equation}
	Let us take $(\sigma_M)_{M>M_0}$ satisfying \eqref{sum} to be of the form
	\begin{equation}
	\sigma_M \sim M^{-1/p} + (M_0/M)^{1/2}\,, \label{sigmaM}
	\end{equation}
	for a suitable choice of implied constant. For $M>M_0$, let $X_M$ denote the event \eqref{final}. Then
	\begin{align}
		& \mathbb{P}_\omega\left[ \left\|\sum_{k \in \Z}\ \frac{\omega_k}{\sqrt{\lambda_k}}e^{2\pi i k x} \right\|_{L^p} > \lambda, \left( \sum_{k \in \Z} \frac{|\omega_k|^2}{\lambda_k}\right)^{1/2} \leq B    \right] \\
		&\leq \mathbb{P}_\omega\left(\cup_{M>M_0} \, X_M \right) \label{1}
		\\
		&\leq \sum_{M>M_0} \sum_{\varphi \in \Xi} \mathbb{P}_\omega \left[\left| \sum_{|k| \sim M} \omega_k\frac{\hat{\varphi}(k)}{\|\varphi\|_{L^2}}\right| \gtrsim \sigma_M M^{1/2+1/p}\lambda \right] \label{3}
		\\
		& \lesssim \sum_{M>M_0} \sum_{\varphi \in \Xi} \exp\left(-cM^{1+2/p}\sigma_M^2\lambda^2\right) \label{4}
		\\
		& \lesssim \sum_{M>M_0} \exp\left(CM-cM^{1+2/p}\sigma_M^2\lambda^2\right) \label{5}
		\\
		&=\sum_{M>M_0} \exp\left(CM-c\left(M+M_0M^{2/p}+2M_0^{1/2}M^{1/2 + 1/p}\right)\lambda^2\right) \label{6} \\
		&\lesssim \sum_{M>M_0}\exp\left(-cM_0M^{2/p}\lambda^2\right) \label{7}
		\\
		\nonumber
		&\lesssim \exp\left(-cM_0^{1+2/p}\lambda^2\right)\,. 
	\end{align}
	Here, \eqref{1} follows from \eqref{final}. \eqref{3} follows from a union bound. \eqref{4} comes from applying Lemma \ref{Hoeff} with $X_i = \omega_i$ and $a_i = \hat{\varphi}(i)/\|\varphi\|_{L^2}$ (so that $Q \sim 1$ and $\|a\|_{\ell^2} \leq 1$ by Plancherel's theorem), and for \eqref{5}, we use property (3) of $\Xi$. \eqref{6} comes from \eqref{sigmaM}. We obtain \eqref{7} from the fact that $\lambda$ is large and noticing that the second term will give a factor less than one. The final inequality follows from the fact we have a geometric series with common ratio equal to $1-\zeta_{M_0}$, with $\zeta_{M_0}>0$. So we have shown \eqref{eq1}, which completes the proof.
\end{proof}

\section{Remarks about the cut-off function $f$}
\label{Appendix_cutoff_function}
In this appendix, we expand on Remark \ref{Remark_AppendixB} \eqref{Remark_Appendix_B_1} and \eqref{Remark_Appendix_B_2}. 
\subsection{Interaction Potentials of Positive Type}
\label{Appendix B1}
For a bounded, real-valued, even interaction potential $w$ of positive type (i.e.\ $\hat{w} \geq 0$ pointwise almost everywhere), we claim we can apply the methods used in the proof of \cite[Theorem 1.8]{FKSS17} to get the result of Theorem \ref{bounded_trace_class_final_convergence_theorem} for $\rho_\#$ defined without a truncation in $\mathcal{N}_{\sharp}$. To do this, we follow the convention from the two and three dimensional cases from \cite{FKSS17} and consider a non-normal ordered quantum interaction, namely
\begin{equation}
\label{nonnormal_quantum_interaction}
\mcW'_\tau := \frac{1}{2}\int dx\,dy\, \varphi^*_\tau(x)\varphi_\tau(x) w(x-y) \varphi^*_\tau(y)\varphi_\tau(y)\,,
\end{equation}
which we note is different to the convention adopted in the rest of the paper. We also define $H'_\tau := H_{\tau,0} + \mcW'_\tau$ in contrast to \eqref{quantum_free_ham}. Applying \eqref{commutation_relations}, we have
\begin{equation*}
\mcW'_\tau = \mcWt + \frac{1}{2\tau}w(0)\mcNt^2\,,
\end{equation*}
where we recall \eqref{quantum_interaction_defn} and \eqref{rescaled_particle_number_defn}. We consider this non-normal ordered interaction since $\mcW'_\tau$ acts on the $n{\mathrm{th}}$ sector of Fock space as multiplication by
\begin{equation}
\label{nth_kernel_nonnormal_interaction}
\frac{1}{2\tau^2} \sum_{i,j=1}^n w(x_i-x_j)\,.
\end{equation}
The key difference from \eqref{mcWt_hn} is \eqref{nth_kernel_nonnormal_interaction} includes the diagonal terms of the sum. The remark follows from showing that if $w$ is of positive type, $\eqref{nth_kernel_nonnormal_interaction} \geq 0$ almost everywhere, since we can apply Proposition \ref{Feynman-Kac_formula} as in the proof of \cite[Proposition 4.5]{FKSS17}. We can further reduce this to showing $\eqref{nth_kernel_nonnormal_interaction} \geq 0$ for $w \in C^\infty$ of positive type by taking $w^\eps := w * \varphi^\eps$ for a standard approximation to the identity $\varphi^\eps$ of positive type, since then $w^\eps \to w$ pointwise almost everywhere.

To see this, recall that for $g \in L^2$, Parseval's theorem implies
\begin{equation}
\label{AppendixB_positive_convolution}
\langle g,w*g \rangle \sim \sum_{k \in \Z} |\hat{g}(k)|^2\hat{w}(k) \geq 0
\end{equation}
since $w$ is of positive type. Taking $g^\eps \in C^\infty$ with $g^\eps \to \sum_{j=1}^n\delta(\cdot-x_j)$ weakly with respect to continuous functions, for $w \in C^\infty$ we have, by \eqref{AppendixB_positive_convolution} 
\begin{align*}
 0  \leq \langle g^\eps,w*g^\eps \rangle &\to \sum_{i,j=1}^n w(x_i-x_j)\,.
\end{align*} 
Letting $\eps \rightarrow 0$ then
yields $\eqref{nth_kernel_nonnormal_interaction} \geq 0$ for $w$ smooth of positive type.
\subsection{General $L^\infty$ interaction potentials}
\label{Appendix B2}
For a general bounded, even, real-valued interaction potential $w$ we show we could have used a Gaussian cut-off rather than a compactly supported one. Notice that since $w \in L^\infty$, there is some $c$ such that $w^c:= w+c \geq 0$ pointwise. Throughout this section, for an object $X_\#$, we use $X_\#^c$ to denote $X_\#$ defined using $w^c$ rather than $w$. Notice that
\begin{align}
	\label{AppendixB_quantum_shifted_interaction}
	\mcW_\tau^{\prime,c} &= \mcW'_\tau + \frac{c}{2}\,\mcNt^2\,, \\
	\label{AppendixB_classical_shifted_interaction}
	\mcW^c &= \mcW + \frac{c}{2}\,\mcN^2\,.
\end{align}
Applying an adapted form of \cite[Theorem 1.8]{FKSS17} for non-normal ordered interactions, we have
\begin{equation}
\label{general_interaction_state}
	\lim_{\tau \to \infty} \frac{\mathrm{Tr}\left(\Theta_\tau(\xi)e^{-H^{\prime,c}_\tau}\right)}{\mathrm{Tr}\left(e^{-H^{\prime,c}_\tau}\right)} = \frac{\int d\mu \, \Theta(\xi)e^{-H^c}}{\int d\mu \, e^{-H^c}}\,.
\end{equation}
We note that the adapted form of \cite[Theorem 1.8]{FKSS17} holds by applying the same proof, but using $\sum_{i,j=1}^n w^c(x_i-x_j) \geq 0$ instead of $\sum_{i,j=1, i \neq j}^n w^c(x_i-x_j) \geq 0$ in the proof of \cite[Proposition 4.5]{FKSS17}.

Rewriting \eqref{general_interaction_state} using \eqref{AppendixB_quantum_shifted_interaction} and \eqref{AppendixB_classical_shifted_interaction} gives
\begin{equation*}
	\lim_{\tau \to \infty} \frac{\mathrm{Tr}\left(\Theta_\tau(\xi)e^{-H'_\tau}e^{-\frac{c}{2} \, \mcNt^2}\right)}{\mathrm{Tr}\left(e^{-H'_\tau}e^{-\frac{c}{2}\,\mcNt^2}\right)} = \frac{\int d\mu \, \Theta(\xi)e^{-H}e^{-\frac{c}{2}\,\mcN^2}}{\int d\mu \, e^{-H}e^{-\frac{c}{2}\,\mcN^2}}\,.
\end{equation*}

\bigskip
\emph{Acknowledgements:}
The authors would like to thank Zied Ammari, J\"{u}rg Fr\"{o}hlich, Sebastian Herr, Antti Knowles, Benjamin Schlein, and Daniel Ueltschi for useful discussions during various stages of the project. They would also like to thank David Brydges, Trishen Gunaratnam, Mathieu Lewin, Phan Th\`{a}nh Nam, Nicolas Rougerie, Gigliola Staffilani, Nikolay Tzvetkov, and Oleg Zaboronski for their helpful comments.
A.R. is supported by the Warwick Mathematics Institute Centre for Doctoral Training, and gratefully acknowledges funding from the University of Warwick. V.S. acknowledges support of the EPSRC New Investigator Award grant EP/T027975/1. The authors thank the referees for their helpful feedback on the manuscript.

\end{document}